\documentclass[a4paper,fleqn,usenatbib]{mnras}

\usepackage{ae,aecompl}
\usepackage{graphicx}
\usepackage{float}
\usepackage[caption = false]{subfig}
\usepackage{color}
\usepackage{multirow}
\usepackage{times}
\usepackage{tikz}
\usetikzlibrary{patterns}
\usepackage{graphicx}	% Including figure files
\usepackage{amsmath}	% Advanced maths commands
\usepackage{amssymb}	% Extra maths symbols
\usepackage[normalem]{ulem}
\usepackage[T1]{fontenc}
\newcommand{\massUnit}[1]{$ h^{-1} \rm{ M_{\odot}}$}
\newcommand{\distUnit}[1]{$ h^{-1} \; \rm {Mpc}$}

\newcommand{\Nexus}{\textsc{NEXUS}}
\newcommand{\nexus}{\textsc{NEXUS+}}

\def\hmpc{~h^{-1}\rm Mpc}

\def\hMsun{~h^{-1}\rm M_\odot}

\newcommand{\revised}[1]{#1} 
\newcommand{\newrevised}[1]{#1} 

\title[Halo spin and shape alignments in the cosmic web]{The Cosmic Ballet: \\
spin and shape alignments of haloes in the cosmic web}

\author[P. Ganeshaiah Veena et al.]
{\parbox{\textwidth}{Punyakoti Ganeshaiah Veena$^{1,2}$\thanks{E-mail:punyakoti.gv@gmail.com},
Marius Cautun$^{3}$,
Rien van de Weygaert$^{1}$,\\
Elmo Tempel$^{2,4}$,   
Bernard J.~T. Jones$^{1}$,
Steven Rieder$^{5}$ and 
Carlos S. Frenk$^{3}$ \vspace{.4cm}}\\
$^{1}$Kapteyn Astronomical Institute, University of Groningen,PO Box 800, 9747 AD, Groningen, The Netherlands\\
$^{2}$Tartu Observatory, University of Tartu, Observatooriumi 1, 61602 T$\tilde{o}$ravere, Estonia \\
$^{3}$Institute for Computational Cosmology, Department of Physics, University of Durham, South Road, Durham, DH1 3LE, UK\\
$^{4}$Leibniz-Institut f$\ddot{u}$r Astrophysik Potsdam (AIP), An der Sternwarte 16, 14482 Potsdam, Germany\\
$^{5}$RIKEN Center for Computational Science, 7-1-26 Minatojima-minami-machi, Chuo-ku, Kobe, 650-0047, Hyogo, Japan
}
\pubyear{2018}

\begin{document}
\label{firstpage}
\pagerange{\pageref{firstpage}--\pageref{lastpage}}
\maketitle
%==========================
% -----ABSTRACT -------
%==========================
\begin{abstract}
	We investigate the alignment of haloes with the filaments of the cosmic web using an unprecedently large sample of dark matter haloes taken from the P-Millennium $\Lambda$CDM cosmological N-body simulation. We use the state-of-the-art NEXUS morphological formalism which, due to its multiscale nature, simultaneously identifies  structures at all scales. 
	We find strong and highly significant alignments, with both the major axis of haloes and their peculiar velocity tending to orient along the filament. However, the spin - filament alignment displays a more complex trend changing from preferentially parallel at low masses to preferentially perpendicular at high masses. 
This ``spin flip" occurs at an average mass of $5\times10^{11}~h^{-1}M_\odot$. This mass increases with increasing filament diameter, varying by more than an order of magnitude between the thinnest and thickest filament samples. We also find that the inner parts of haloes have a spin flip mass that is several times smaller than that of the halo as a whole.
These results confirm that recent accretion is responsible for the complex behaviour of the halo spin - filament alignment. Low-mass haloes mainly accrete mass along directions perpendicular to their host filament and thus their spins tend to be oriented along the filaments. In contrast, high-mass haloes mainly accrete along their host filaments and have their spins preferentially perpendicular to them. Furthermore, haloes located in thinner filaments are more likely to accrete along their host filaments than haloes of the same mass located in thicker filaments.
\end{abstract}
\begin{keywords}
	large-scale structure of Universe - galaxies: haloes - methods: numerical
\end{keywords}
%================================
% -----BODY OF THE PAPER-------
%================================
\section{Introduction}
Starting from almost uniform initial conditions, the Universe has evolved over billions of years to contain a wealth of structure, from small-scale virialized objects, such as haloes and galaxies, to tens-of-Megaparsec-sized structures, such as super-clusters and filaments \citep{Peebles1980,Oort1983,Springel2006,Frenk2012, Tempel2014, Tully2014}. All these are embedded in the so-called cosmic web, a wispy weblike spatial arrangement consisting of dense compact clusters, elongated filaments, and sheetlike walls, amidst large near-empty void regions \citep{bond1996, vdw2008}. \revised{This} pattern is marked by prominent anisotropic features, a distinct multiscale character, a complex spatial connectivity
and a distinct asymmetry between voids and overdense regions. 
\revised{The large-scale web} is shaped by the large-scale tidal field, which itself is generated by the
inhomogeneous distribution of matter. Within this context, the cosmic web is the most salient manifestation of the anisotropic nature
of gravitational collapse, and marks the transition from the primordial (Gaussian) random field to highly nonlinear structures that have fully
collapsed into haloes and galaxies. 

The same tidal field that shapes the cosmic web is also the source of angular momentum build-up in collapsing haloes and galaxies. This is neatly encapsulated by Tidal Torque Theory (TTT), which explain how in the linear stages of evolution the tidal field torques the non-spherical collapsing protohaloes to generate a net rotation or spin \citep{hoyle1949, peebles1969, doroshkevich1970, white1984}. Specifically, this occurs due to a differential alignment between the inertia tensor of the protohalo and the local gravitational tidal tensor. TTT posits a direct correlation between halo properties such as angular momentum, shape and the large-scale tidal field at their location (see \citealt{schaffer2009} for a review). For example, linear TTT predicts that the halo spin is preferentially aligned with the direction of secondary collapse (\cite{Lee2001}, but see \cite{joneswey2009}), and thus the spin is perpendicular on the direction of slowest collapse, which corresponds to the filament ridge \citep{efstathiou1979,barnes1987,Heavens1988,Lee2001,porciani2002,porciani2002TTT2,Lee2004}. This alignment is mostly imprinted at the time of turn-around, when the protohaloes are the largest, and is expected to be preserved during the subsequent non-linear collapse of the protohaloes into virialized objects. 

Large cosmological simulations have shown that the alignments of halo shape and spin with their surrounding mass distribution are not as straightforward as predicted \revised{by the simplified TTT framework described above}. The correlations present in the linear phase of structure formation are preserved in the case of halo shapes, which are strongly oriented along the filament in which the haloes are embedded, with the alignment strength increasing with halo mass \citep{Altay2006,aragon2007,Brunino2007,hahn2007}. In contrast, the spin of haloes shows a more complex alignment with their host filament. This was first pointed out by \citet{aragon2007}, and shortly thereafter by \citet{hahn2007}, which have shown that the spin - filament alignment is mass-dependent, with low- and high-mass haloes having a preferential parallel and perpendicular alignment, respectively. 
This result has since been reproduced in multiple cosmological simulations with and without baryons \citep{hahn2010,codis2012,libeskind2013,trowland2013,Dubois2014,romero2014,Wang2017a}. The alignment has been confirmed by observational studies, most outstandingly so in the finding by \citet{tempel2013} that massive elliptical galaxies tend to have their spin perpendicular to their host filaments while the spin of less massive
bright spirals has a tendency to lie parallel to their host filaments \citep[see also][]{jones2010,elmonoam2013,Zhang2013,Zhang2015,Hirv2017}. 
%\revised{\citet{codis2015} \sout{have explained how the halo spin - filament alignment can be understood in the TTT framework. The key is that filaments form only in certain large scale tidal field configurations, in which the alignment between the inertia tensor and the tidal field follows a particular distribution that is different from the general expectation. This explains, at least qualitatively, the dichotomy in spin-filament alignment between low- and high-mass haloes. }}
The transition mass from halo spins preferentially perpendicular to preferentially parallel to their host haloes is known as the \emph{spin flip} mass. While most studies agree on the existence of such a transition mass, they report highly disparate values for the spin flip mass that spread over more than an order of magnitude in halo mass, from ${\sim}0.5$ to ${\sim}5\times10^{12}$ \massUnit{}. Furthermore, the spin flip mass varies with the smoothing scales used to identify the large-scale filaments, being higher for larger smoothing scales \citep{codis2012,aragon2014,romero2014}, and decreases at higher redshifts \citep{codis2012,Wang2018a}. It suggests that the mechanisms responsible for the tendency of low-mass haloes to have their spins oriented along their host filaments are complex, being both time and environment dependent.

Previous works have posited a diverse set of explanations for the spin flip phenomenon, with most responsible processes having to do with the nature of halo late-time mass accretion, the so-called secondary accretion \citep{bertschinger1985}. \revised{A theoretical solution is provided by \citet{codis2015}, who explain the dichotomy in spin-filament alignment between low- and high-mass haloes within the TTT framework,.
%who have shown that by including the cosmic web environmental effects in the classical TTT framework, the phenomenon of spin flip results as a natural outcome. 
The key is that filaments form only in certain large scale tidal field configurations, in which the alignment between the inertia tensor and the tidal field follows a particular distribution that is different from the general expectation. %This explains, at least qualitatively, the dichotomy in spin-filament alignment between low- and high-mass haloes. 
\citeauthor{codis2015} and \citet{Laigle2015} have suggested also that this is due to the vorticity distribution inside filaments (for galaxies, see \citealt{Pichon2011}). They have claimed that the filament cross-section can be split into four quadrants, each with an opposite vorticity sign. Low-mass haloes typically reside in one of the four quadrants and thus acquire a spin along the filament, while high-mass haloes overlap multiple vorticity quadrants and acquire a spin that is preferentially perpendicular on their host filament.}  \citet{Welker2014} have shown that massive galaxies tend to have their spin perpendicular to their filament due to an excess of mergers along the filament direction, while low-mass galaxies tend to be aligned along their filament due to having undergone none or many fewer mergers. However, \citet{bett2012,Bett2016} have shown that more than $75\%$ of changes in halo spins are due to accretion of small substructures or flyby encounters, and not due to major mergers.
% \revised{ \citet{Laigle2015} \sout{and} \citet{codis2015} \sout{have put forward another explanation, that the halo spin - filament alignment and its mass dependence is due to the vorticity distribution inside filaments. They have claimed that the filament cross-section can be split into four quadrants, each with an opposite vorticity sign. Low-mass haloes typically reside in one of the four quadrants and thus acquire a spin along the filament, while high-mass haloes overlap multiple vorticity quadrants and acquire a spin that is preferentially perpendicular on their host filament.} }
On the other hand, \citet{Wang2017a,Wang2018a} have explained the spin - filament alignment in terms of the formation time of haloes and their migration time from sheets into filaments. Low-mass haloes accrete most of their mass at high redshift, while residing in sheets, while high-mass objects undergo most of their growth at low redshift, when they are embedded in filaments.  

In this study, we carry out a systematic analysis of the alignment between the spin and shape of haloes and the orientation of the filaments in which the haloes reside. We employ one of the largest cosmological simulations available, P-Millennium, which is characterized by a large volume and very high mass resolution, with the large dynamic range being critical for our goal of understanding how the large-scale cosmic web influences small-scale phenomena, such as spin and shape orientations of haloes. We identify the cosmic web using the state-of-the-art \Nexus{} technique, which employs a multiscale formalism to identify in one go both prominent and tenuous filaments \citep[][see \citealt{libeskind2018} for a comparison to other web detection methods]{cautun2013}. We employ two \Nexus{} variants, \nexus{} and \Nexus{}\_velshear, which identify the web on the basis of the density and the velocity shear fields, respectively. These two \Nexus{} variants show the largest difference between their identified filamentary network \citep{cautun2014} and comparing the halo - filament alignments between the two method reveals key details about the processes behind the halo - filament alignments and their dependence on halo mass.  

Our analysis involves two major new themes which have not been studied in the literature and which we show to be indispensable for understanding the halo - filament alignments. First, we study the properties of the entire halo as well as those corresponding to different inner radial cuts. The latter is highly relevant since: i) galaxies are very strongly aligned with the inner region of the halo, and only poorly with the full halo \citep{Bailin2005,Tenneti2014,Velliscig2015,Shao2016,Chisari2017}  
, and ii) recent accretion is mainly deposited in the outer regions of the halo \citep{salvador-sole1998, wechsler2002, tasitsiomi2004,wang2011} and thus the alignments of the inner regions trace the alignment of the full halo at high redshift. The second novel features involves studying the halo spin - filament alignment as a function of filament properties to find that the spin flip mass shows a very strong dependence on filament thickness.

\begin{figure*}
   \mbox{\hskip -0.4cm \includegraphics[width=1.06\textwidth]{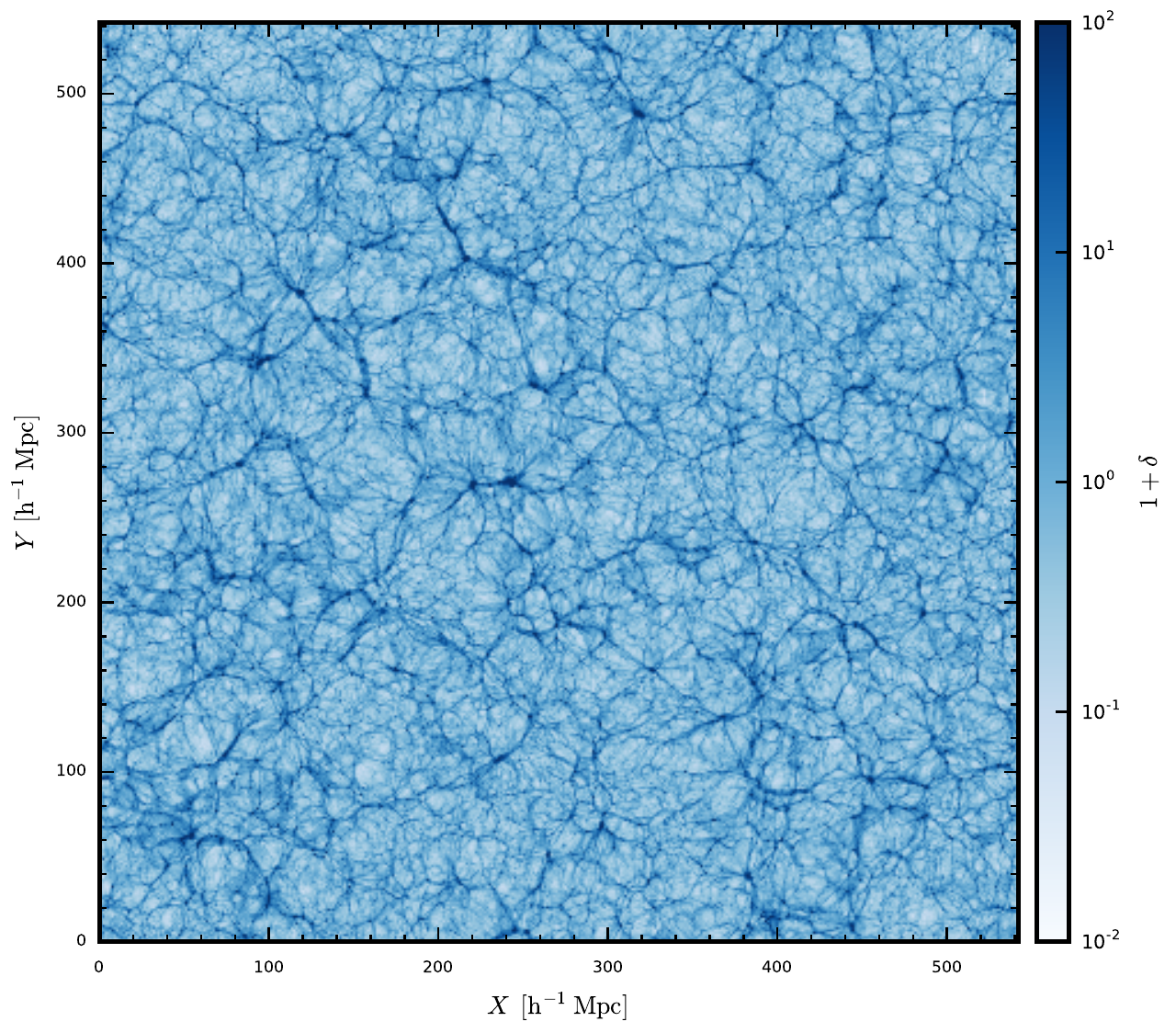}}
	\vspace{-.7cm}
    \caption{A $2\hmpc$ slice trough the $z=0$ density field of the P-Millennium simulation. The width and height of the figure corresponds to the side length of the simulation. The colour bar indicates the density contrast, $1 + \delta$.
	}
    \label{fig:density_field}
\end{figure*}

The layout of the paper is as follows: \autoref{sec:filament_population} introduces the cosmological simulations and the \Nexus{} formalism used to identify the cosmic web; \autoref{sec:halo_population} describes the halo catalogues, how we calculate halo spins and shapes, and presents a detailed comparison of the halo population in filaments between our two web finders;  \autoref{sec:spin_alignment} presents the main results regarding the halo spin - filament alignment;  \autoref{sec:shape_alignment} studies the alignment between the shape of haloes and their host filaments; in \autoref{sec:accretion} we present a detailed discussion on how secondary accretion is likely to be the main process that shapes the halo spin - filament alignment; and we end with a summary and discussion of our main results in ~\autoref{sec:discussion}. 

\section{Filament Population}
\label{sec:filament_population}
Our analysis is based on a high resolution simulation with an unsurpassed dynamic range, Planck-Millennium, which we introduce in this section. Here, we also describe the filament identification procedure, which is based on two different versions of the MMF/\Nexus{} cosmic web detection algorithm: one starting from the density field and the other from the velocity shear field. By comparing the two filament populations, we hope to identify supplementary information on the processes that affect the alignment of halo angular momentum with the large scale structure. 
 
\subsection{Simulation}
For this study we used the Planck-Millennium high resolution simulation (hereafter P-Millennium; \citealt{McCullagh2017,Baugh2018}), 
which is a dark matter only N-body simulation of a standard $\Lambda$CDM cosmology. It traces structure formation in a periodic box of $542.16\hmpc$ side length using $5040^3 $ dark matter particles, each having a mass of $1.061 \times 10^{8}~{h}^{-1}\rm{M}_{\odot} $. The cosmological parameters of the simulation are those obtained from the latest Planck survey results \citep{planck2014}: the density parameters are $\Omega_{\Lambda} =0.693 $, $ \Omega_{\rm M} =  0.307$, %$\Omega_{\rm b} = 0.0455$, 
the amplitude of the density fluctuations is $\sigma_8=0.8288$, and the Hubble parameter is $ h= 0.6777$, where $h = H_{0}/100$ $ \rm{ km \; s^{-1} Mpc^{-1}} $ and $H_0$ is the Hubble's constant at present day.
In the analysis presented here we limit ourselves to the mass distribution at the current epoch, $z=0$.

Due to its large dynamic range and large volume, the P-Millennium simulation is optimally suited for investigating the issue of angular momentum acquisition and the relation between
spin and web-like environment over a large range of halo masses. P-Millennium simulates the formation nearly 7.5 million well resolved haloes over three orders of magnitude in halo mass, which is critical for the success of this work.
This is especially the case for the alignment between halo spin and filament orientation, which is a subtle effect \citep[e.g. see][]{aragon2007, libeskind2013}, and for robustly characterising the dependence of this alignment on halo mass, which is one of the principal aspects addressed in this paper. Besides its importance for identifying the subtle dynamical effects underlying the spin transition, the large  volume of P-Millennium allows us to fully take into account the large-scale tidal forces responsible for the generation of halo angular momentum and for the formation of the cosmic web.

A visual illustration of the mass distribution in the P-Millennium simulation is shown in \autoref{fig:density_field}. It shows a slice of $2\hmpc$ width through the entire simulation box, with the white-blue colour scheme representing the density contrast,
\begin{equation}
1+\delta({\mathbf{x}},t)\,=\,\frac{\rho({\mathbf{x}})}{\rho_u}\,,
\end{equation}
where $\rho({\mathbf{x}})$ and $\rho_u$ denote the local and background mean density. Clearly visible is the intricate structure of the cosmic web,
with its visual appearance dominated by elongated medium to high density filaments and low-density voids. The image
illustrates some of the characteristic properties of the cosmic web, such as the complex and pervasive connectivity of the filamentary network. We also recognize the multiscale structure of the web: the dominant thick filaments, which are often found in high density regions bridging the cluster mass haloes and the thin, tenuous filamentary tendrils that branch out from the thick ones. These thin filaments typically have lower densities and pervade the low-density void regions.
Note that in a two-dimensional slice like the one shown in \autoref{fig:density_field}, it is difficult to make a clear distinction between
filaments and cross-sections through planar walls \citep{cautun2014}. However, the more moderate density of the walls means that they
would not correspond to the most prominent high-density ridges seen in the slice. 

\subsection{Filament detection}
\label{detecting_filaments}
We use the MMF/NEXUS methodology for identifying filaments in the P-Millennium simulation. The MMF/NEXUS multiscale morphology technique \citep[][]{aragon2007MMF,cautun2013} performs the morphological identification of the cosmic
web using a \textit{Scale-Space formalism} that ensures the detection of structures present at all scales. The formalism 
consists of a fully adaptive framework for classifying the matter distribution on the basis of local variations in the density, velocity or gravity fields, which are encoded in the Hessian matrix. Subsequently, a set of morphological filters is used to classify the spatial 
distribution of matter into three basic components: the nodes, filaments and walls of the cosmic web. The outcome of the identification 
procedure is \revised{a set of diverse and complex} cosmic web components, from the prominent features present in overdense regions 
to the tenuous networks pervading the cosmic voids.

The \Nexus{} version of the MMF/\Nexus{} formalism \citep{cautun2013,cautun2014} builds upon the original Multiscale Morphology Filter
\citep{aragon2007,aragon2007MMF} algorithm and was developed with the goal of obtaining a more robust and more physically motivated environment classification method.
The full \Nexus{} suite of cosmic web identifiers \citep[see][]{cautun2013} includes options for a range of cosmic web tracers, such as the raw
density, the logarithmic density, the velocity divergence, the velocity shear and the tidal force fields. \Nexus{} has incorporated these 
options in a versatile code for the analysis of cosmic web structure and dynamics following the realization that they represent key physical aspects that shape the cosmic mass distribution. 

The goal of our analysis of halo - filaments alignments is to understand the role of large scale
tidal forces in the acquisition of angular momentum in haloes. The dominant tidal field effects and the large scale peculiar velocity flows are
expected to be related to the most prominent web-like structures. This motivates us to employ two methods for identifying the cosmic web filaments, on the basis of their signature in the shear or velocity fields. By contrasting the alignments of the halo spin with the two filament populations, we seek to disentangle the contribution of local small-scale forces from those of larger-scale ones.

\begin{figure*}
    \includegraphics[width=\textwidth]{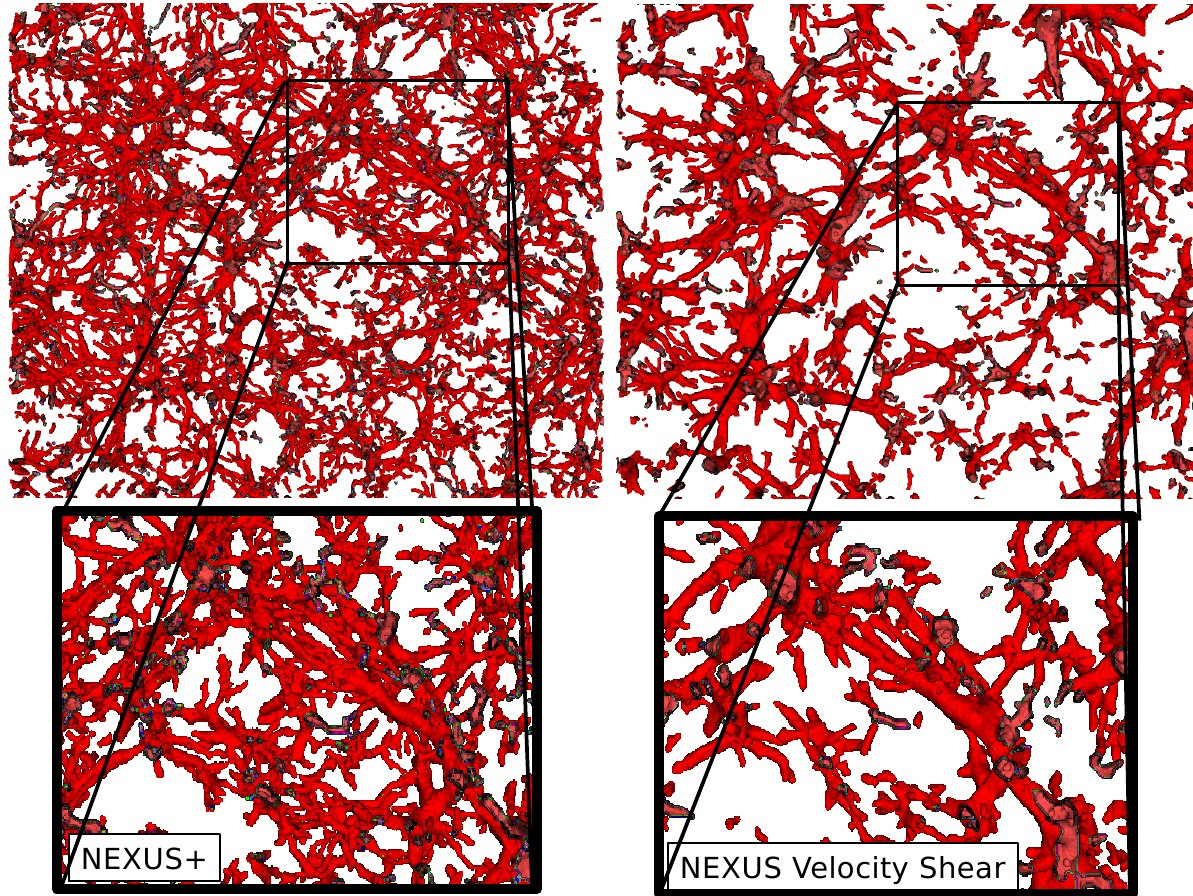}
    \vskip -0.0cm
	\caption{ \textit{Left panels}: Filaments detected by the \nexus{} method, which identifies filaments in the density field. \textit{Right panels}: For the same volume as in the left panels, filaments detected by the NEXUS\_velshear method, which identifies filaments in the velocity shear field. The top row shows a $20\hmpc$ slice of $300 \times 300 \; (h^{-1} \rm{Mpc})^2$ size across. The bottom row shows a zoom-in into a smaller region of this slice. NEXUS\_velshear identifies typically only the thick filaments, whereas \nexus{} identifies even the thin  and tenuous tendril like filaments in low-density regions. }
	\label{fig:filaments_illustration}
    \vskip -0.1cm
\end{figure*}

\subsubsection{MMF/NEXUS}
A major advantage of the MMF/NEXUS formalism is that it simultaneously pays heed to two crucial 
aspects of the web-like cosmic mass distribution: the morphological identity of structures and the multiscale character of the distribution. The first aspect is recovered by calculating the local Hessian matrix, which reveals the existence and identity of morphological web components. The second, equally important, aspect uses a scale-space analysis to uncover the multiscale nature of the web, which is 
a manifestation of the hierarchical evolution of cosmic structure formation. 

The scale-space representation of a data set consists of a sequence of copies of the data at different resolutions 
\citep{florack1992,lindeberg1998}. A feature searching algorithm is applied to all of these copies, and the features are extracted in a 
scale independent manner by suitably combining the information from all the copies. A prominent application of scale-space analysis involves 
the detection of the web of blood vessels in a medical image \citep{sato1998,li2003}, which bears a striking similarity to the structural patterns seen on 
Megaparsec scales. The MMF formalism has translated, extended and optimized the scale-space methodology to 
identify the principal morphological elements in the cosmic mass and galaxy distribution.  

The outcome of the MMF/\Nexus{} procedure is a volume-filling field which specifies at each point the local morphological signature: node, filament, wall or void. The MMF/\Nexus{} methods perform the environment detection by applying their formalism first to
nodes, then to filaments and finally to walls. Each volume element is assigned a single environment characteristic by requiring that filament regions 
cannot be nodes and that wall regions cannot be either nodes or filaments. The remaining regions are classified as voids.

The basic setup of MMF/\Nexus{} is to define a four-dimensional scale-space representation of the input tracer
field $f(\mathbf{x})$. In nearly all implementations this is achieved by means of a Gaussian filtering of $f(\mathbf{x})$ over a set of scales $[R_0,R_1,...,R_N]$. 
\begin{equation}
    f_{R_n}(\mathbf{x}) = \int \frac{{\rm d}^3k}{(2\pi)^3} e^{-k^2R_n^2/2} \hat{f}(\mathbf{k})  e^{i\mathbf{k}\cdot\mathbf{x}} ,
    \label{eq:filtered_field}
\end{equation}
\noindent where $\hat{f}(\mathbf{k})$ is the Fourier transform of the input field $f(\mathbf{x})$. 

Subsequently, the Hessian, $H_{ij,R_n}(\mathbf{x})$, of the filtered field is calculated via
\begin{eqnarray}
    H_{ij,R_n}(\mathbf{x})&\,=\,&R_n^2 \; \frac{\partial^2f_{R_n}(\mathbf{x})}{\partial x_i\partial x_j}\,,
    \label{eq:hessian_general}
\end{eqnarray} 
where the $R_n^2$ term is as a renormalization factor that has to do with the multiscale nature of the algorithm. When expressed in Fourier space, the Hessian becomes
\begin{eqnarray}
    \hat{H}_{ij,R_n}(\mathbf{k})&\,=\,& -k_ik_j R_n^2 \hat{f}(\mathbf{k}) e^{-k^2R_n^2/2}\,.
    \label{eq:hessian_general_Fourier}
\end{eqnarray} 

While in principle there are an infinite number of scales in the scale-space formalism, in practice 
our implementation uses a finite number of filter scales, restricted to the range of $[0.5,4.0]\hmpc$. This range has been predicated on
the expected relevance of filaments for understanding the properties of the haloes in our sample, which have masses in
the range $5 \times 10^{10}$ to $ 1 \times 10^{15} h^{-1}M_{\odot}$ (see next section). The upper filter scale of $4\hmpc$ allows the identification of the most massive filaments, while the lower filter scale allows for the detections of thin and tenuous filaments that host the  occasional isolated low-mass haloes. 

The morphological signature is contained in the local geometry as specified 
by the eigenvalues of the Hessian matrix, $h_1 \le h_2 \le h_3$. The eigenvalues are used to assign to every point, $\mathbf{x}$, a node, filament and wall characteristics which are determined by a set of morphology filter functions \citep[see][]{aragon2007,cautun2013}.
The morphology filter operation consists of assigning to each volume element and at each filter scale an environment signature, $\mathcal{S}_{R_n}(\mathbf{x})$. Subsequently, for each point, the environmental signatures 
calculated for each filter scale are combined to obtain a scale independent signature, $\mathcal{S}(\mathbf{x})$, which is defined as the maximum signature over all scales,
\begin{equation}
    \mathcal{S}(\mathbf{x}) = \max_{\rmn{levels\;}n} \mathcal{S}_{R_n}(\mathbf{x})\,.
    \label{eq:total_response}
\end{equation} 
The final step in the MMF/\Nexus{} procedure involves the use of criteria to find the threshold signature that identifies 
valid structures. Signature values larger than the threshold correspond to real structures while the rest 
are spurious detections. 
For nodes, the threshold is given by the requirement that at least half of the nodes should be virialized. For
filaments and walls, the threshold is determined on the basis of
the change in filament and wall mass as a function of signature.
The peak of the mass variation with signature delineates the most
prominent filamentary and wall features of the cosmic web \newrevised{(for more details and for a study of different threshold values for the environment signature see \citealt{cautun2013})}.
%The various implementations of the MMF/NEXUS technique can differ in the
%definition of the detection threshold. 

\subsubsection{\nexus{} and \Nexus{} velocity shear}
\label{sec:filament}
In our study, we use two \Nexus{} methods for identifying filament populations. The first, the \nexus{} algorithm,
is based on the local geometry of the density field. The strongly non-Gaussian nature of the non linearly evolved density field is marked by density ranges over many orders of magnitude. Simply applying a Gaussian smoothing can wash out the anisotropic nature of the matter distribution, especially close to high density peaks. This can be alleviated by applying a Log-Gaussian filter \citep{cautun2013}, which consists of three steps: (1) calculate the density logarithm, $\log\left(1+\delta(\mathbf{x})\right)$, (2) apply a Gaussian smoothing to $\log\left(1+\delta(\mathbf{x})\right)$, and (3) calculate the smoothed overdensity, $\delta_{\rm smooth}(\mathbf{x})$, from the smoothed density logarithm. Subsequently, \nexus{} calculates the Hessian matrix of the Log-Gaussian smoothed density field using Eq.~\eqref{eq:hessian_general}.
The Hessian eigenvalues, $\chi_{1,+}\le \chi_{2,+} \le \chi_{3,+}$, and eigenvectors, $\mathbf{e}_{i,+}$, determine the local shape and directions of the mass distribution. For example, a filamentary feature corresponds to $\chi_{1,+}<0$, $\chi_{2,+}<0$ and $|\chi_{1,+}|\simeq |\chi_{2,+}| \gg|\chi_{3,+}|$. The orientation of the filament is indicated by the eigenvector $\mathbf{e}_{3,+}$, while the sectional plane is defined by the eigenvectors $\mathbf{e}_{1,+}$ and $\mathbf{e}_{2,+}$. \revised{See the top panel of \autoref{fig:haloes_in_web_zoomIn} for a visual illustration of the filament orientation.}

\bigskip
The second method, \Nexus{}$\_$velshear, identifies the cosmic web through its dynamical signature, that is using the shear of the velocity flow induced by the gravitational forces that drive the growth of cosmic structure. The velocity shear is the symmetric part of the velocity gradient\footnote{Sometimes the velocity shear is defined
as the traceless symmetric part of the velocity gradient. Here, we include the divergence part of the velocity flow that
indicates the expansion or contraction of a mass element.}, with the $ij$ component defined as:
\begin{equation} 
    \sigma_{ij}(\mathbf{x}) = \frac{1}{2H} \left( \frac{\partial v_j}{\partial x_i} + \frac{\partial v_i}{\partial x_i}\right) \,,
    \label{eq:velocity_shear}
\end{equation}
where $v_i$ is the $i$ component of the velocity. In this definition, the velocity shear is normalized by the Hubble constant, $H$. To keep a close parallel to the cosmic web definition based on the density field, we apply the \Nexus{} formalism to the negative velocity shear, i.e. to $-\sigma_{ij}(\mathbf{x})$. This is motivated by linear theory, where the velocity shear is determined by the linear velocity growth factor times the negative gravitational tidal field. 

The morphological identity and the principal directions at a given location are determined by the eigenvalues, $\chi_{1,\sigma}\le \chi_{2,\sigma} \le \chi_{3,\sigma}$, and the eigenvectors, $\mathbf{e}_{i,\sigma}$, of the Hessian matrix calculated from the negative velocity shear. Similarly to \nexus{}, a filament is marked by $\chi_{1,\sigma}<0$, $\chi_{2,\sigma}<0$ and $|\chi_{1,\sigma}| \simeq |\chi_{2,\sigma}| \gg|\chi_{3,\sigma}|$, that is 
contraction along the first two directions and small contraction or dilation along the third direction. The filament orientation is given by the
third eigenvector of the shear field, $\mathbf{e}_{3,\sigma}$. 

In this sense, \Nexus{}\_velshear follows the same cosmic web classification philosophy as the (monoscale) V-web algorithm \citep{hoffman2012,libeskind2018}. The crucial difference between the two is that \Nexus{}$\_$velshear takes into
account the multiscale nature of the velocity field.

\subsection{Density- versus shear-based filaments}
There are several intriguing  differences in filament populations identified by \nexus{} and \Nexus{}$\_$velshear. Both procedures identify the most prominent
and dynamically dominant arteries of the cosmic web. These massive filaments, with diameters of the order of $5\hmpc$, may extend over vast
lengths, sometimes over tens of Megaparsec. They are the main transport channels in the large scale universe, along which matter, gas and
galaxies flow towards higher density mass concentrations. As such, they can nearly always be identified with pairs of massive and compact clusters, whose
strong tidal forces give rise to very prominent and massive filaments \citep{bond1996,colberg2005,vdw2008,bos2016}. They are nearly always
located on the boundaries of large voids. These filaments have a dominant contribution to the large scale tidal and velocity field \citep{rieder2018}, with their dynamical imprint being recognizable as a distinct shear pattern in the velocity flow.

The contrast between \Nexus{}$\_$velshear and \nexus{}, described in detail in \citet{cautun2013,cautun2014}, is illustrated in \autoref{fig:filaments_illustration}, which compares the two filamentary networks in a slice of $20\hmpc$ thickness and of $300\times300 ~ (h^{-1} \rm{Mpc})^2$ in area.
While the prominent and massive filaments are identified by both methods, \nexus{} manages to identify many more thin filamentary structures that illustrate the multiscale character of the cosmic filamentary network.

A second major difference between the two web finders is due to the non-linear velocity shear field having a larger scale coherence (i.e. being more non-localized) than the density field. This is due to the difference in the non-linear power spectra between velocity shear and density, with the former decreasing faster on small scales \citep{bertjain1994,jainbert1994,bondmyers1996,weygaert2002,romano2007}. 
Gravity, and hence tidal fields, are integrals over the density field. Hence they also manifest themselves at a distance from the source (the density fluctations) that generated them. Shear, as with the velocity field itself, is similar: it results from the action of gravity (the tidal field) over time.  Hence, while you are outside the generating source, you still see the imprint of the tidal field on the velocity field. \footnote{It is precisely this fact which is central to using the gravitational lensing shear field as a tracer of the source. And thus we need not be amazed it is also seen in the \Nexus$\_$velshear filament results: they are thicker than the equivalent density identified filaments.}

For tides, and shear, this means you can have the signature for a filament or a node while far removed from the object, even way into the voids. Which is indeed what you see.   We need not be amazed that it is also seen in the \Nexus$\_$velshear filament results: they are thicker than the corresponding filaments identified from the density field.
Because of this, the \Nexus$\_$velshear filaments are typically thicker than their \nexus{} counterparts, and thus the \Nexus$\_$velshear filaments tend to include matter and haloes in the immediate vicinity that would visually be more likely to be identified as part of the wall or void regions surrounding the \nexus{} filaments.

\begin{figure*}
  	\vskip -0.0truecm
  	\centering
  	\begin{tabular}{ @{} c c @{} }
  	(a) The entire distribution of haloes. &
  	(b) \parbox[c]{.8\columnwidth}{ Haloes assigned to filaments by both NEXUS+ and NEXUS$\_$velshear. } \\[.1cm]
  	\includegraphics[height=\columnwidth]{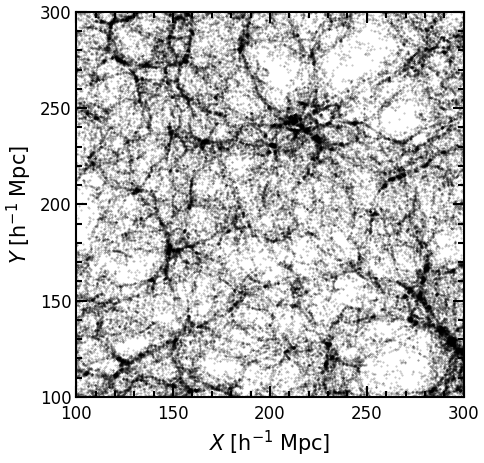} &
  	\includegraphics[height=\columnwidth]{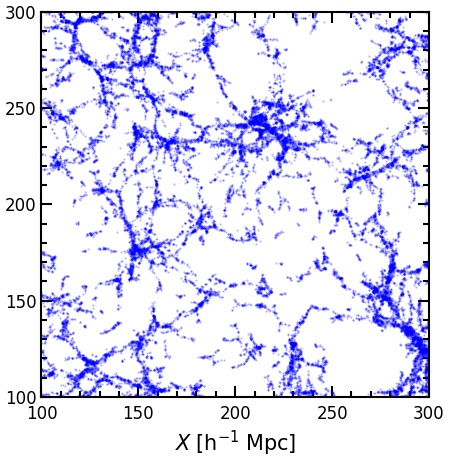} \\[.3cm]
  	(c) Haloes assigned to filaments by NEXUS+. &
  	(d) Haloes assigned to filaments by NEXUS$\_$velshear. \\[.1cm]
  	\includegraphics[height=\columnwidth]{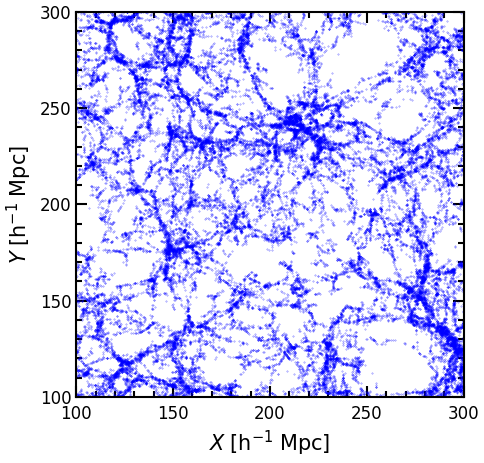} &
  	\includegraphics[height=\columnwidth]{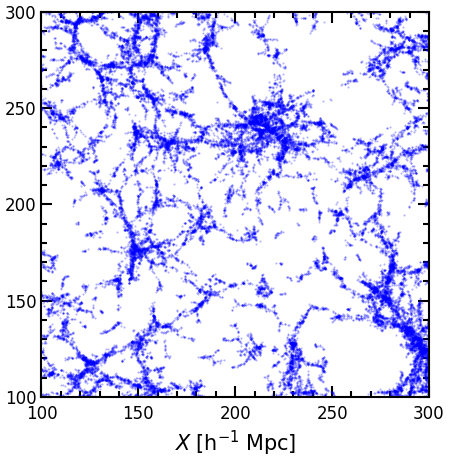}
  	\end{tabular}

	\vskip -0.3cm
	\caption{The distribution of haloes in a $20\hmpc$ slice of the P-Millennium simulation. Each dot represents a halo more massive than $3.2\times 10^{10}~$\massUnit{}~. It shows: all the haloes (top-left panel), the haloes residing in \nexus{} filaments (bottom-left) and the haloes residing in NEXUS$\_$velshear filaments (bottom-right). The haloes classified as residing in filaments by both methods are shown in the top-right panel.
}
	\label{fig:haloes_in_web_zoomOut}
\end{figure*}

\begin{figure*}
  	\vskip -0.0truecm
  	\centering
  	\begin{tabular}{ @{} c c @{} }
  	(a) Haloes assigned to filaments by NEXUS+. &
  	(b) Haloes assigned to filaments by NEXUS$\_$velshear. \\[.1cm]
  	\includegraphics[height=\columnwidth]{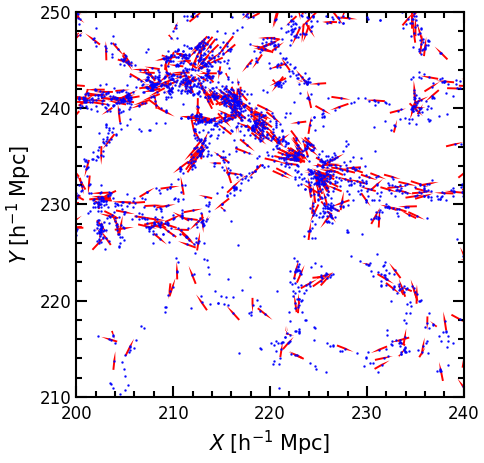} &
  	\includegraphics[height=\columnwidth]{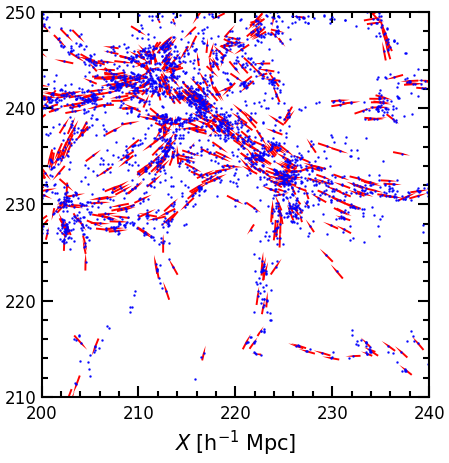} \\[.3cm]
  	(c) \parbox[c]{.8\columnwidth}{ Haloes assigned to filaments by NEXUS+, but not by NEXUS$\_$velshear. } &
  	(d) \parbox[c]{.8\columnwidth}{ Haloes assigned to filaments by NEXUS$\_$velshear, but not by NEXUS+. } \\[.1cm]
  	\includegraphics[height=\columnwidth]{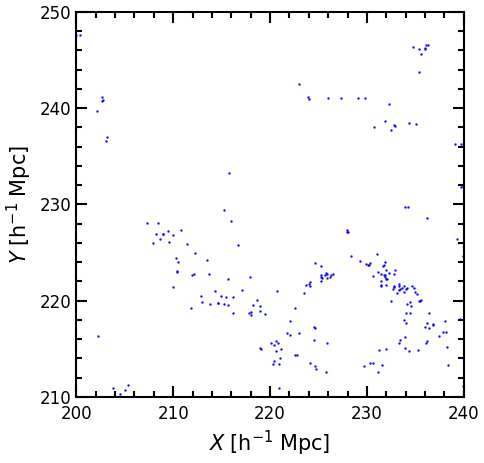} &
  	\includegraphics[height=\columnwidth]{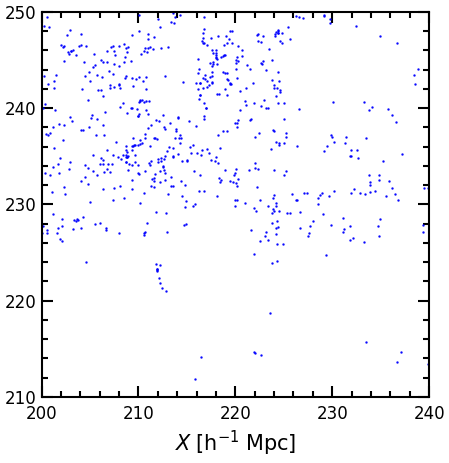}
  	\end{tabular}
	\vskip -0.3cm
	\caption{ Comparison of haloes assigned to filaments by NEXUS+ and NEXUS$\_$velshear. It shows a subregion of the volume shown in \autoref{fig:haloes_in_web_zoomOut} selected to enclose a massive filament. The thickness of the slice is $10\hmpc$.
%\sout{\revised{which is the same as in \autoref{fig:haloes_in_web_zoomOut}}}.
The four panels show: all NEXUS+ filament haloes (top-left), all NEXUS$\_$velshear filament haloes (top-right), haloes assigned to filaments \textit{only} by NEXUS+ (bottom-left) and haloes assigned to filaments \textit{only} by NEXUS$\_$velshear (bottom-right). \revised{The red lines in the top two panels depict the orientation of the NEXUS+ and NEXUS$\_$velshear filaments. The filament orientation is shown at the position of a random sample of 20\% of the haloes in the slice.}
    The contrast between both methods is substantial: NEXUS+ traces small filaments and tendrils whose minor dynamical impact eludes detection by NEXUS$\_$velshear. Furthermore, the prominent filaments detected by NEXUS$\_$velshear are substantially thicker than their NEXUS+ counterparts.
}
\label{fig:haloes_in_web_zoomIn}
\end{figure*}

\bigskip
An even more detailed and insightful illustration of the differences between the \nexus{} and \Nexus{}$\_$velshear filamentary networks is provided by studying the halo distribution. \autoref{fig:haloes_in_web_zoomOut} and \autoref{fig:haloes_in_web_zoomIn} depict the spatial distribution of haloes assigned to filaments by the two methods.
The overall impression is one of \nexus{} identifying a sharper outline of the cosmic web, while it includes a wide spectrum of small-scale
filamentary features that are not seen in the \Nexus{}$\_$velshear web-like network.
While \Nexus{}$\_$velshear identifies the massive filamentary arteries, it does not recover the small-scale tendrils branching out from these dominant structures or the complex network of tenuous filaments in
low-density regions. The large dynamic range of the \nexus{} procedure, however, does recognize and identify these small filaments.
On the other hand, the prominent \Nexus{}$\_$velshear  
filaments have a considerable number of haloes assigned to them that lie in the dynamical influence region of the filaments but that may in fact
be located in low density boundary regions. As a result, the \Nexus{}$\_$velshear filaments are more massive and broader than their
\nexus{} equivalents. 

\begin{table}
	\centering
	\caption{The population of P-Millennium haloes more massive than $ 3.2~\times~10^{10}$~\massUnit{} assigned to filaments by the NEXUS+ and NEXUS$\_$velshear web identification methods. 
  The columns specify: (1) method name, (2) the number of haloes assigned to filaments, (3) the fraction of the total halo population, (4) the number of \emph{common} haloes assigned to filaments by both methods, and (5) the number of \emph{exclusive} haloes assigned to filaments by \textit{only} one method.
  }
 	\label{table:number_of_haloes}
 	\begin{tabular}{ @{}l cccc@{}} % four columns, alignment for each
 \hline
 \\[-0.5cm]
 \hline
 Method & Number & Fraction & Common  & Exclusive  \\[.05cm]
  &  $[~\times 10^6~]$  & $[~\%~]$ &  $[~\times 10^6~]$  &  $[~\times 10^6~]$  \\
 \hline
  \\ [-.2cm]
 % \ \\
 \nexus{} & $2.80$ & 36.7 & \multirow{ 2}{*}{$2.36$} & $0.43$ \\[.2cm]
 % \ \\
 \Nexus$\_$velshear & $2.47$ & 32.6 &  & $0.10$ \\[.1cm]
 % \ \\
 \hline
	\end{tabular}
\end{table}

\section{Halo population}
\label{sec:halo_population}
The halo catalogue has been constructed by first identifying Friends-of-Friends (FOF) groups using a linking length of $0.2$ times the mean dark matter particle separation. The FOF groups were further split into bound structures using the SUBFIND algorithm \citep{springel2001}, which first associates potential subhaloes to local dark matter density peaks and then progressively discards particles that are not gravitationally bound to these substructures. 
For each FOF group, SUBFIND identifies the most massive subhalo as the main halo of the group. Our study uses only these main haloes. We define the halo radius, $R_{200}$, as the radius of a sphere located at the halo centre that encloses a mean density $200$ times the critical density of the universe. Then, the halo mass, $M_{200}$, is the mass contained within $R_{200}$.

We limit our analysis to haloes more massive than $3.2 \times 10^{10}~$\massUnit{}, which is motivated by the condition that the structure of a halo is resolved with a sufficiently large number of particles. Following \cite{bett2007}, we select haloes resolved with at least 300 dark matter particles within $R_{200}$. The P-Millennium contains $3.76 \times 10^{6}$ such main haloes which represent a very large and statistically
representative sample. This enables us to characterize the alignment between halo properties and the cosmic web directions to an unprecedented extent.

For all the haloes above our mass threshold limit, we calculate physical properties such as angular momentum and shape. Unless specified otherwise, these properties are calculated using all the gravitationally bound dark matter particles inside the halo radius, $R_{200}$. In order to gain deeper insight, we also calculate properties for the inner region of all haloes. We use two different radial cuts corresponding to the radii that enclose $10$ and $50\%$, respectively, of the halo particles. We refer to these radial cuts as the inner $10\%$ and $50\%$ of the halo, while when describing the full halo properties we denote that as the entire halo. The inner radial cuts are motivated by the observation that recent mass accretion is mainly deposited on the outer regions of a halo \citep{wang2011}, and thus, by studying the inner halo, we can probe how recent mass accretion, which is often anisotropic \citep[e.g.][]{Vera-Ciro2011,Shao2018}, may be affecting halo shape and spin.

\subsection{Cosmic web environment}
\label{sec:halo_environ}
We split the halo population into node, filament, wall and void samples according to the web environment identified at the location of the halo. We do so for both the \Nexus{}$\_$velshear and \nexus{} web classification schemes. In general, many of the same haloes are assigned to nodes and filaments by both methods, but there are also differences (see \autoref{table:number_of_haloes}), which we discuss in more details shortly.

In the present study, we focus on main haloes residing in filaments. The statistics of filament haloes in P-Millennium are presented in \autoref{table:number_of_haloes}. The filaments contain roughly 35\% of the main haloes, with \nexus{} identifying a slightly larger fraction of filament haloes. Both methods assign roughly the same haloes to filaments, with 96\% of the \Nexus{}$\_$velshear filament haloes also residing in \nexus{} filaments. For \nexus{}, 84\% of its filament haloes are in common with the \Nexus{}$\_$velshear ones, while the remaining 16\% corresponds  to haloes that populate filamentary tendrils in underdense regions. 
 
In \autoref{fig:haloes_in_web_zoomOut} we illustrate the similarities and differences in the distribution of filament haloes identified by the two web finders. For this, we show the full halo distribution (top-left panel) as well as the haloes inside \Nexus{}$\_$velshear and \nexus{} filaments inside a $200 \times 200\hmpc$ region, of $20\hmpc$ in width. Visually, we find that both methods are successful in recovering the most prominent filaments and also some of the less conspicuous ones, although it is more difficult to visually assess the latter due to the larger slice thickness. The haloes in \nexus{} filaments (bottom-left-hand panel) trace a sharp and intricate network with prominent filamentary arteries, as well as a substantial web of thinner tenuous branches and minor filaments in low-density areas. In contrast, the \Nexus{}$\_$velshear filament haloes (bottom-right-hand panel) have a rather different character, tracing mostly thick filaments. 

The comparison between \nexus{} and \Nexus$\_$velshear filaments reveal that the latter are considerably thicker. This is a reflection of the non-local character of the velocity shear field, which, compared to \nexus{}, leads to assigning to the same filament haloes that are found at larger distances from the filament spine. The extent of this effect can be best appreciated in the top-right panel, which shows the distribution of \textit{common} haloes, that is the ones assigned to filaments by both \nexus{} and \Nexus{}$\_$velshear. The common filament haloes have almost the same appearance, although thinner and sharper, as the ones residing in the \Nexus{}$\_$velshear filaments. This clearly illustrates that \nexus{} finds the \Nexus{}$\_$velshear filaments and that it assigns them a smaller thickness. 

To have a more detailed comparison between the filament haloes identified by the two web finders, \autoref{fig:haloes_in_web_zoomIn}
zooms in on to a $40 \times 40\hmpc$ region centred on a prominent filamentary network. The figure shows the distribution of
filament haloes in and around a junction of many prominent filaments which are found around a concentration of cluster-mass haloes. This region is certainly one of the most dynamically active areas of the cosmic web and is expected to be strongly influenced by the substantial tidal forces resulting from the highly anisotropic distribution of matter in the region.

The contrast between the two web finders is substantial. \nexus{} includes small filaments and tendrils whose minor dynamical impact on the velocity shear field eludes detection
by the NEXUS$\_$velshear method. The top row of \autoref{fig:haloes_in_web_zoomIn} provides a telling visualization of this effect, with \nexus{} pointing out many thin low-density filaments around the main filamentary mass concentrations. This can also be observed in the bottom row of \autoref{fig:haloes_in_web_zoomIn}, which shows the \textit{exclusive} filament haloes, that is the haloes assigned to filaments by only one of the two methods. \Nexus{}$\_$velshear misses the halo population of minor filaments while identifying thicker prominent filaments, which may even include haloes that \nexus{} assigns to underdense void regions.

\begin{figure}
	\mbox{\hskip -0.3cm\includegraphics[width = 1.05\columnwidth]{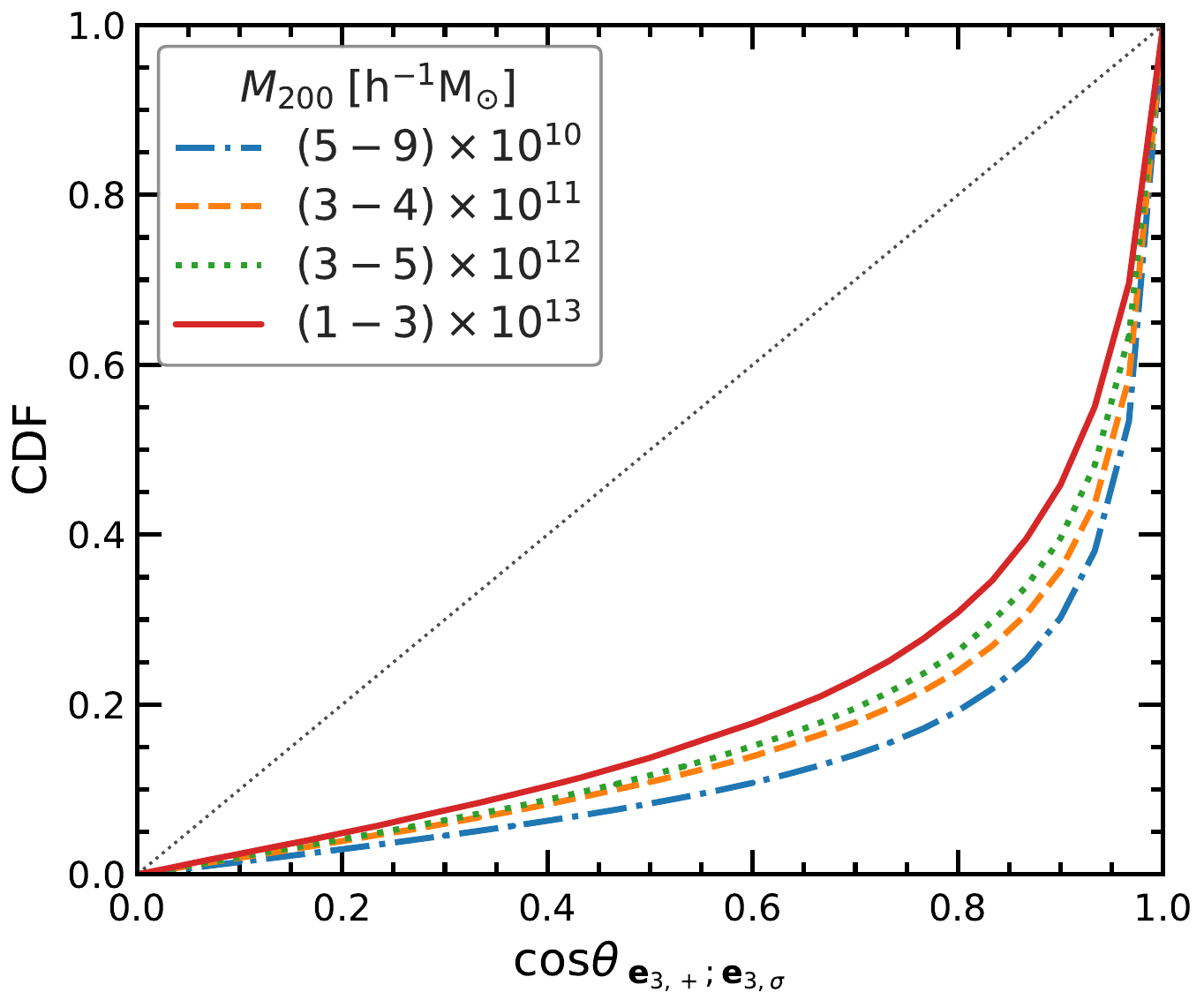}}
    \vskip -.2cm
	\caption{ \revised{The CDF of the angle between the orientation of filaments identified using NEXUS+ and NEXUS\_velshear. The curves correspond to the alignment of the two filament types at the positions of different mass haloes. The NEXUS+ and NEXUS\_velshear filaments are mostly aligned (compare to the expectation for random alignment shown in dotted grey), with the strength of the alignment slightly decreasing for high mass haloes.} 
    }
	\label{fig:nexus_velshear_alignment}
\end{figure} 

\revised{The directions of \nexus{} and \Nexus{}\_velshear filaments are illustrated in the top two panels of \autoref{fig:haloes_in_web_zoomIn}. This shows that the orientations assigned by the two web finders match well with the visually inferred local direction of the filamentary network. The \nexus{} and \Nexus{}\_velshear filament orientations are nearly parallel as can be seen from \autoref{fig:nexus_velshear_alignment}. The figure shows the misalignment angle between \nexus{} and \Nexus{}\_velshear filament axes, which was calculated at the position of each halo that is assigned to both filament types. The \nexus{} and \Nexus{}\_velshear filaments are well aligned over the entire halo mass range, with a median misalignment of  ${\sim}20^{\circ}$. The alignment shows a small dependence on halo mass, with higher mass haloes having slightly lower alignment between the two filament types.}

\begin{figure} 
	\includegraphics[width = \columnwidth]{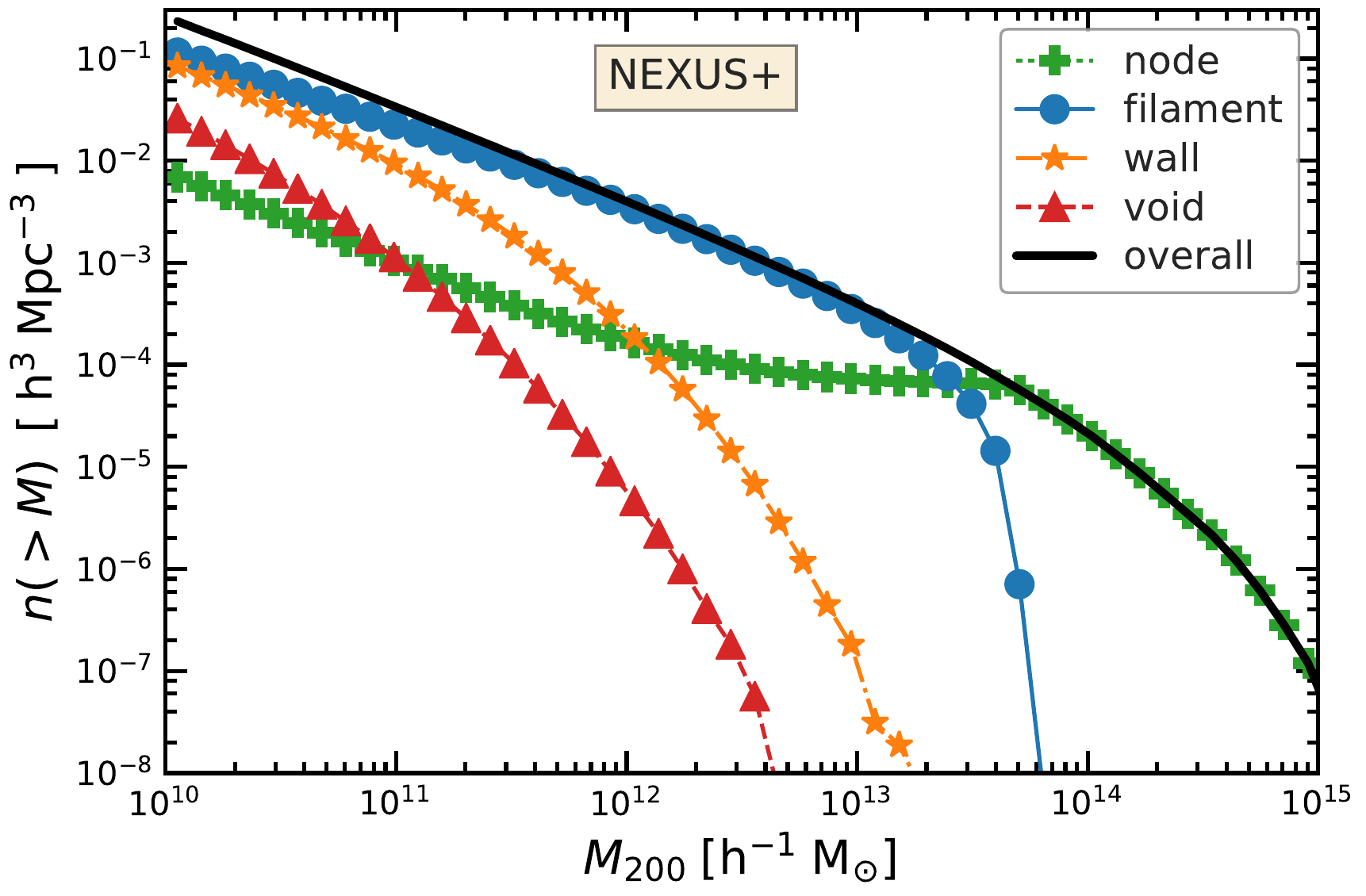}
    \\[-0.75cm]
    {\tikz \fill [white] (0,0) rectangle (1.\linewidth,1.01cm);}
    \\[-1.cm]
	\includegraphics[width =\columnwidth]{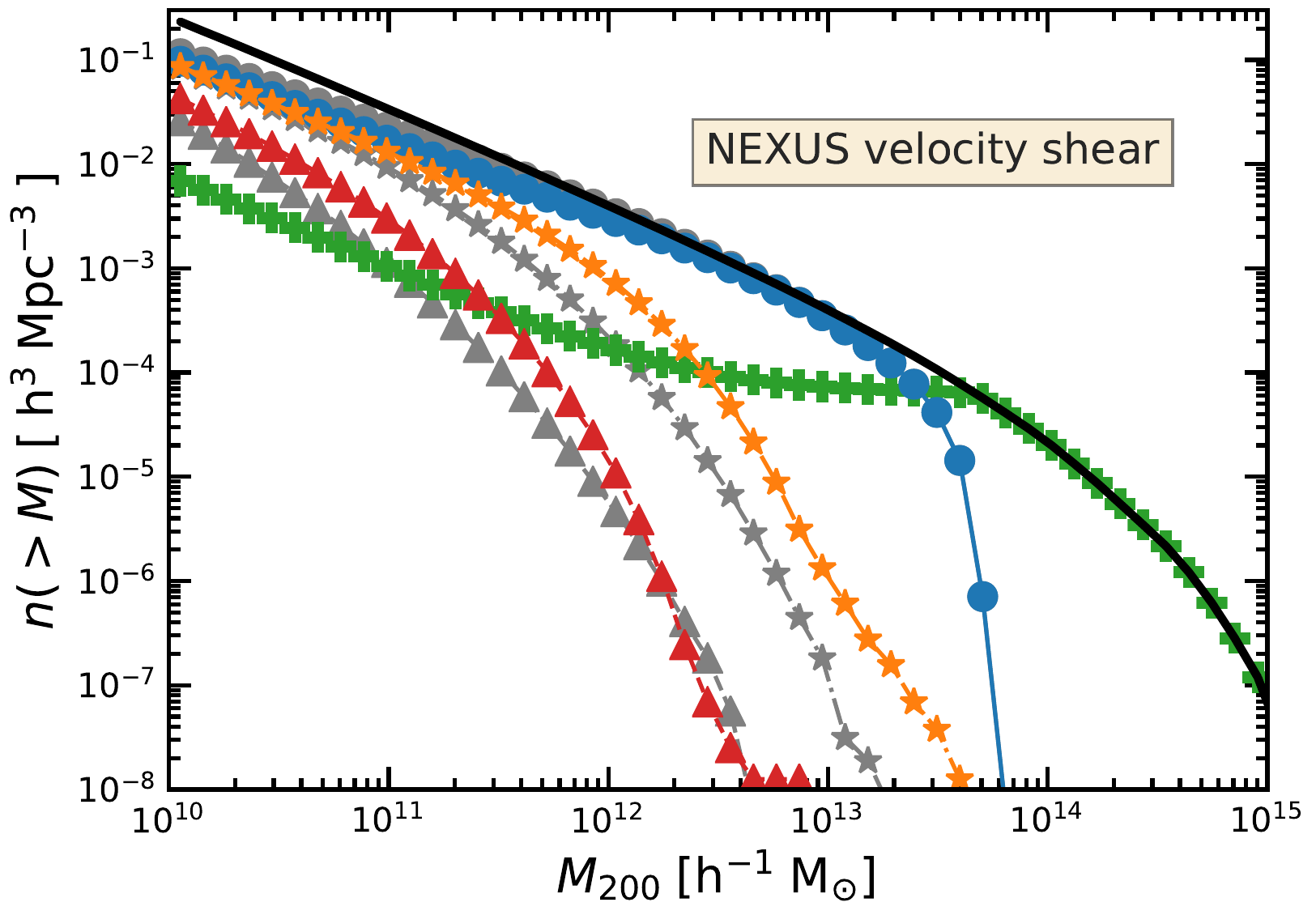}
    \\[-0.75cm]
    {\tikz \fill [white] (0,0) rectangle (1.\linewidth,1.01cm);}
    \\[-1.cm]
    \includegraphics[width =\columnwidth]{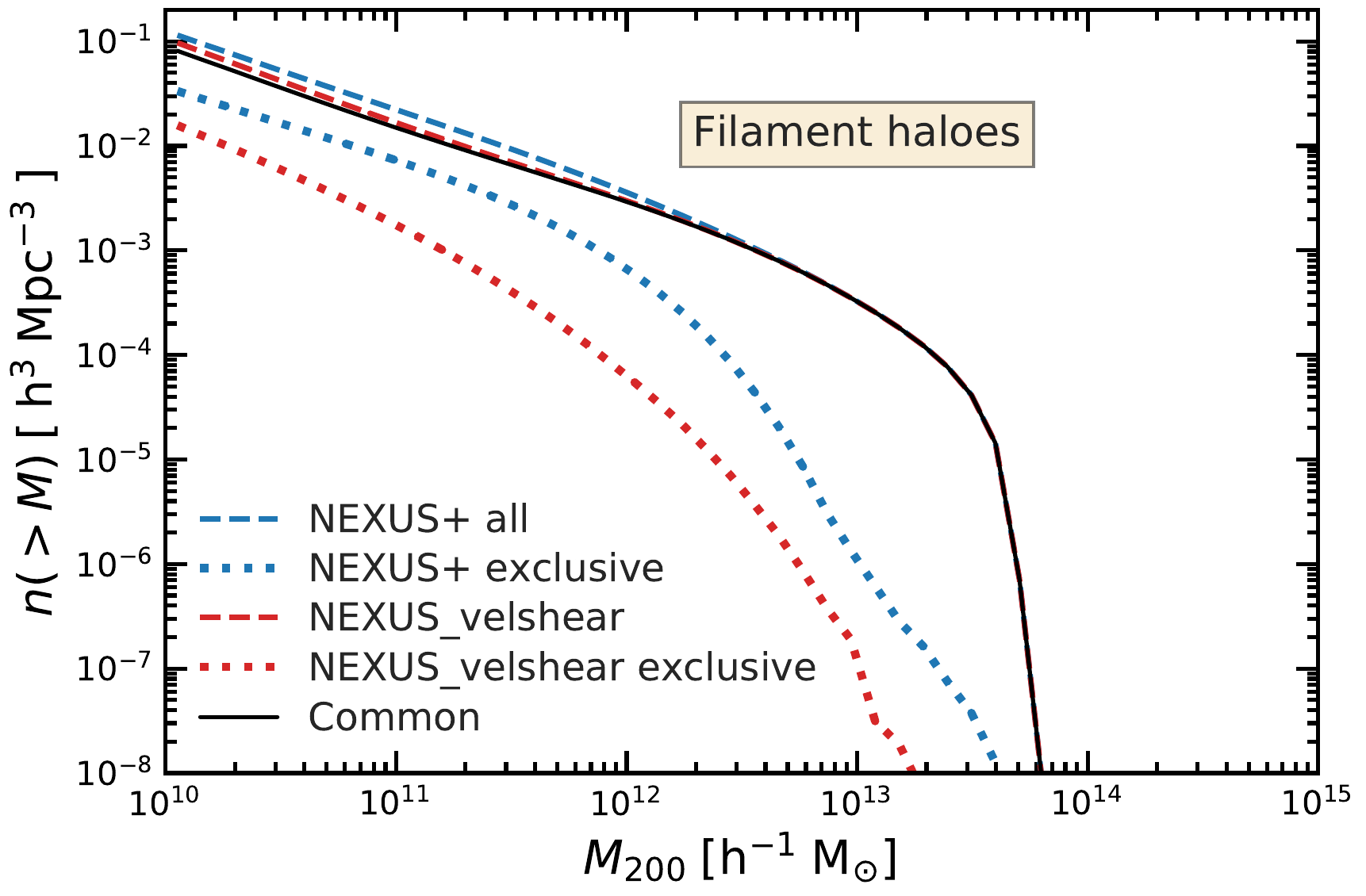}
    \vskip -.2cm
	\caption{Cumulative halo mass function in the different cosmic web environments of the P-Millennium simulations at $z = 0$. \textit{First panel}: Environments detected using NEXUS+. \textit{Second panel}: Environments detected using NEXUS$\_$velshear, \revised{with the grey curves showing the NEXUS+ results from the top panel. \textit{Third panel}: A closer comparison of the halo mass function in NEXUS+ and NEXUS\_velshear filaments. The common sample corresponds to haloes that reside in both filament types and it comprises most of the filament halo population. The exclusive sample consists of haloes assigned to only one of the two filament types.}
    }
	\label{fig:halo_mass_functions}
\end{figure}

\subsection{Halo mass function}
A first aspect of the connection between web-like environment and the halo distribution concerns how haloes populate the different cosmic web environments. This is shown in \autoref{fig:halo_mass_functions}, where we present the (cumulative) mass function of haloes segregated according to the environment in which they reside. Here we show the population of main haloes with at least 100 particles, i.e. $M_{200}>1.1\times10^{10}~{h}^{-1}\rm{M}_{\odot}$, however, for the rest of the paper, we limit the analysis to objects at least three times as massive. The halo mass function represents the number density, $n(>M)$, of haloes with a mass in excess of $M$,
\begin{equation}
	n(>M)\,=\,\int_{M}^\infty\,\frac{{\rm d} n}{{\rm d} \log M}\,\frac{{\rm d}M}{M} \,,
\end{equation}
where ${\rm d} n/{\rm d} \log M$ denotes the specific mass function, that is the number density of haloes of mass $M$ per logarithmic mass bin. \autoref{fig:halo_mass_functions} shows the halo mass function split according to web environments for both the \nexus{} (top panel) and the \Nexus{}$\_$velshear (middle panel) methods. We note that the identifications of node environments using the velocity shear field poses challenges \citep{cautun2013}, which are due to the presence of a substantial level of vorticity in these highly
multi-stream regions that is not accounted by the velocity shear field. To deal with this limitation, following \citet{cautun2013}, we augmented the \Nexus{}$\_$velshear scheme such that the node identification is done using the density field, which is the procedure used by \nexus{}.

\autoref{fig:halo_mass_functions} shows that the halo mass function depicts a substantial difference between environments \citep[also see e.g.][]{cautun2014,libeskind2018}: the most massive haloes reside at nodes of the web while most lower mass objects are predominantly found in filaments. While there are some differences in details, in particular concerning the higher mass tails
of the void and wall halo mass functions, overall the halo populations segregated by environment are very similar in both the \nexus{} and \Nexus{}$\_$velshear web finders. 

Except for the most massive objects, we find that the majority of haloes are found in filaments. The exception concerns the objects with masses in excess of $M \approx 10^{13.5} \rm{M_{\odot}}$, which are almost exclusively found in nodes. The mass function for void haloes is strongly shifted to lower masses, and has a significantly lower amplitude than that for filament or wall haloes. This is to be expected, since voids represent the lowest density regions and are mostly populated by low mass haloes. This agrees with observations which reveal that most void galaxies are typically faint and have low stellar masses \citep[see e.g.][]{kreckel2011, kreckel2012}.
A similar trend is seen for haloes residing in the membranes of the cosmic web, i.e. walls, though less extreme than for void galaxies. Haloes more
  massive than $10^{12.0} \rm{M_{\odot}}$ are hardly found in walls, nearly all of them residing in filaments. It explains, amongst others, why
  walls are so hard to trace in magnitude-limited galaxy surveys \citep[see also][]{cautun2014}.
Overall, the halo mass functions in \nexus{} environments are the same as in their \Nexus{}$\_$velshear equivalents, with only minor differences. \revised{In the second panel of \autoref{fig:halo_mass_functions}, we can notice that the} \Nexus{}$\_$velshear allocates somewhat more haloes of all masses to voids
  and walls, and thus slightly fewer haloes to filaments. \revised{The bottom panel of \autoref{fig:halo_mass_functions} compares in detail the filament mass function identified by the two web finders. The common sample represents the majority of the filament halo population. This is the case in particular for NEXUS\_velshear, for which the exclusive sample is nearly a factor of ten less numerous at all masses. The NEXUS+ exclusive sample is more sizeable, consisting of  ${\sim}30\%$ of the low mass haloes found in \nexus{} filaments.} This is a direct reflection of the fact that \nexus{} identifies the small and tenuous filamentary
  tendrils, which are largely ignored by \Nexus{}$\_$velshear. These less prominent features contain mostly low mass haloes \citep{cautun2014}, which explain why the differences between \nexus{} and \Nexus{}$\_$velshear are mostly seen for low mass haloes.

\subsection{Halo shape}
\label{sec:shape}
We determine the shape of a halo by calculating its moment of inertia tensor, $\mathbf{I}_{ij}$ \citep{haarlem1993,bett2007,araya2008}.
For a halo that contains $N$ particles,
the moment of inertia with respect to the centre of mass 
is defined as 
\begin{equation}
\mathbf{I}_{ij} = \sum_{k=1}^{N} m_k r_{k,i}r_{k,j}\,,
\end{equation}
where $m_k$ is the mass of the $k$-{th} particle, and $r_{k,i}$ is the particle position along the $i$-{th} coordinate axis with respect to the halo centre of mass.
\noindent The inertia tensor is a $3 \times 3$ symmetric matrix that can be diagonalised to calculate its eigenvalues, $s_a \geq s_b \geq s_c$, and eigenvectors, $\mathbf{v}_a$, $\mathbf{v}_b$ and $\mathbf{v}_c$. The shape of the halo is commonly described in terms of the axes ratios ${b}/{a}$ and ${c}/{a}$, where $a=\sqrt{s_a}$, $b=\sqrt{s_b}$ and $c=\sqrt{s_c}$ denote the major, intermediate and minor halo axes, respectively.
A perfectly spherical halo has $b/a = c/a =1$, a prolate one has a major axis significantly longer than the intermediate and minor axis, $ c \approx b << a $, while an oblate one has a much smaller minor
axis than the other two, $c << b \approx a$. The orientation of the halo is specified by the corresponding eigenvector, with $\mathbf{v}_a$, $\mathbf{v}_b$ and $\mathbf{v}_c$ pointing along the major, intermediate and minor axes, respectively.

\begin{figure}
	%\mbox{\hskip -0.3cm\includegraphics[width = 1.05\columnwidth]{figures/cosmic_ballet_fig6.pdf}}
    \includegraphics[width = \columnwidth]{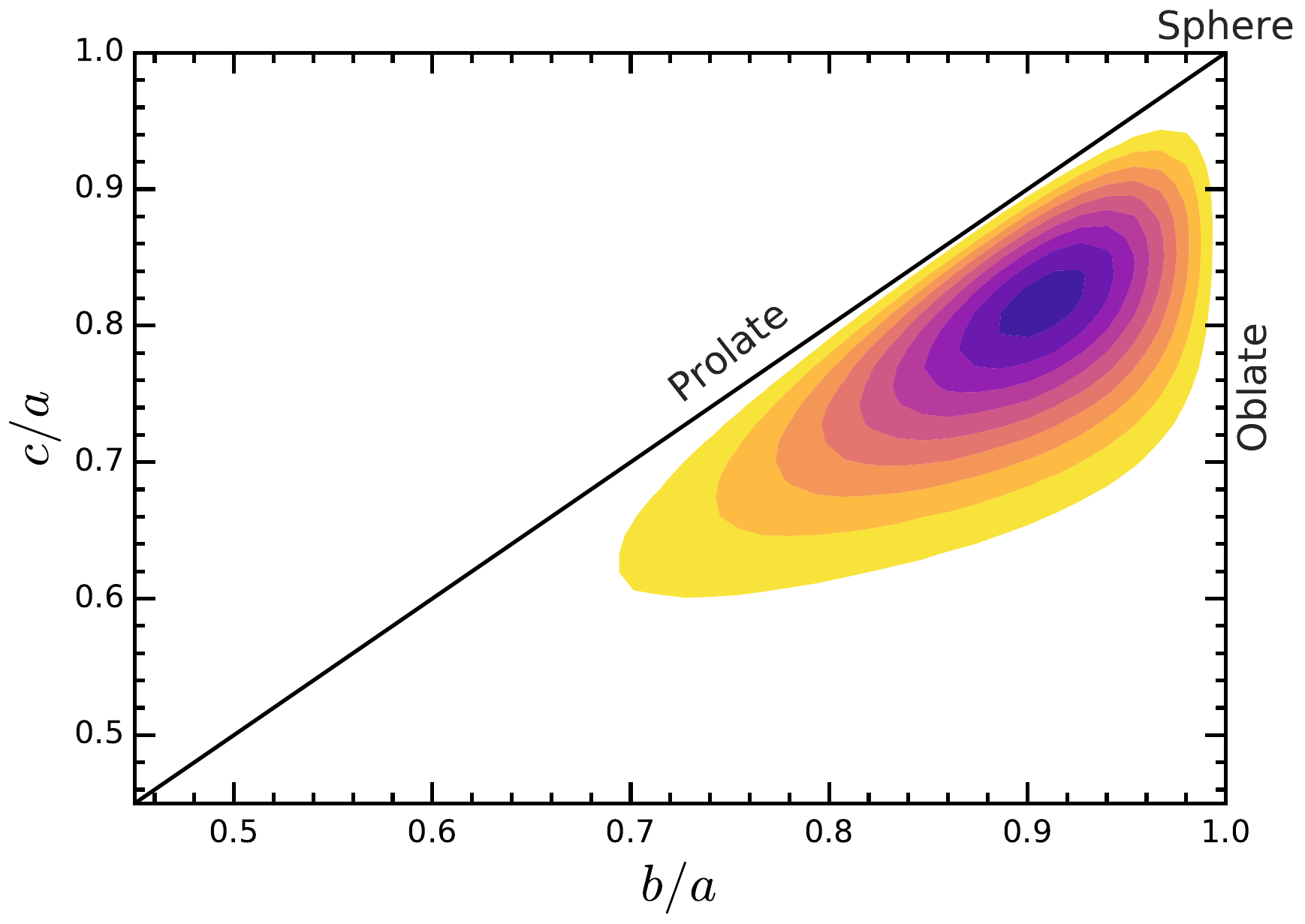}
    \includegraphics[width = 0.97\columnwidth]{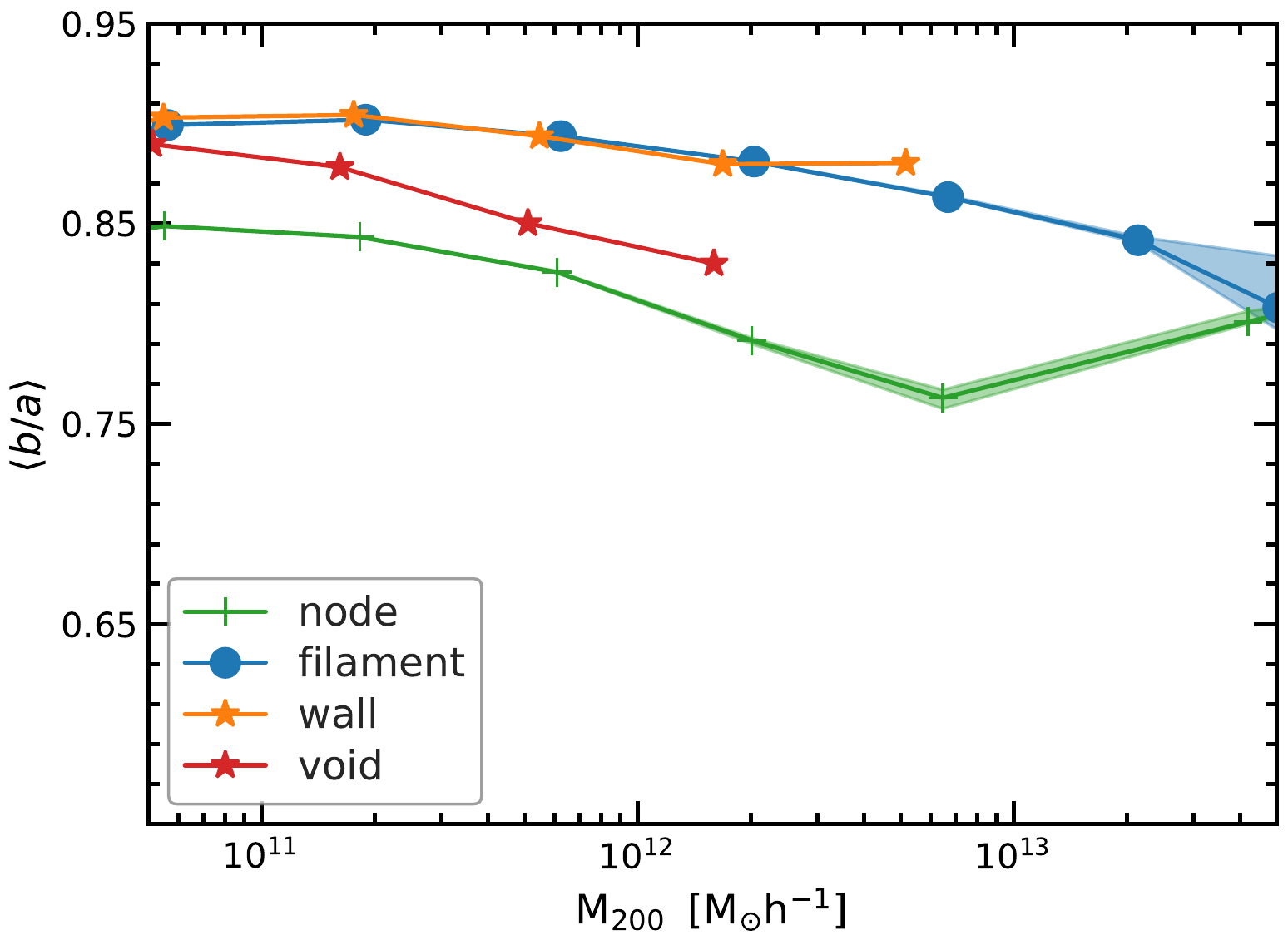}
    \\[-0.75cm]
    {\tikz \fill [white] (0,0) rectangle (1.\linewidth,1.01cm);}
    \\[-1.cm]
  \includegraphics[width = 0.97\columnwidth]{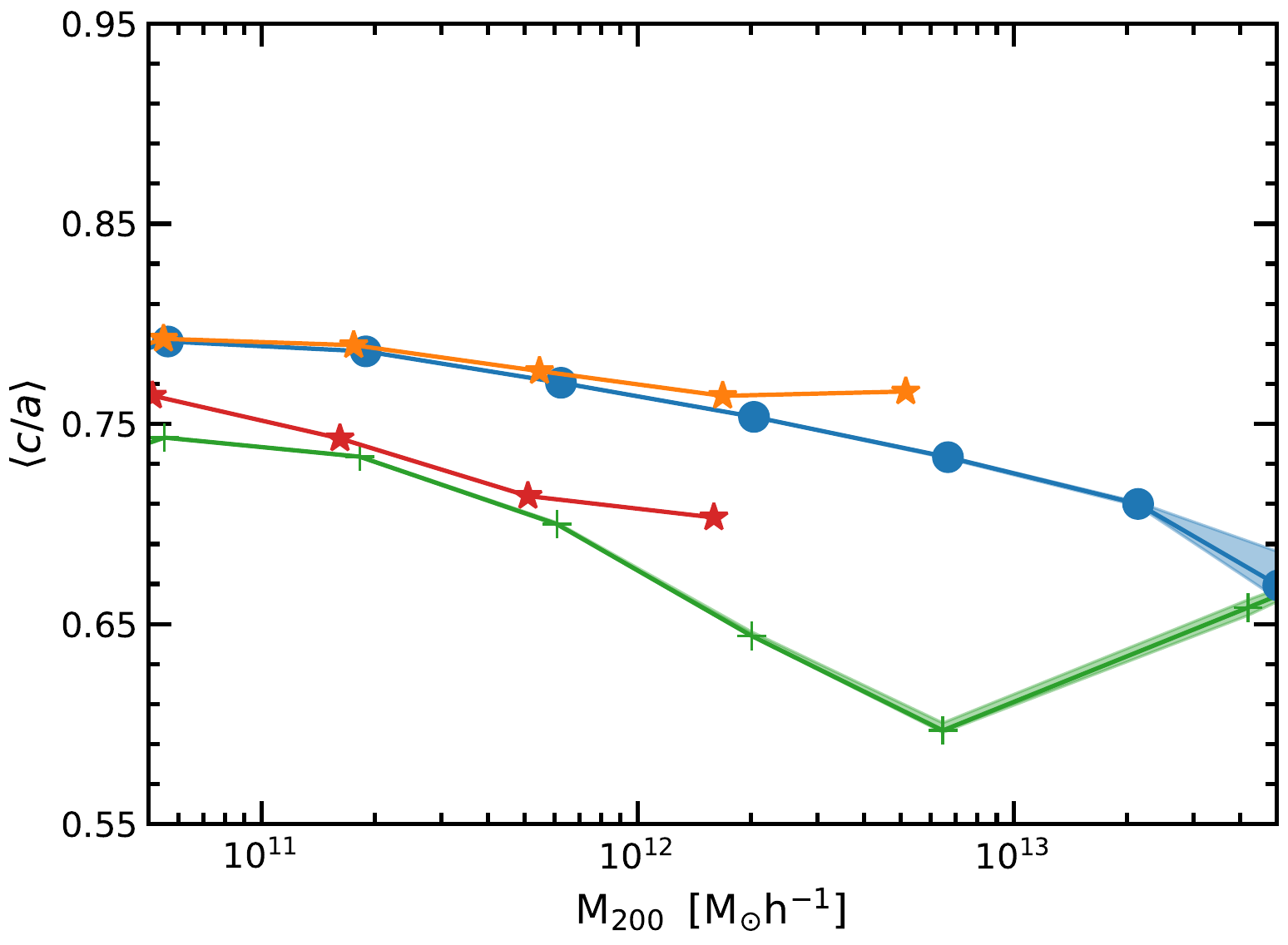}
    \vskip -.2cm
	\caption{ \revised{\textit{Top panel:}} The distribution of halo shapes in P-Millennium in terms of the axes ratios $c/a$ versus $b/a$, where $a$, $b$ and $c$ are the major, intermediate and minor axes. The coloured regions represent contour lines of the density of points, with darker colours corresponding to higher densities. We also show the point of perfect sphericity, $b/a=c/a=1$, and the two axes corresponding to purely oblate (flattened) and prolate (elongated) haloes. \revised{\textit{Middle and bottom panels:} The median axis ratios, $\left\langle b/a \right\rangle $ and $\left\langle c/a \right\rangle$, as a function of halo mass for haloes in different cosmic web environments. The shaded region indicate the $1\sigma$ error. Both axis ratios shows a weak dependence on halo mass and a stronger variation with web environment.}
  	}
	\label{fig:halo_shape_contour}
\end{figure}

\revised{The top panel of} \autoref{fig:halo_shape_contour} shows the halo shape distribution in P-Millennium, which is in good agreement with previous studies \citep[e.g.][]{bett2007}. Overall, the haloes are triaxial, with a clear trend towards a roundish - but never perfectly spherical - shape. Most haloes have $c/a> 0.8$ and $b/a > 0.9$. They also have a slight tendency towards a prolate shape. \revised{The halo shapes show a small, but statistically significant, variation with the web environment in which a halo resides. This is clearly in indicated in the middle and bottom panels of \autoref{fig:halo_shape_contour}, which shows the median halo shape axis ratios, $b/a$ and $c/a$, as a function of halo mass and environment. Haloes in nodes and voids are more flattened than haloes residing in filaments and walls \citep{hahn2007,romero2014}.}

\subsection{Halo angular momentum}
\label{angular_momentum}
The angular momentum -- or spin -- of the halo is defined as the sum over the angular momentum of the individual particles
that constitute the halo, 
\begin{equation}
\mathbf{J} =  \sum\limits_{k=1}^{N} m_k \left( \mathbf{r}_{k} \times \mathbf{v}_{k}  \right),
\end{equation}
where $\mathbf{r}_{k}$ and $\mathbf{v}_{k} $ are the position and velocity of the  $k$-{th} particle with respect to the halo centre of mass. 

For each halo, we calculate the angular momentum for the entire virialized halo, as well as for inner halo regions consisting of the inner $10\%$ and $50\%$ of the halo particles. This yields 
3 angular momenta, $\mathbf{J}_{100}$, $\mathbf{J}_{50}$ and $\mathbf{J}_{10}$ for each halo. We are interested in two aspects of the
halo angular momenta: its amplitude and its orientation (i.e. the spin direction). 

\subsubsection{Spin parameter $\lambda$}
The angular momentum amplitude, $J=|\mathbf{J}|$, is usually expressed in terms of a dimensionless spin parameter, $\lambda_P$, introduced by
\cite{peebles1969},
\begin{equation}
  \lambda_P\,=\,\frac{J |E|^{1/2}}{GM^{5/2}}\,,
  \end{equation}
where $J$, $E$ and $M$ are the total angular momentum, energy and mass of the halo, and $G$ is Newton's constant. The spin parameter $\lambda_P$
quantifies the extent of coherent rotation of a halo (or any self-gravitating system). A value of unity of the parameter means that a self gravitating
system is supported by rotation \citep{paddy1993}, a value closer to zero would imply that it hardly has coherent rotation and that the system is
dispersion-supported.
\begin{figure}
	\mbox{\hskip -0.5cm 
    \includegraphics[width=1.05\columnwidth]{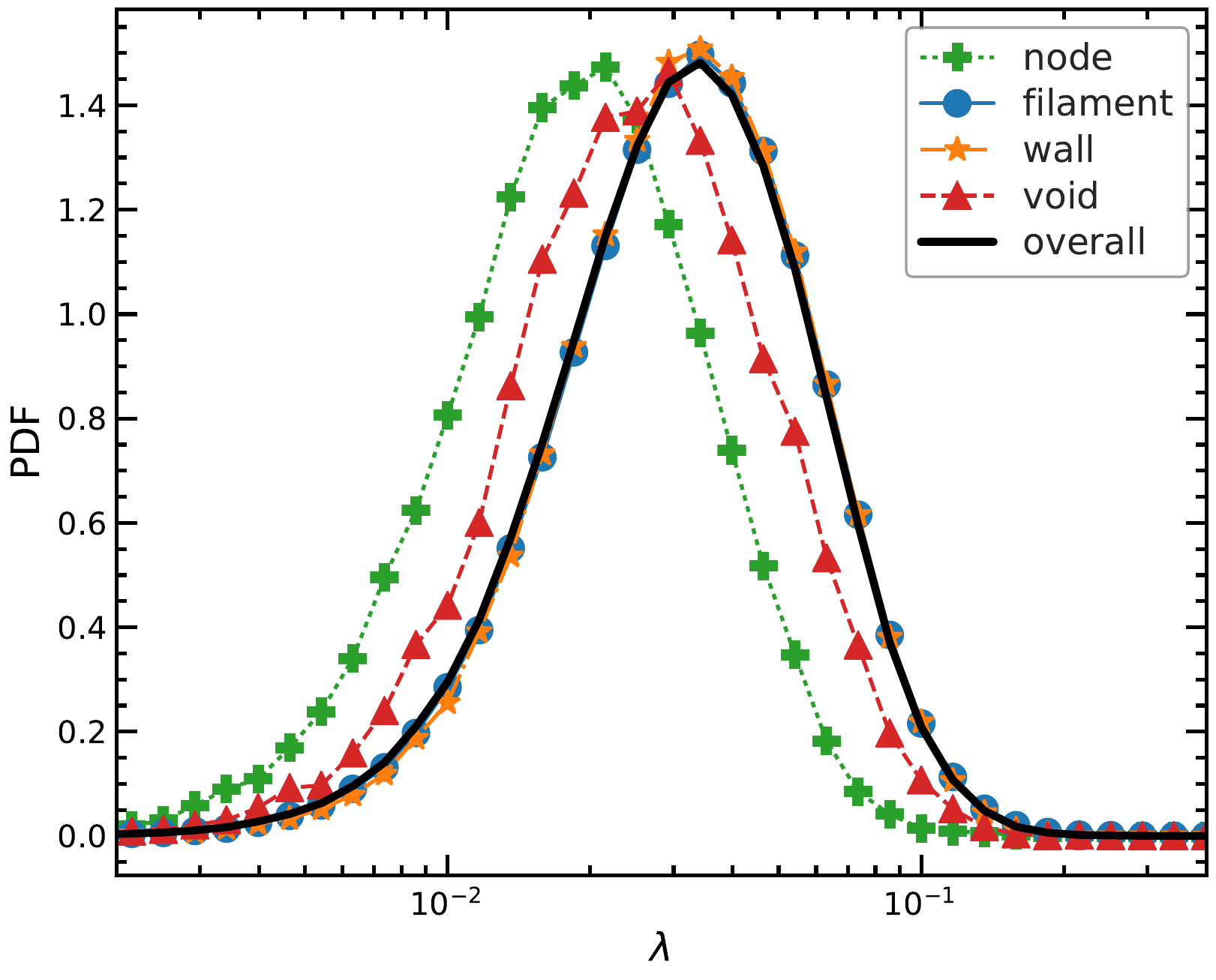}}
    \vskip -.2cm
	\caption{The distribution of halo spins segregated according to the NEXUS+ environment in which a halo resides: nodes or clusters (solid line with crosses), filaments (dotted line with dots), walls or sheets (solid line with star symbol) and voids (dashed line with triangles). The results are calculated using haloes in the mass range $[3,5] \times 10^{11}\hMsun$.
    }
	\label{fig:bullock_spin_pdf}
\end{figure}

We use an alternative definition of the spin parameter introduced by \cite{bullock2001}. The Bullock spin parameter $\lambda$ has a more
practical definition, in particular when considering a subvolume of a virialized sphere, and it is also easier to calculate.
For a region enclosed within a sphere of radius $R$, this spin parameter is defined as 
\begin{equation}
\lambda = \frac{J}{\sqrt{2}MVR},
\end{equation}
where $V$ is the halo circular velocity at radius $R$ and $J$ the angular momentum within this radius. This spin parameter definition
reduces to the standard Peebles parameter $\lambda_P$ when measured at the virial radius of a truncated isothermal halo. The spin parameters
$\lambda_P$ and $\lambda$ are in fact very similar for typical NFW haloes \citep{nfw1997, bullock2001}, having a roughly lognormal distribution
with a median value of $\lambda \approx 0.05$ \citep{efstathiou1979, barnes1987}. 

\bigskip
In order to determine if halo spin amplitude is correlated to the web environment in which a halo resides, \autoref{fig:bullock_spin_pdf} shows the probability distribution functions (PDF) of the Bullock spin parameter for halo samples split according to the \nexus{} environment in which a halo is  located.
We observe a clear segregation between the
rotation properties of haloes in different web environments, with filament and wall haloes having on average the largest spin, while node haloes are the slowest rotation objects. For all environments, the PDF is close to a lognormal distribution, but with its peak value slightly shifted, from $\lambda = 0.035$ for filaments and walls, to $\lambda = 0.030$ for voids and $\lambda = 0.020$ for node haloes.

\autoref{fig:bullock_spin_pdf} clearly reveals the influence of cosmic environment
on the spin parameters of haloes, with filament and wall haloes showing a significantly stronger coherent rotation than their counterparts residing in nodes, which have a more prominent dispersion-supported character. Interestingly, this is similar with the morphology-density relation \citep{dressler1980} found in observations, with early type galaxies dominating the galaxy population of clusters while the late-type spirals dominating the filamentary and wall-like ``field'' regions. 

% We speculate that this could be a result of \revised{violent relaxation, %\citep{LyndenBell1967} 
% the re-distribution of particle orbits due to mergers that take place frequently in these high density environments.\sout{ vorticity present in these regions. While all web environments have some vorticity.} In addition to this, all web environments have some vorticity and is largest in the nodes of the cosmic web. Large scale vorticity can increase or decrease the halo spin \citep[see e.g.][]{Laigle2015}. One possible explanation for the lower spin of node haloes is the disruption of their orderly rotation by the large vorticity present in their environment.}
% %which arises due to accretion of mass onto clusters along filaments. This could be the key factor disrupting the orderly rotation within haloes. 
% %The lack of such a high turbulent environment in filaments would thus make it appropriate to select filament haloes and study the spin alignments without the influence of such tertiary environmental effects. 

\subsubsection{Spin orientation}
When calculating the alignment of halo spin with the web directions, we make use of the spin direction of each halo, which is defined as
\begin{equation}
  {\mathbf{e}}_J\,=\,\frac{{\mathbf J}}{|J|}\,.
\end{equation}
We apply this relation for each of the three radial cuts for the radial extent, i.e. for the radii including $10\%$, $50\%$ and $100\%$ of the mass of the halo.

\section{Spin Alignment Analysis}
\label{error}
\label{sec:spin_alignment}
Here we study the alignment between the halo spins and the orientation of the filaments in which the haloes are embedded. The filament orientation corresponds to the direction along the filament spine, which is given by the ${\mathbf e}_{3,+}$ and ${\mathbf e}_{3,\sigma}$ eigenvectors for the \nexus{} and \Nexus{}$\_$velshear methods, respectively (for details see \autoref{sec:filament} and \autoref{fig:schematic_cos}). Furthermore, we limit our analysis to filament haloes, which are the dominant population of objects.

\subsection{Alignment analysis: definitions}
\begin{figure}
	\includegraphics[width=\columnwidth]{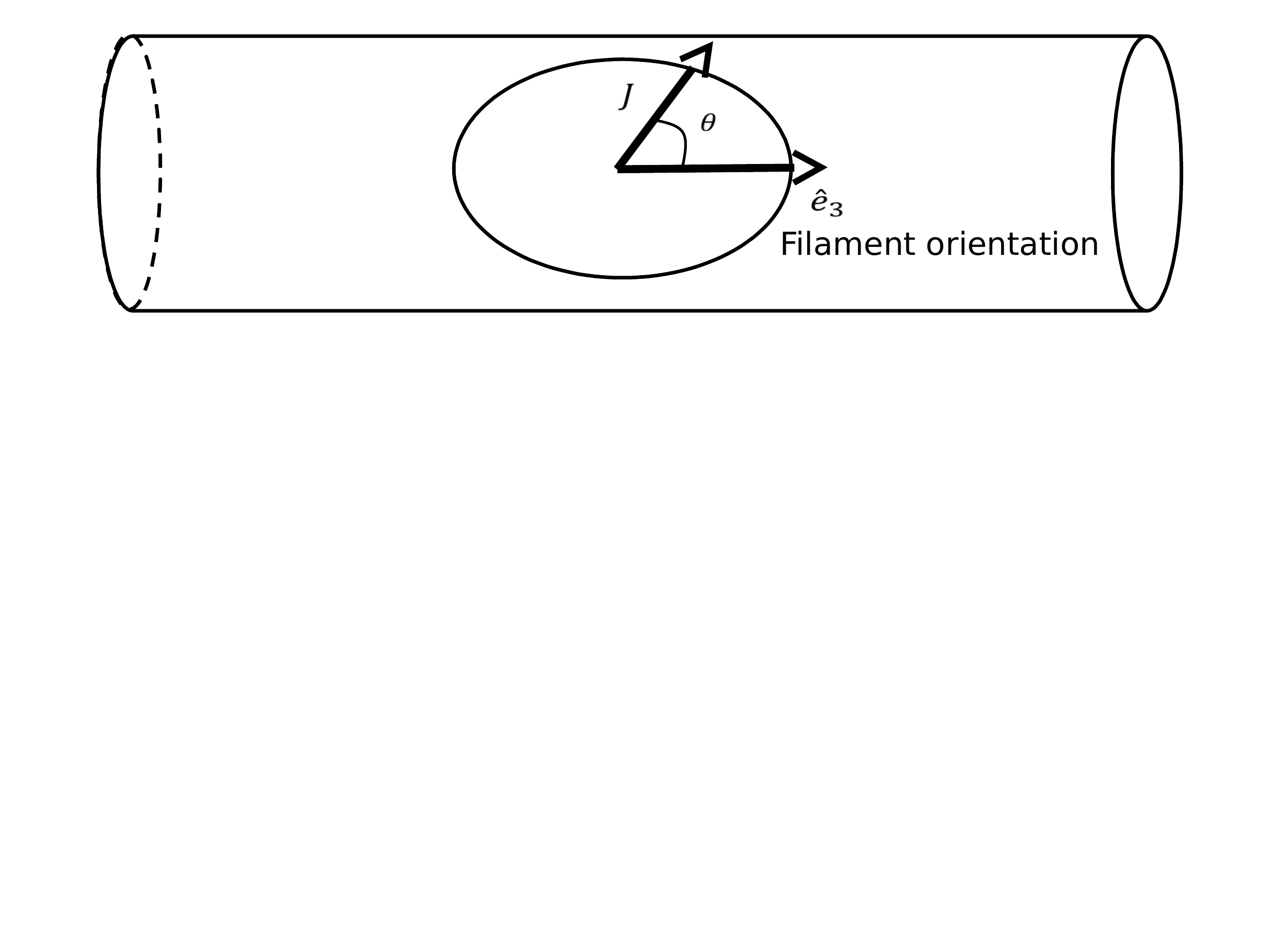}
	\vspace{-4.2cm}
	\caption{ Schematic representation of the alignment angle $\theta$ between the angular momentum of a halo, $\mathbf{J}$, and the filament orientation, $\mathbf{e}_3$. The cylinder represents the filament and the ellipse depicts the halo residing in it. 
    A value of $\cos \theta \sim 1 $ corresponds to the halo spin direction being parallel with the filament, while $\cos \theta \sim 0 $ corresponds to a perpendicular configuration.
    }
	\label{fig:schematic_cos}
\end{figure}

We define the alignment angle as the angle between the direction of a halo property, which can be spin, shape or velocity, and the orientation of the filament at the position of the halo. A diagrammatic illustration of the
alignment angle $\theta$ is shown in \autoref{fig:schematic_cos}, with the ellipse representing a halo and the cylinder the local
stretch of the filament. For a given halo vector property ${\mathbf h}$, the halo-filament alignment angle is
\begin{equation}
	\mu_{hf}\, \equiv \,\cos \theta_{\mathbf{h}, \mathbf{e}_3} = \left| \frac{\mathbf{h} \cdot \mathbf{e}_3}{|\mathbf{h}| |\mathbf{e}_3 |} \right|\,,
	\label{cos}  
\end{equation}
which is the normalized scalar product between the halo and filament orientations. We take the absolute value of the scalar product since filaments have an orientation and not a direction, that is both $\mathbf{e}_3$ and $-\mathbf{e}_3$ correspond to a valid filament orientation. Note that the symbol, $\mu_{hf} \equiv \cos \theta_{\mathbf{h},\mathbf{e}_3}$, denotes the cosine of the alignment angle, however, for simplicity, we refer to it both as the alignment parameter and as the alignment angle.  

A halo property that is parallel to the filament orientation corresponds to $\mu_{hf}=1$, while a property that is perpendicular to the filament orientation corresponds to $\mu_{hf}=0$. A random or isotropic distribution of alignment angles corresponds to a uniform distribution of $\mu_{hf}$ between 0 and 1, which provides a useful reference line for evaluating deviations from isotropy. In the case of a distribution of alignment angles for a halo population, we refer to that sample as being \textit{preferentially parallel} if the median alignment angle is larger than 0.5. Conversely, that sample is \textit{preferentially perpendicular} if the median alignment angle is lower than 0.5. Following this, and somewhat arbitrary, we consider that $\mu_{hf}=0.5$ marks the transition between a preferentially parallel, median $\mu_{hf}>0.5$, and a preferentially perpendicular, median $\mu_{hf}<0.5$, alignment.

\begin{figure*}
	\includegraphics[width=\textwidth]{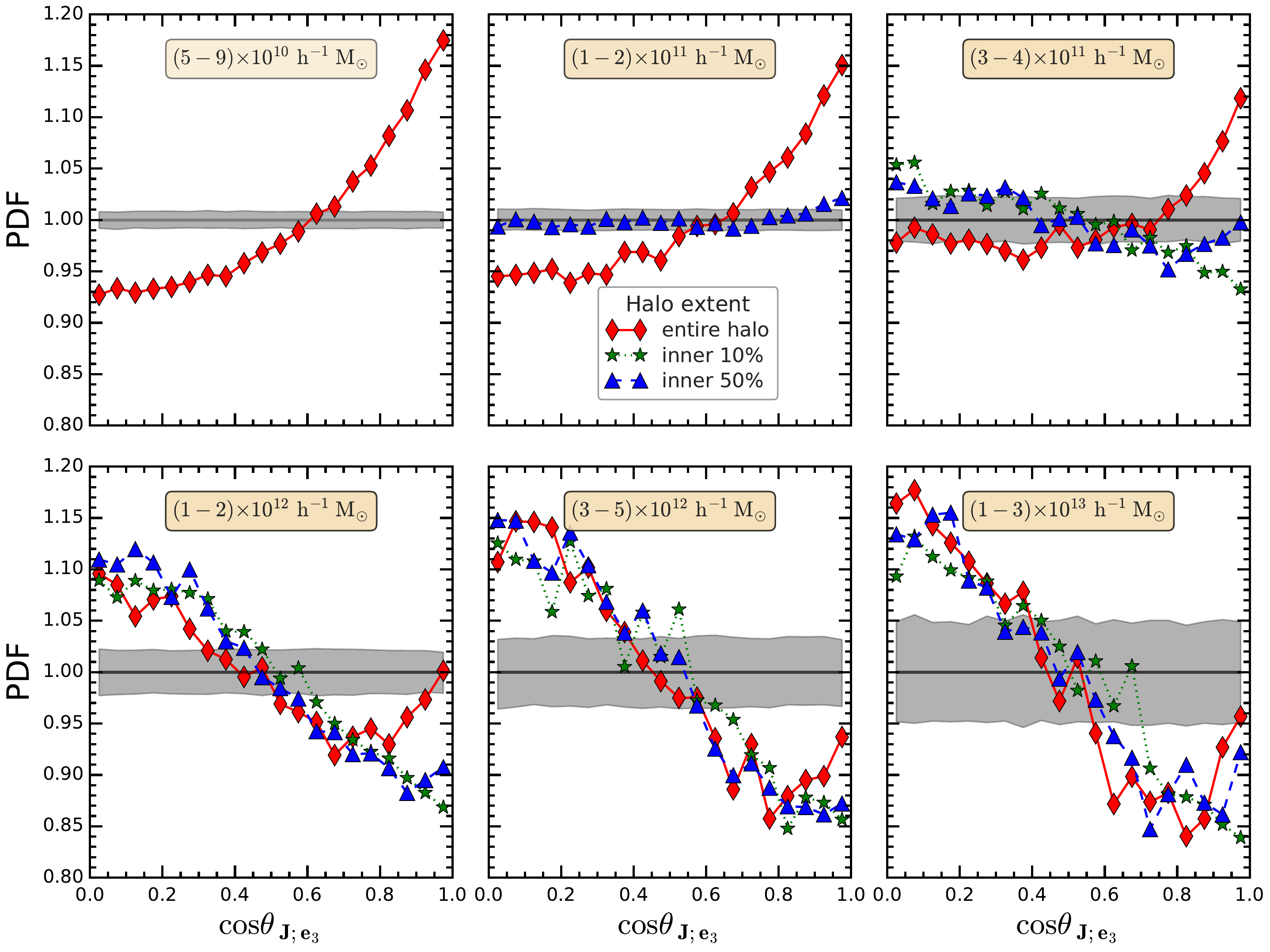}
    \vskip -.2cm
    \caption{The distribution of alignment angles, $\cos\theta_{\mathbf{J};\mathbf{e}_3}$, between the halo angular momentum, $\mathbf{J}$, and the filament orientation, $\mathbf{e}_3$, for haloes residing in filaments identified using the NEXUS velocity shear method. Each panel shows the PDF for a different range in halo mass, $M_{200}$. Each panel (except the two for the lowest halo mass) shows the alignment between the filament orientation and the angular momentum calculated using different radial cuts: entire halo (red rhombus symbols), and the inner regions that contain 50\% (blue triangles) and 10\% (green stars) of the particles. The horizontal line shows the mean expectation in the absence of an alignment signal and the grey shaded region shows the 1-sigma uncertainty region given the sample size. The alignment distribution depends on halo mass, with the spin of massive haloes being preferentially perpendicular and that of low mass haloes being preferentially parallel to the filament orientation. Furthermore, at low masses the alignment depends on the inner radial cut used for calculating the halo spin. 
   }
    \label{fig:multipanel_ang_mom_velshear}
\end{figure*}

\begin{figure}
	\mbox{\hskip -0.4truecm\includegraphics[width=1.05\columnwidth]{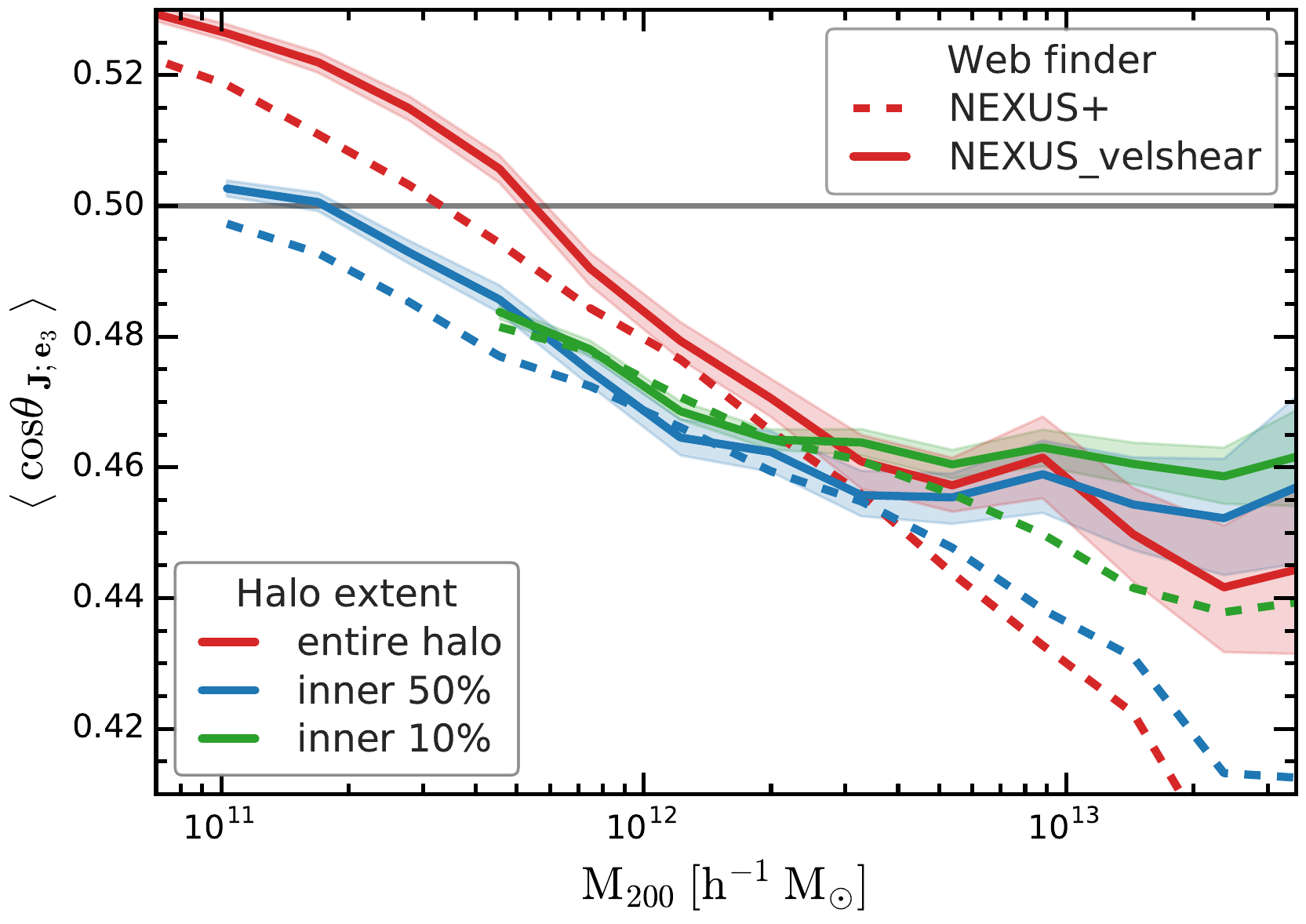}}
    \\[-0.85cm]
    {\tikz \fill [white] (0,0) rectangle (1.\linewidth,1.01cm);}
    \\[-1.cm]
	\mbox{\hskip -0.4truecm\includegraphics[width = 1.07\columnwidth]{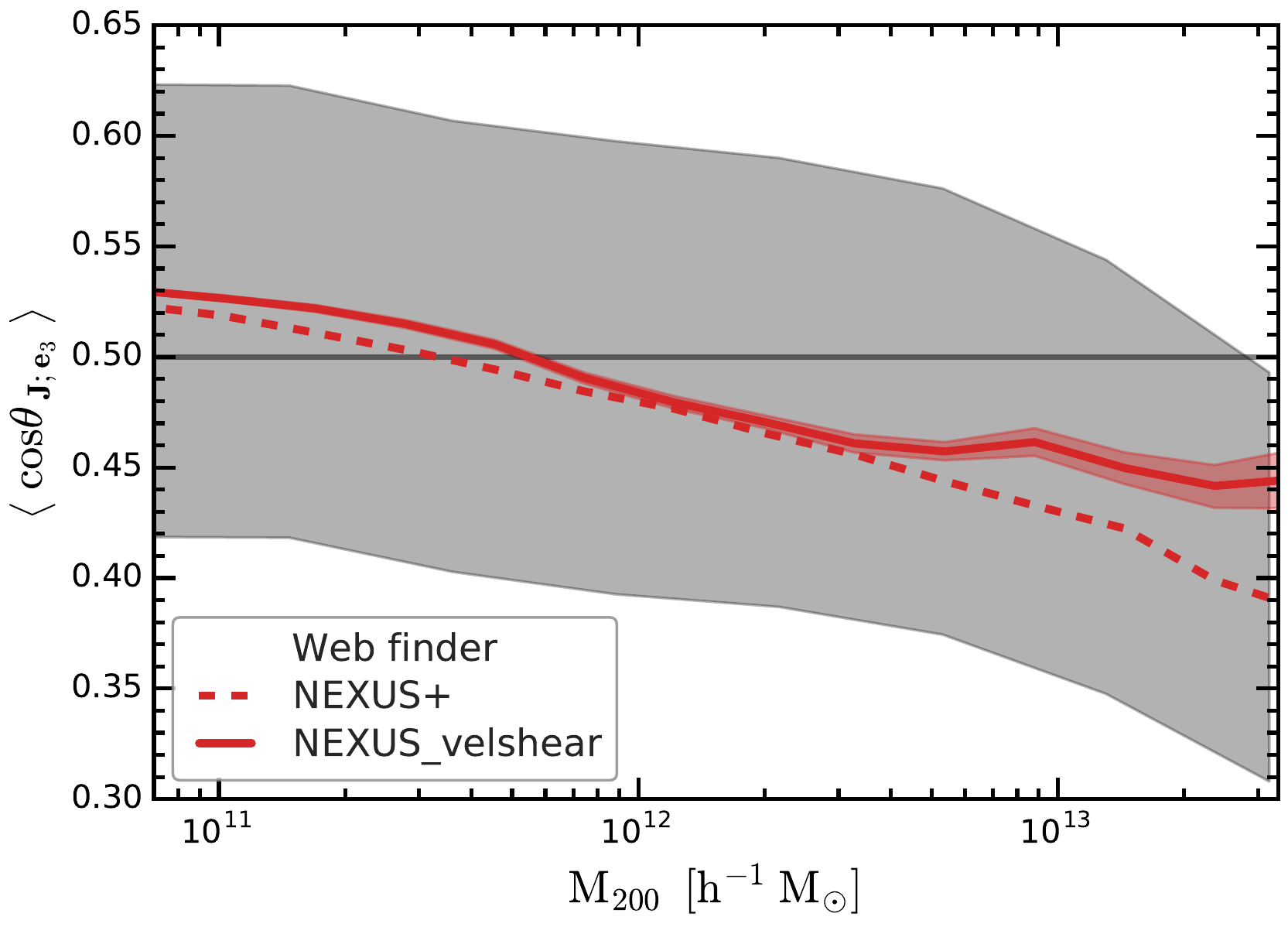}} 
    \vskip -.2cm
	\caption{ Median alignment angle, $\left\langle \cos\theta_{\mathbf{J};\mathbf{e_3}} \right\rangle$, between the angular momentum and filament orientation as a function of halo mass. It shows the alignment with the filament orientation identified by NEXUS+ (dotted line) and NEXUS velocity shear (solid line). The various colours correspond to different angular momentum definitions using the entire halo and using the innermost 50\% and 10\% of the particles. The coloured shaded region around each lines gives the 2-sigma bootstrap uncertainty in determining the median, which we only show for solid lines. The bottom panel (note the different y-axis) shows the 40 to 60 percentiles of the $\cos \theta_{\mathbf{J};\mathbf{e_3}}$ distribution, which is indicated via the grey shaded region.
    }
	\label{fig:median_ang_mom}
\end{figure}

We use bootstrapping to estimate the uncertainty in the distribution of alignment angles. For each distribution, we generate 1000 bootstrap realizations and compute the distribution and median values for each realizations. These are then used to estimate $1$ and $2\sigma$ uncertainty intervals. We apply this procedure for estimating PDF uncertainties (e.g. see \autoref{fig:multipanel_ang_mom_velshear}) as well as for calculating the error in the determination of the median value (e.g. see \autoref{fig:median_ang_mom}).

\subsection{Halo spin alignment: statistical trends}
\autoref{fig:multipanel_ang_mom_velshear} gives the distribution of the halo spin alignment angle, i.e. of  $\mu_{Jf}=\cos (\theta_{\mathbf{J};\mathbf{e}_3}) $,
between the halo spin directions and the filament orientation at the position of the haloes. The panels of the figure correspond to haloes of different masses. The figure shows the alignment only for \Nexus{}$\_$velshear filaments, but a nearly identical result is found for \nexus{} filaments. We study the alignment of the entire halo, as well as for inner radial cuts that contain 50\% and 10\% of the halo mass. In each case, we require at least 300 particles to determine the halo spin, which is why the spin for the 50\% and 10\% inner radial cuts is shown only for haloes more massive than $1$ and $3\times10^{11}~h^{-1} \rm M_{\odot}$, respectively.

For haloes in each mass range, we find that the alignment angle has a wide distribution, taking values over the full allowed range from $\cos\theta=0$ up to $\cos\theta=1$ (note that the y-axis only goes from 0.8 to 1.2). Nonetheless, the distribution is clearly different from an isotropic one, which is the case even when accounting for uncertainties due to the finite size of the sample, which are shown as the grey shaded region around the isotropic expectation value. The spin directions of low-mass haloes show an excess probability to have $\cos \theta_{\mathbf{J};\mathbf{e}_3} \simeq 1$, which indicates a tendency to be preferentially parallel to the filament spine. In contrast, high-mass haloes show an opposite trend, with an excess of objects with $\cos \theta_{\mathbf{J};\mathbf{e}_3} \simeq 0$, i.e. tendency to be preferentially perpendicular to the filament axis.  
\revised{To summarize, while we find a wide distribution of halo spin - filament orientations, there is a statistically significant excess of haloes that, depending on mass, have their spin preferentially parallel or perpendicular to their host filaments.}

The nature of the spin - filament alignment depends on halo mass. Many low-mass haloes, with masses in the range $M_{200}=(5 - 9) \times 10^{10}  h^{-1} \rm M_{\odot}$ (top left-hand panel of \autoref{fig:multipanel_ang_mom_velshear}) have alignment angles, $ \cos \theta \gtrsim 0.8$, which indicates their tendency to orient parallel to the filament spine. On the other hand, evaluating the alignment in the subsequent panels, which correspond to increasing halo mass, we observe a systematic shift from preferentially parallel to preferentially perpendicular configurations. For example, haloes with masses of $(3-4) \times 10^{11}~h^{-1} \rm{M_{\odot}}$ show a considerably weaker parallel alignment excess, while for halo masses of $(1-2) \times 10^{12}~h^{-1} \rm{M_{\odot}}$ and higher, most haloes have an alignment angle $ \cos \theta \lesssim 0.3$. 

The spin - filament alignment depends not only on halo mass, but also on the radial extent in which the halo spin direction is calculated. This is illustrated in \autoref{fig:multipanel_ang_mom_velshear}, which shows the spin-filament alignment calculated using the inner most 10\% and 50\% of the halo mass. While for high halo masses, $M_{200}>1\times10^{12}~h^{-1} \rm{M_{\odot}}$, the inner and the entire halo spins are aligned to the same degree with their host filament, at lower masses, $M_{200}<5\times10^{11}~h^{-1} \rm{M_{\odot}}$, the inner halo spin shows no preferential alignment. This is in contrast to the entire halo spin, which is preferentially parallel to the filament spine. The most remarkable contrast between the inner and outer halo spin orientations is found for objects in the mass range $(3-4) \times 10^{11} ~h^{-1} \rm M_{\odot}$ (third panel of \autoref{fig:multipanel_ang_mom_velshear}).  While the inner halo spin
has a slight tendency for a perpendicular alignment, the entire halo spin is oriented preferentially along the filament spine. 

In summary, the halo spin - filament alignment is mass dependent: low-mass haloes have a preferentially parallel alignment, while haloes of Milky Way mass and more massive have a preferentially perpendicular alignment. The latter fits with the tidal torque theory (TTT) which predicts that halo spin directions are perpendicular on the filament in which the haloes reside \citep{lee2000}. However, the spin - filament alignment of low mass haloes is opposite to the predictions of TTT. The picture is further complicated since the alignment of low-mass haloes depends on the radial extent used for calculating their spin, with the alignment changing from preferentially perpendicular in the inner region, which agrees with TTT predictions, to preferentially parallel in the outer region. The inner region consists of mostly early accreted mass while the converse is true for the outer region. This suggests that the initially induced halo spin during
the linear evolution phase \citep{peebles1969} is substantially modified by subsequent mass accretion stages. Particularly outstanding in this respect is the contrast between low- versus high-mass haloes, with the spin - filament alignment of the latter being less disturbed by recent accretion. 

\subsection{The spin flip}
We now proceed to study in more detail the dependence on halo mass of the halo spin - filament alignment. This is shown in \autoref{fig:median_ang_mom}, where we plot the median spin - filament alignment angle, $\left\langle \mu_{Jf} \right\rangle =\left\langle \cos\theta_{\mathbf{J};\mathbf{e}_3} \right\rangle$, calculated using narrow ranges in halo mass. To assess the statistical robustness of the median alignment angle, we show the $2\sigma$ uncertainty in the median value, which is calculated using bootstrap sampling. The uncertainty range is small, especially at low masses, which is due to the large number of haloes in each mass range. For clarity, we only show the uncertainty in the alignment with \Nexus{}$\_$velshear filaments, but roughly equal uncertainties are present in the alignment with \nexus{} filaments. The threshold between preferentially parallel and perpendicular alignments corresponds to $\left\langle \cos\theta_{\mathbf{J};\mathbf{e}_3} \right\rangle = 0.5$ and is shown with a horizontal solid line in \autoref{fig:median_ang_mom}.

\autoref{fig:median_ang_mom} shows a clear and systematic trend between halo mass and the median spin - filament alignment angle: the alignment angle, $\left\langle \cos\theta_{\mathbf{J};\mathbf{e}_3} \right\rangle$ increases with decreasing halo mass. This trend is visible for both \nexus{} and  \Nexus$\_$velshear filaments, although the exact median angles vary slightly between the two methods. Especially telling is the transition from preferentially perpendicular alignment at high masses to a preferentially parallel alignment at low masses, which takes place at $M_{200}=5.6$ and $3.8 \times 10^{11}~{h}^{-1} \rm{M}_{\odot}$ for \Nexus{}$\_$velshear and \nexus{} filaments, respectively. This transition is known as \emph{spin flip} and has been the subject of intense study \citep{aragon2007,hahn2010,codis2012,trowland2013}. The exact value of the spin flip halo mass varies between studies, and, as we found here, it varies between the two web finders employed here. In the next subsection we investigate this difference in more detail.

\autoref{fig:median_ang_mom} also shows the spin - filament alignment for the inner halo, whose strength and mass dependence is different from that of the entire halo. The difference between the inner and entire halo spin alignment is most pronounced for low mass haloes, in line with the conclusions of \autoref{fig:multipanel_ang_mom_velshear}. For example, the spin of the inner 10\% of the halo mass shows little mass dependence for high masses, after which it slowly increases from preferentially perpendicular towards preferentially parallel alignment with the filament spin. However, due to the limited resolution of the simulation (we need at least 300 particles in the inner 10\% region of the halo), we cannot probe if there is a spin flip and at what halo mass it takes place. However, for the spin - filament alignment of the inner 50\% of the halo mass, we just resolve the spin flip, which takes place at masses a factor of~ ${\sim}3$ times lower than the spin flip of the entire halo. 

The systematic nature of the \emph{spin flip} is a clear indication of the significant role played by additional physical processes not captured by TTT in  determining the final angular momentum of haloes. The spin - filament alignment of high-mass haloes is, at least qualitatively, in agreement with TTT, so it is unclear what is the effect, if any, of additional processes not included in TTT. In contrast, the alignment of low-mass haloes is contrary to TTT predictions, suggesting that the spin acquired during the linear evolution phase, which is well described by TTT, gets modified by additional phenomena that result in a gradual transition towards spins aligned with the filament spine. The major keys to the dynamics of this process are to be found in the contrast between the
spin of the inner and outer halo regions, as well as in the variation of the alignment strength between different regions of the filamentary
network, which is the topic of the next subsection.

\subsection{Spin alignment and the nature of filaments}
\label{sec:different_fila_population}
Here we investigate how the spin - filament alignment varies with the filament properties, focusing on two crucial aspects. First, we study what explain the small, but statistically significant, variation in \emph{spin flip} mass between the \nexus{} and \Nexus{}$\_$velshear filaments (see \autoref{fig:median_ang_mom}). Secondly, we study if the halo spin - filament alignment is sensitive to the filament type in which a halo is located, focusing on prominent versus tenuous filaments.

\begin{figure} 
	\mbox{\hskip-0.4truecm\includegraphics[width=1.05\columnwidth]{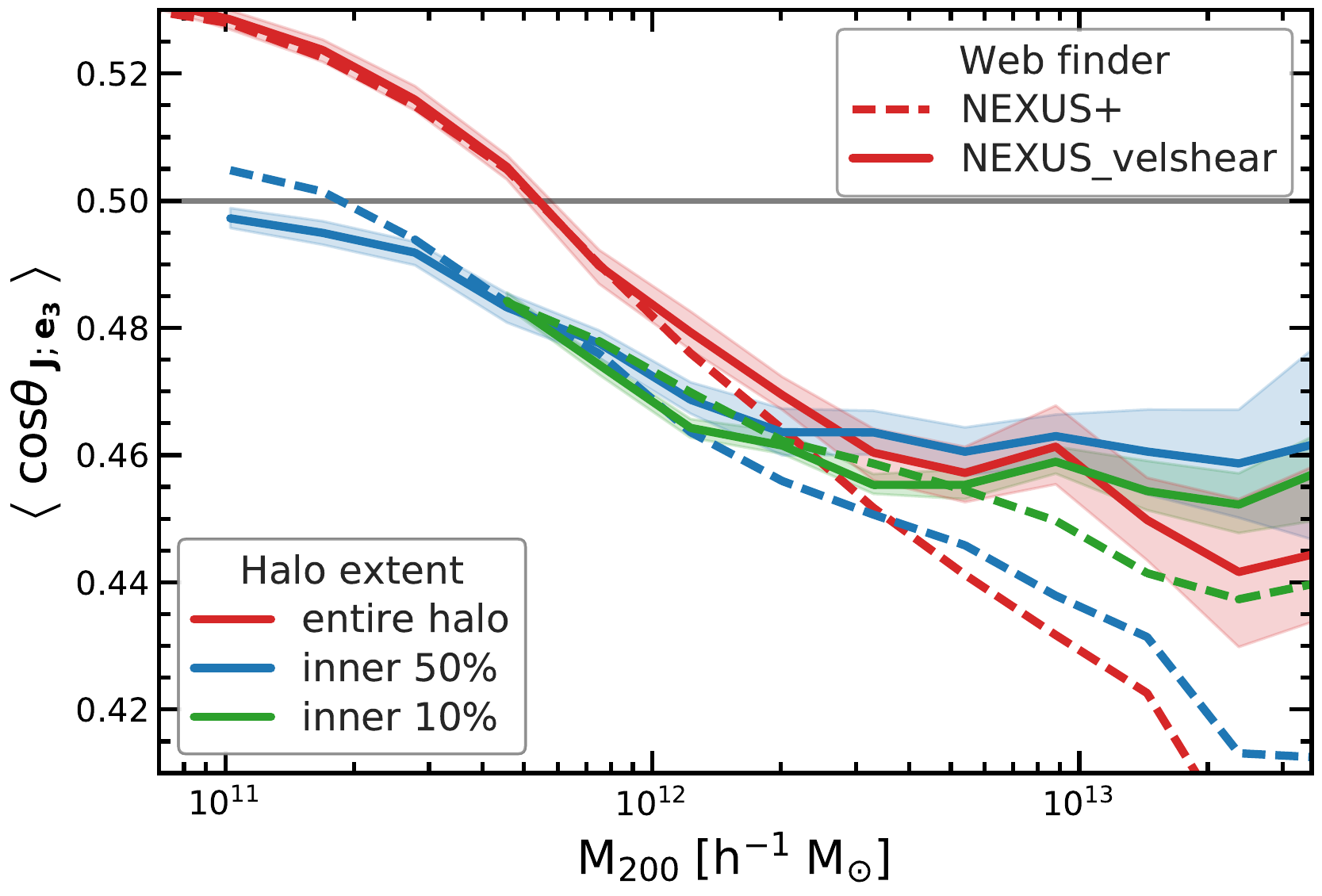}}
	\vskip -0.25cm
	\caption{ Median alignment angle, $\left\langle \cos\theta_{\mathbf{J};\mathbf{e_3}} \right\rangle$, between halo spin and filament orientation for \textit{common} haloes found to reside in both NEXUS+ and NEXUS velocity shear filaments. Note that for $M_{200}<10^{12}\hMsun$ the alignment strength of the entire halo (solid and dashed red curves) is independent of the filament identification method implying that the differences seen in \autoref{fig:median_ang_mom} are due to the two web finders assigning somewhat different haloes to filaments.
    }
	\label{fig:common halos}

	\vskip 0.3cm
	\mbox{\hskip -0.4truecm\includegraphics[width=1.05\columnwidth]{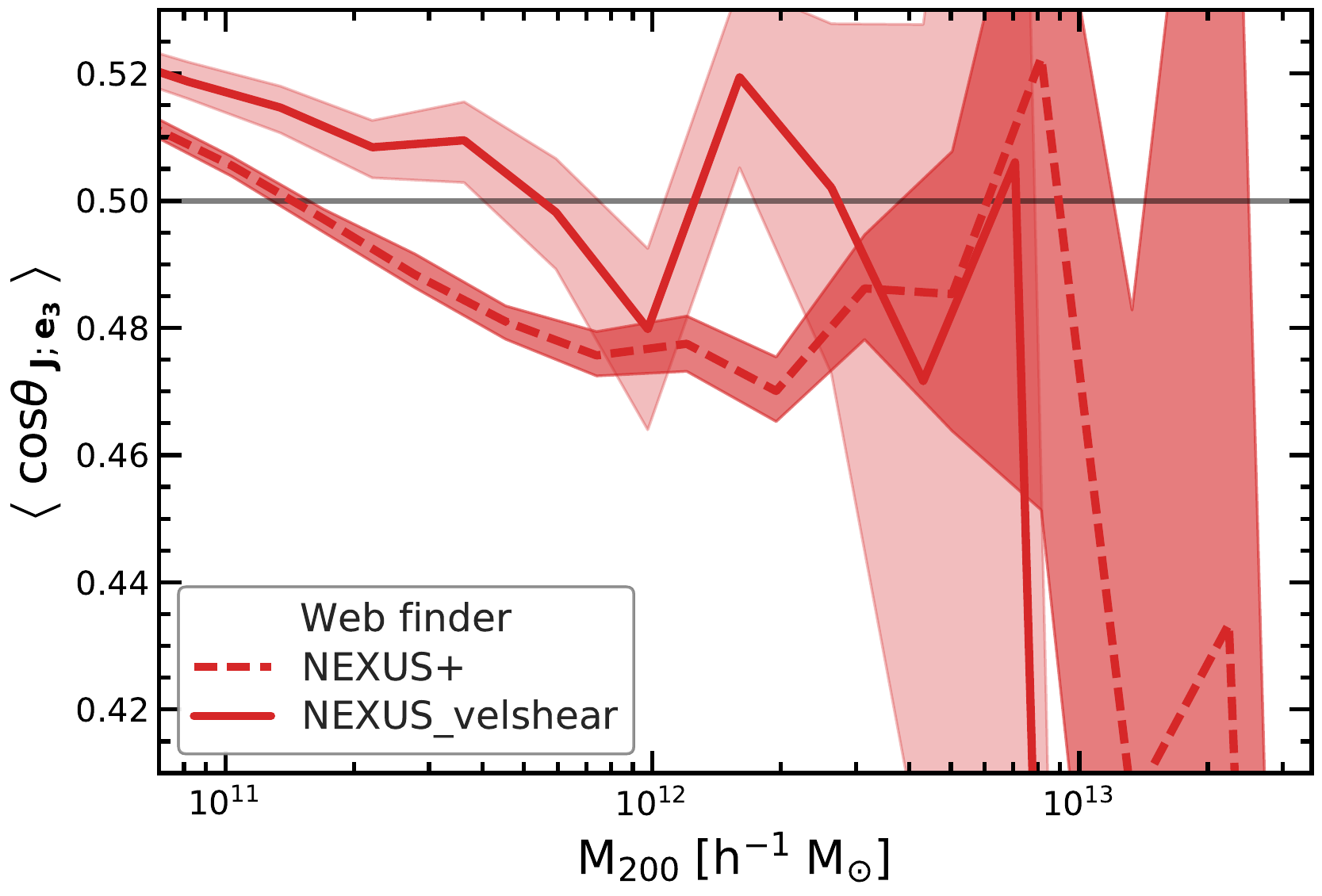}}
	\vskip -0.25cm
	\caption{ Median alignment angle, $\left\langle \cos\theta_{\mathbf{J};\mathbf{e_3}} \right\rangle$, between halo spin and filament orientation for \textit{exclusive} haloes, that is haloes that reside \textit{only} in NEXUS+ (dashed line) or \textit{only} in NEXUS$\_$velshear (solid line) filaments. For clarity, we only show the alignment of the entire halo spin.
    For NEXUS$\_$velshear, the exclusive haloes have roughly the same spin flip mass as the total population of filament haloes. In contrast, for \nexus{}, the exclusive haloes have a ${\sim}4$ times smaller spin flip mass than the total population of filament haloes.
    } 
	\label{fig:unique_halos}
\end{figure}

%=============================================
% -------TABLE FOR TRANSITION MASS-------
%=============================================
\begin{table}
\centering
\caption{\revised{The values of the spin flip mass which determines the transition of the halo spin - filament alignment from preferentially parallel at low halo masses to preferentially perpendicular at high halo masses. We specify the spin flip masses for the two filament population studied here as well as for various halo sub-samples. We also give the spin flip mass for NEXUS+ filaments of different thickness.}}
\label{table:spin_flip_mass}
\begin{tabular}{ @{}l cc@{}} 

\hline
 \\[-0.5cm]
 \hline
Web finder & Halo subsample &  Spin flip mass \\
  & &  $[~\times 10^{11} \rm{M_\odot}~]$    \\
 \hline
\multirow{3}{*}{NEXUS+}  & all & $3.3$  \\ %[.2cm]
 & common & $5.4$ \\
 & exclusive & $1.4$ \\[.5cm]
 \multirow{3}{*}{NEXUS\_velshear}  & all & $5.4$  \\
 & common & $5.4$ \\
 & exclusive & $5.6$   \\
 \hline \\[-.2cm]
% & Filament diameter & \\
 & $[h^{-1}\rm{Mpc}]$ & \\
% \hline \\
 \multirow{3}{*}{NEXUS+ filament thickness}  & $(0 - 2)$ & $1.8$  \\
 & $(2-4)$ & $5.2$ \\
 & $( \; > \;4)$ & $18$ \\
 \hline
\end{tabular}
\end{table}

\subsubsection{\nexus{} vs. \Nexus{}$\_$velshear filaments.}

There are two sources of difference between the two filament populations. First, even if a \nexus{} filament overlaps with a \Nexus{}$\_$velshear one, they do not necessarily have the same orientation, since the filament orientation is given by the eigenvectors of the density gradient and velocity shear fields, respectively. However, the density gradient and velocity shear are reasonably well aligned, with a median alignment angle of ${\sim}22$ degrees \citep{Tempel2014a}. Secondly, the two filaments contain different halo populations. As we discussed in \autoref{sec:halo_environ}, \nexus{} filaments include many thin filamentary tendrils, either branching off from more prominent filaments or residing in low density regions. These tenuous structures, which are mostly populated by low-mass haloes, are not identified by \Nexus{}$\_$velshear. In contrast, the \Nexus{}$\_$velshear formalism includes a fair number of haloes far from the ridge of prominent filaments \citep[see][]{cautun2014}; these haloes would typically be assigned by \nexus{} to neighbouring low-density areas (see \autoref{fig:haloes_in_web_zoomIn}).

\autoref{fig:common halos} studies the impact of halo population on the spin - filament alignment. It shows the halo mass dependence of the spin - filament alignment for \textit{common} haloes, which are haloes that are assigned to both \nexus{} and \Nexus{}$\_$velshear filaments (see \autoref{fig:haloes_in_web_zoomOut} for an illustration of the spatial distribution of these haloes). For masses, $M_{200}\leq10^{12}~{h}^{-1} \rm{M}_{\odot}$, the common haloes have the same median spin - filament alignment angle for both web finders, to the extent that the curves almost perfectly overlap each other. This translates into an agreement on the spin flip transition mass, at $M_{200} = 5 \times 10^{11} ~{h}^{-1} \rm{M}_{\odot}$. 
This result demonstrates that there are no fundamental differences between the spin - filament alignment of low-mass common haloes, whether the filaments are identified by \nexus{} or \Nexus{}$\_$velshear methods. 

The story is different for haloes more massive than $10^{12}~{h}^{-1} \rm{M}_{\odot}$, where the spin - filament alignment of common haloes is the same as that of the full filament population. In particular, while both web finders find that halo spins are preferentially perpendicular on their host filaments,
the spin - filament alignment using \nexus{} orientations is stronger (i.e. more perpendicular) than that using \Nexus{}$\_$velshear orientations, and this discrepancy increases at higher halo masses. This is a manifestation of the differences in orientations between \nexus{} and \Nexus{}$\_$velshear filaments \revised{(see \autoref{fig:nexus_velshear_alignment})}, with \nexus{} being able to recover better the orientation of filaments around massive haloes. These haloes, due to their high mass, affect the mass flow around themselves and thus locally change the large-scale velocity shear field. In turn, this diminishes the ability of the \Nexus{}$\_$velshear web finder to recover the orientation of the large-scale filaments.
More massive haloes change the velocity flow to a larger extent and to larger distances, which explains why the difference between the two filament finders increases at higher halo masses.

\autoref{fig:unique_halos} studies the mass-dependence of the spin - filament alignment of \emph{exclusive} haloes, that is haloes assigned exclusively to \nexus{} or to \Nexus{}$\_$velshear filaments. We focus our discussion on haloes with $M_{200} \leq 2\times10^{12}~{h}^{-1} \rm{M}_{\odot}$ since the exclusive halo sample contains a small number of higher mass objects, which is a consequence of the fact that most massive haloes are assigned to filaments by both methods. In contrast to common haloes, which reside typically in the central region of prominent filaments, the exclusive halo population is very different between the two web finders. 

The \nexus{} exclusive sample, which consists of haloes in tenuous filamentary tendrils, shows preferentially perpendicular alignments down to very low masses, with the spin flip mass being ${\sim}1\times10^{11}~{h}^{-1} \rm{M}_{\odot}$. This transition mass is much lower than the corresponding mass of all the \nexus{} filament haloes, which is ${\sim}3\times10^{11}~{h}^{-1} \rm{M}_{\odot}$. \revised{The spin flip mass for haloes in different filament populations is presented in \autoref{table:spin_flip_mass}.} Thus, the spin - filament alignment depends on filament properties, with same mass haloes being more likely to have a preferentially perpendicular configuration if they reside in a thinner filament (the next sub-section discusses this trend in more detail). 

The \Nexus{}$\_$velshear exclusive sample, which consists of mostly haloes found at the outskirts of prominent filaments, shows a spin flip mass of ${\sim}6\times10^{11}~{h}^{-1} \rm{M}_{\odot}$. This spin flip mass is a factor of $6$ times higher than that of \nexus{} exclusive haloes, and thus substantiates the hypothesis that the spin alignment depends on the nature of filaments. Furthermore, the spin flip of the \Nexus{}$\_$velshear exclusive sample has the same value as that of the \Nexus{}$\_$velshear common sample (see \autoref{fig:common halos}). Both samples reside in the same filaments, but the former is typically found in the outskirts, that is farther from the filament spine. Thus, comparing the two \revised{suggests that the spin flip mass} does not vary strongly with the distance from the filament spine.

\begin{figure}
	\mbox{\hskip -0.4truecm\includegraphics[width=1.05\columnwidth]{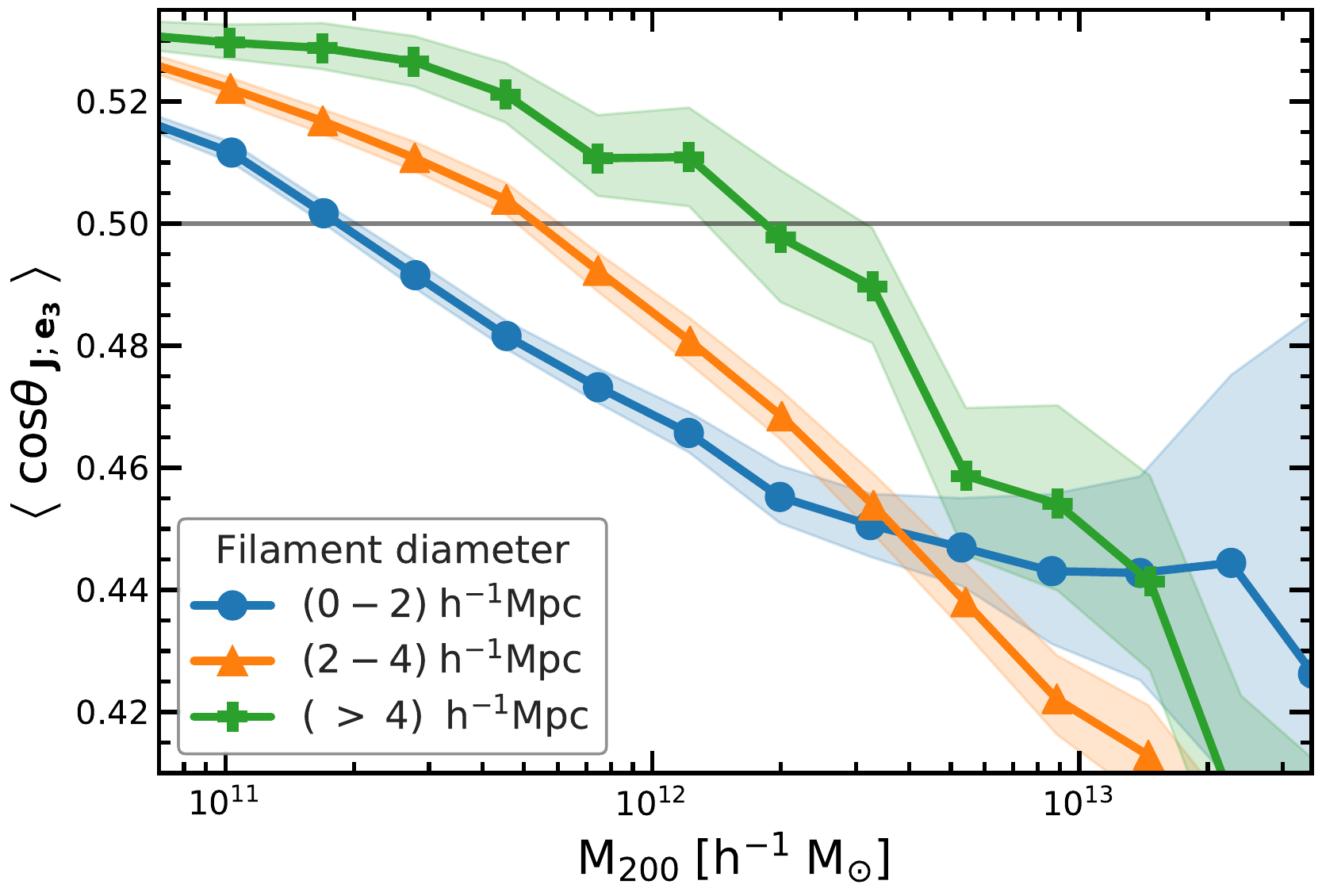}}
	\vskip -0.25cm
	\caption{ Median alignment angle, $\left\langle \cos\theta_{\mathbf{J};\mathbf{e_3}} \right\rangle$, between halo spin and filament orientation when splitting the sample according to the filament diameter. The three curves show the alignment signal for haloes in filaments with diameters: (0 - 2)$\hmpc$ (solid line with circles), (2 - 4)$\hmpc$ (solid line with triangle symbols) and more than $4\hmpc$ (solid line with crosses). The transition halo mass from preferentially parallel to preferential perpendicular alignment increases with increasing filament diameter. 
	}
	\label{fig:filament thickness}

	\mbox{\hskip -0.4truecm\includegraphics[width=1.05\columnwidth]{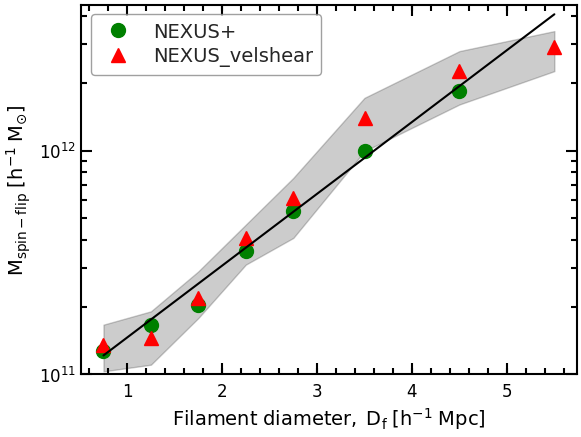}}
	\vskip -0.25cm
	\caption{\revised{The dependence of the spin flip mass, $M_{\rm spin-flip}$, on the filament diameter, $D_{\rm f}$, in which the haloes reside. The two symbols correspond to NEXUS+ (circles) and NEXUS\_velshear (triangles) filaments. The grey shaded region shows the $2\sigma$ error in the determination of $M_{\rm spin-flip}$ for NEXUS\_velshear filaments. The spin flip mass dependence on $D_f$ is well fitted by \autoref{eq:Mspin_filament_diameter}, with the solid line showing the best fit.} 
    %The fit parameters are $m = 0.32$  and $c =  10.8 $. 
	}
	\label{fig:flip_mass_fila_thickness}
\end{figure}

\subsubsection{Alignment \& filament thickness}
\label{sec:filament thickness}
We now carry out a detailed investigation of the hypothesis proposed in the previous sub-section that the spin - filament alignment depends on the nature of filaments. In particular, we study if the alignment of same mass haloes depends on the thickness of the filament in which the haloes are embedded. 
This is shown in \autoref{fig:filament thickness}, where we present the mass-dependence of the spin - filament alignment for halo subsamples split according to the diameter of their host filament. The filament diameter was determined following the \citet{cautun2014} prescription. 
\revised{In a first step, we compress the filaments to their central spine. This involves an iterative procedure where for each iteration step all filament voxels are shifted closer to the filament centre until resulting into a very thin curve, which is the filament spine. In a second step, for each voxel along the filament spine we find the number of neighbouring voxels within a radius of $R=2\hmpc$. 
Then, the filament diameter, $D_{\rm f}$, at that point is given by the diameter of a cylinder of length, $2R$, that has the same volume as the total volume of the neighbouring voxels. The filament diameter associated to each halo is the one corresponding to the voxel in which the halo is located.} 
For simplicity, we focus the analysis on the alignment of the entire halo spin.

\autoref{fig:filament thickness} shows an immediately obvious trend: over nearly the entire mass range, the spin - filament alignment angle, $\left\langle \cos\theta_{\mathbf{J};\mathbf{e_3}} \right\rangle$, is systematically lower for haloes in thin filaments than for those in thick filaments. Thus, same mass haloes tend to have their spin more perpendicular to the filament spine if they reside in a thinner filament. In particular, it is striking the systematic variation in the spin flip transition mass, which varies by an order of magnitude between different filaments: from $ 1.8 \times 10^{11}~h^{-1}M_{\odot}$ for the thinnest filaments to $1.8 \times 10^{12}~h^{-1}M_{\odot}$ for the thickest filaments \revised{(see \autoref{table:spin_flip_mass})}. 
\revised{This is clearly shown in \autoref{fig:flip_mass_fila_thickness} where we show the dependence of the spin-flip mass, $M_{\rm spin-flip}$, on filament diameter. We find that both NEXUS+ and NEXUS\_velshear filaments of the same thickness have approximatively the same spin flip mass. This mass increases systematically with filament diameter, $D_{\rm f}$, and is well described by the linear functional form
\begin{equation}
	\log_{10} M_{\rm spin-flip} / (h^{-1}M_{\odot}) = m D_{\rm f} + c,
    \label{eq:Mspin_filament_diameter}
\end{equation}
with the best fitting parameters having the values $m=0.32~\rm{h}\;\rm{Mpc}^{-1}$ and $c=10.8$. 
}
%The fit parameters are $m = 0.32$  and $c =  10.8 $. 

Thicker filaments are more massive since they typically contain a higher mass per unit length \citep{cautun2014}, and we  expect that they form in regions with a strong tidal field. Thus, we would expect that thicker filaments would host halo spins that are more perpendicular on their filament spine than in the case of thinner filaments. This is opposite to the results of \autoref{fig:filament thickness} and suggests that additional processes, like mergers and secondary or late mass accretion, have a substantial impact on the orientation of halo spins.

%===========================
% --FIG:SHAPE PDF,MEDIAN--
%===========================
\begin{figure}
	\mbox{\hskip -0.4truecm\includegraphics[width = 1.05\columnwidth]{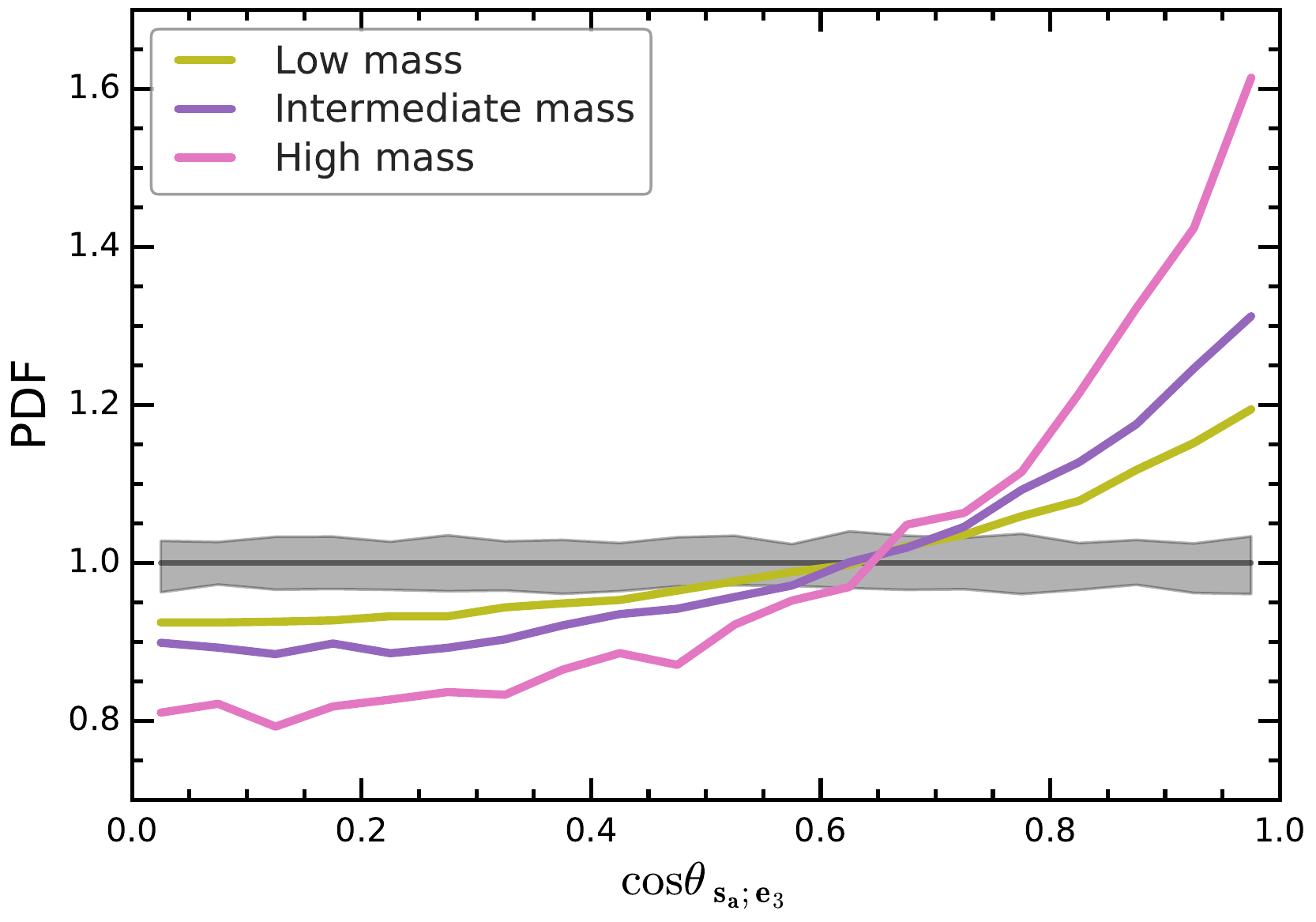}}\\
	\vskip -0.3cm
	\mbox{\hskip -0.4truecm\includegraphics[width = 1.05\columnwidth]{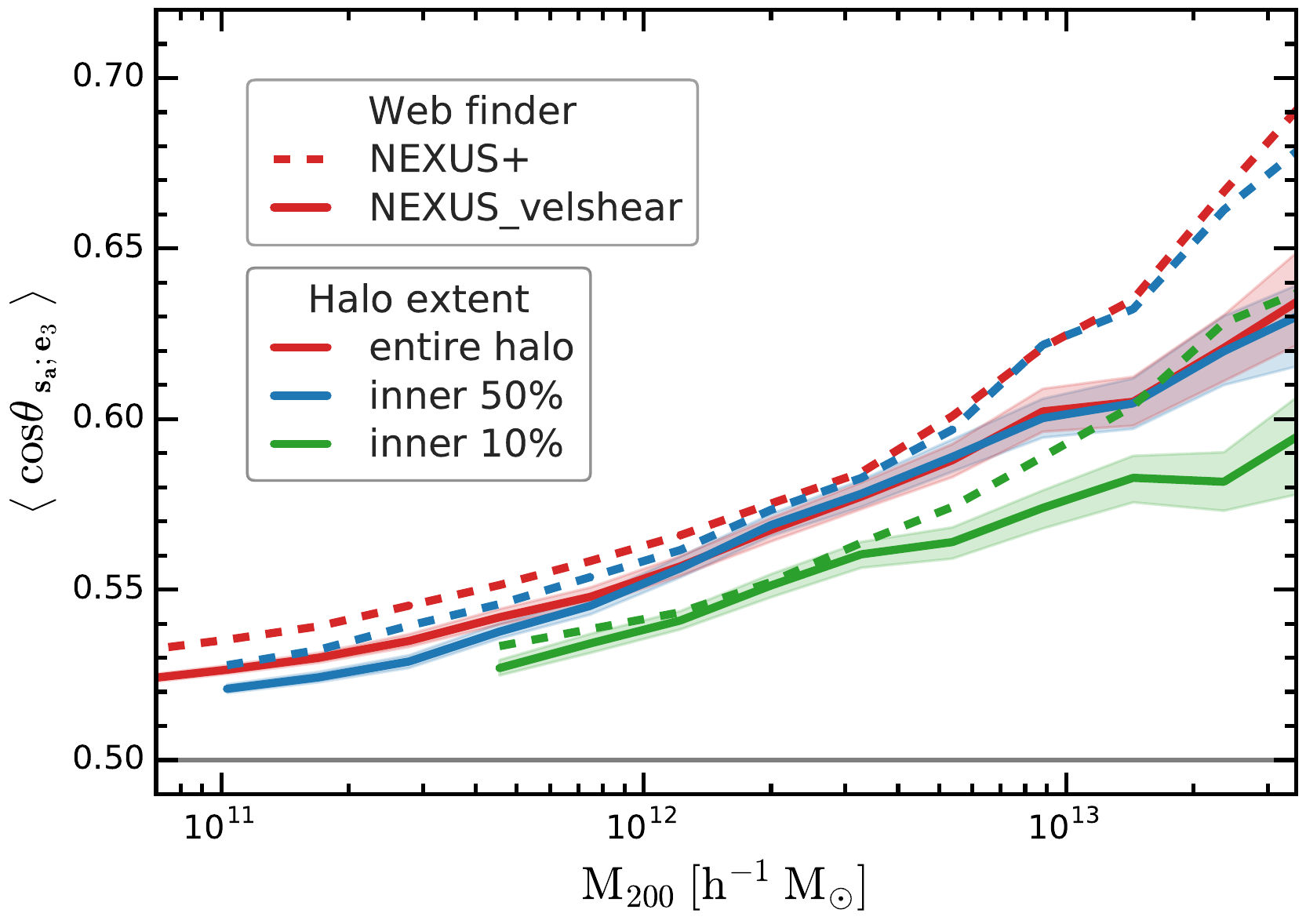}}\\
	%\vskip -0.3cm
        \\[-1.2cm]
    {\tikz \fill [white] (0,0) rectangle (1.\linewidth,1.01cm);}
    \\[-1.cm]
	\mbox{\hskip -0.4truecm\includegraphics[width = 1.05\columnwidth]{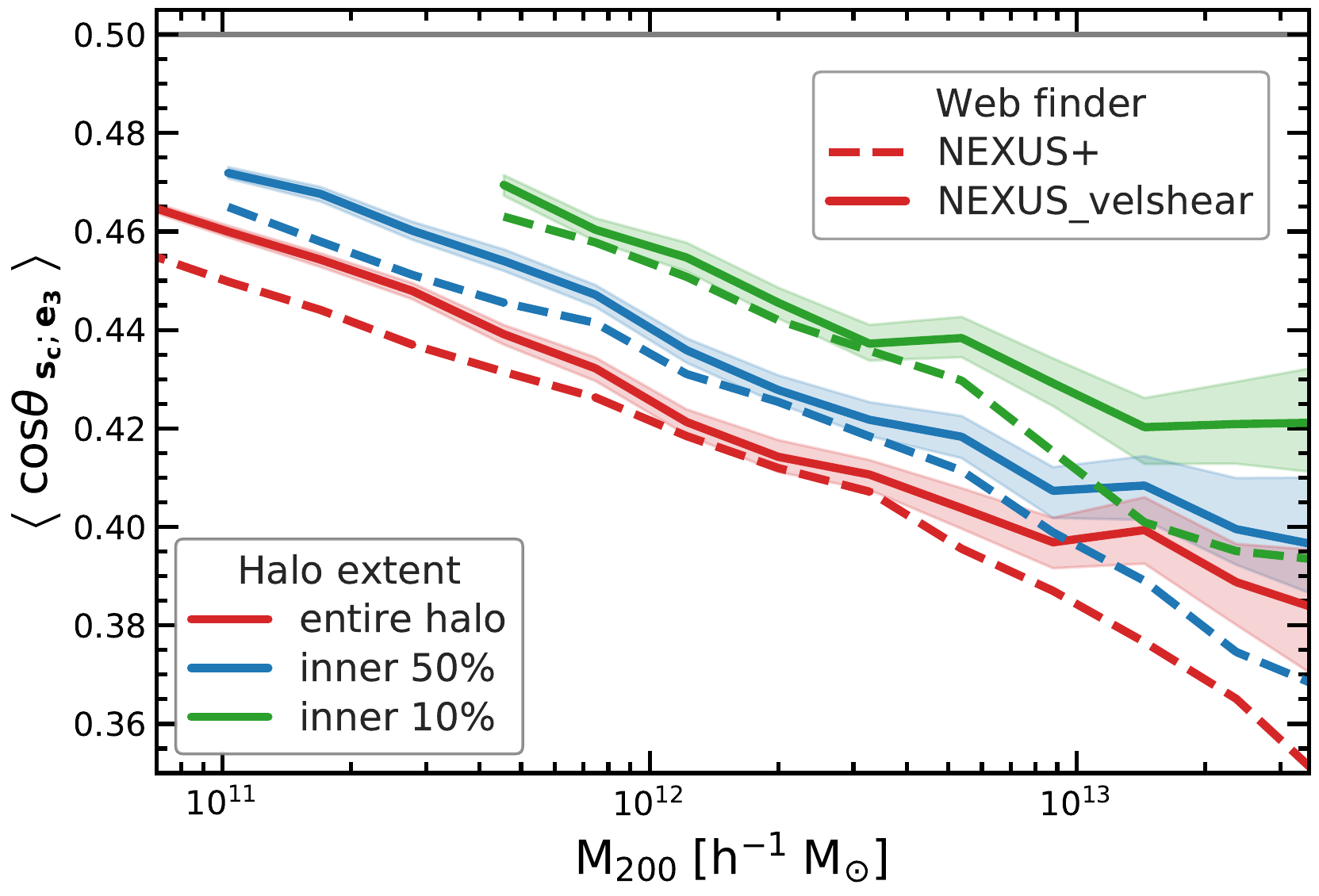}}
	\vskip -0.3cm
  	\caption{ The alignment of halo shape with the filament orientation. \textit{Top panel:} distribution of alignment angle, $\cos \theta_{\mathbf{s_a};\mathbf{e_3}}$, between the halo long axis and the \nexus{} filament orientation for haloes in three mass ranges: low mass, $M_{200} = (5-9) \times 10^{10}$ \massUnit{}, intermediate mass, $ (3-5) \times 10^{11}$ \massUnit{}, and high mass, $ (3-5) \times 10^{12}$ \massUnit{}. \textit{Centre panel:} median alignment angle, $\left\langle\cos \theta_{\mathbf{s_a};\mathbf{e_3}}\right\rangle$, between halo long axis and filament orientation as a function of halo mass. It gives the median for both \nexus{} and \Nexus{}$\_$velshear filaments as well as for different radial extents of the halo. It shows the strong tendency of haloes to have their long axis oriented along that of filaments and that the alignment becomes larger for higher halo masses. The trends are similar for the entire halo, as well as for the inner parts of the haloes. \textit{Bottom panel:} same as centre panel, but for the median alignment angle, $\left\langle\cos \theta_{\mathbf{s_c};\mathbf{e_3}}\right\rangle$, between halo short axis and filament orientation.
    }
	\label{fig:median_shape_alignment}
\end{figure}

\section{Halo shape alignment}
\label{sec:shape_alignment}
The alignment of the halo shape with the large-scale mass distribution represents a complementary aspect to the spin - filament alignment. Here, we focus on two aspects related to the orientation of haloes: 
\begin{itemize}
\item{} the halo shape - filament alignment, and
\item{} the halo shape - halo spin alignment.
\end{itemize}
\noindent The shape and orientation of a halo is specified in terms of its three principal axes $a$, $b$ and $c$, and the corresponding eigenvectors (see \autoref{sec:shape}). Of particular interest are the longest  and the shortest axes. The longest axis, $a$, is the one that specifies the orientation along which the main body of the halo is pointing. The shortest axis, $c$, is preferentially oriented in the same direction as the halo spin and their mutual misalignment reflects the history of the angular momentum acquisition by the halo. It is also of interest to see in how far the shortest halo axis emulates the halo spin - filament alignment.

\subsection{Halo shape - filament alignment}
Already in the initial Gaussian field there is a strong correlation between the shape of peaks and the surrounding cosmic matter distribution \citep{weyedb1996, rossi2009}. For example, an emerging filament is defined by a primordial configuration of the tidal or velocity shear fields with one expanding and two contracting directions, with the former corresponding to the filament axis \citep[see][]{weyedb1996, rossi2009}. As pointed out by \cite{bond1996}, this identification is the principal reason why prominent filaments form between and connect pairs of massive clusters
\citep[see][for an extensive theoretical description]{vdw2008}.

Following the non-linear collapse and build-up of haloes, we wish to see in how far the alignment between the shape of haloes and the filaments in which they reside still reflects the primordial alignment. To this end, we evaluate the angle between the principal axes of haloes, $\mathbf{v_a}$, $\mathbf{v_b}$, and $\mathbf{v_c}$ (see section~\ref{sec:shape}), and the filament orientation, $\mathbf{e_3}$, which specifies the direction along the ridge of the filament (see section~\ref{sec:filament}). 
Similar to the spin - filament alignment, we characterize the shape - filament alignment in terms of the cosine of the angle between halo shape principal axes and the filament orientation (see eq. \ref{cos}).  

\autoref{fig:median_shape_alignment} reveals the tendency of the halo shape to be aligned with the filament ridge. The top panel of the figure shows the distribution of alignment angles, $\cos \theta_{\mathbf{s}_a;\mathbf{e}_3}$, between the halo major axis and the filament orientation. For all halo mass bins, the alignment angle distribution is broad, reflecting the wide range of halo - filament orientations. At the same time, the plot shows an excess of objects with $\cos \theta_{\mathbf{s}_a;\mathbf{e}_3} \simeq 1$, which reveals the tendency of the major axis of haloes to be aligned preferentially parallel to their host filaments \citep{hahn2007,Shao2016}. The alignment is mass dependent, being most pronounced for high-mass haloes. We further investigate the mass dependence in the middle panel of \autoref{fig:median_shape_alignment}, where we show the median alignment angle, $\left\langle\cos \theta_{\mathbf{s_a};\mathbf{e_3}}\right\rangle$, as a function of halo mass. It shows how high-mass haloes are strongly aligned with their host filaments, while the lowest mass ones show a much weaker, almost random, alignment with their host filament. The major axis - filament alignment is the largest for the entire halo, and becomes weaker when considering inner halo radial cuts. This is expected, since the outer region of the halo consists of mostly recently accreted mass, which fall in preferentially along the filaments in which a halo is embedded \citep{Aubert2004,Libeskind2005,Rieder2013}. Late time accretion is most anisotropic in higher mass haloes, which explain the mass dependence of the major axis - filament alignment \citep{Kang2015,Wang2018a}. \autoref{fig:median_shape_alignment} also shows that the haloes are aligned to almost identical degrees to both NEXUS$\_$velshear and \nexus{} filaments, with the shape - filament alignment being slightly stronger in the latter case, especially at high halo masses.

The bottom panel of \autoref{fig:median_shape_alignment} illustrates the alignment between the halo minor axis and the filament ridge. Unsurprisingly, the minor axis of haloes is preferentially perpendicular on their host filament, with the alignment being the strongest for the highest mass objects. 
%\revised{\sout{Furthermore, the minor axis is perpendicular to the filament to a larger extent than the major axis is parallel to the filament. For example, haloes of mass, $M_{200}=10^{13}$ \massUnit{}, embedded in \nexus{} filaments have alignment angles for the major and minor axes of $\theta_{\mathbf{s_a};\mathbf{e_3}}= 50.9^\circ$ and $90^\circ - \theta_{\mathbf{s_c};\mathbf{e_3}} = 22.3^\circ$, respectively, with the latter clearly much smaller.}}

The plots of \autoref{fig:median_shape_alignment} show that the level of alignment of halo shapes with the tidal field has increased considerably with respect to that present in the primordial Gaussian field. This is due to non-linear evolution, which has lead to substantial changes in the orientation of haloes \citep[see][]{haarlem1993}. There is also a rather strong and
systematic increase in alignment as a function of halo mass: more massive haloes are more strongly oriented along the filaments in which they reside. This
may be partially a reflection of primordial conditions, in which the tidal shear at a given location is more strongly correlated with the orientation
of more substantial peaks \citep{bbks,weyedb1996}. More important, however, may be the subsequent anisotropic nature of the accretion of mass and
substructure \citep{haarlem1993,Shao2018}, which, since it takes places mostly along filaments, amplifies the halo shape - filament alignment.

\begin{figure}
	\mbox{\hskip -0.4truecm\includegraphics[width = 1.05\columnwidth]{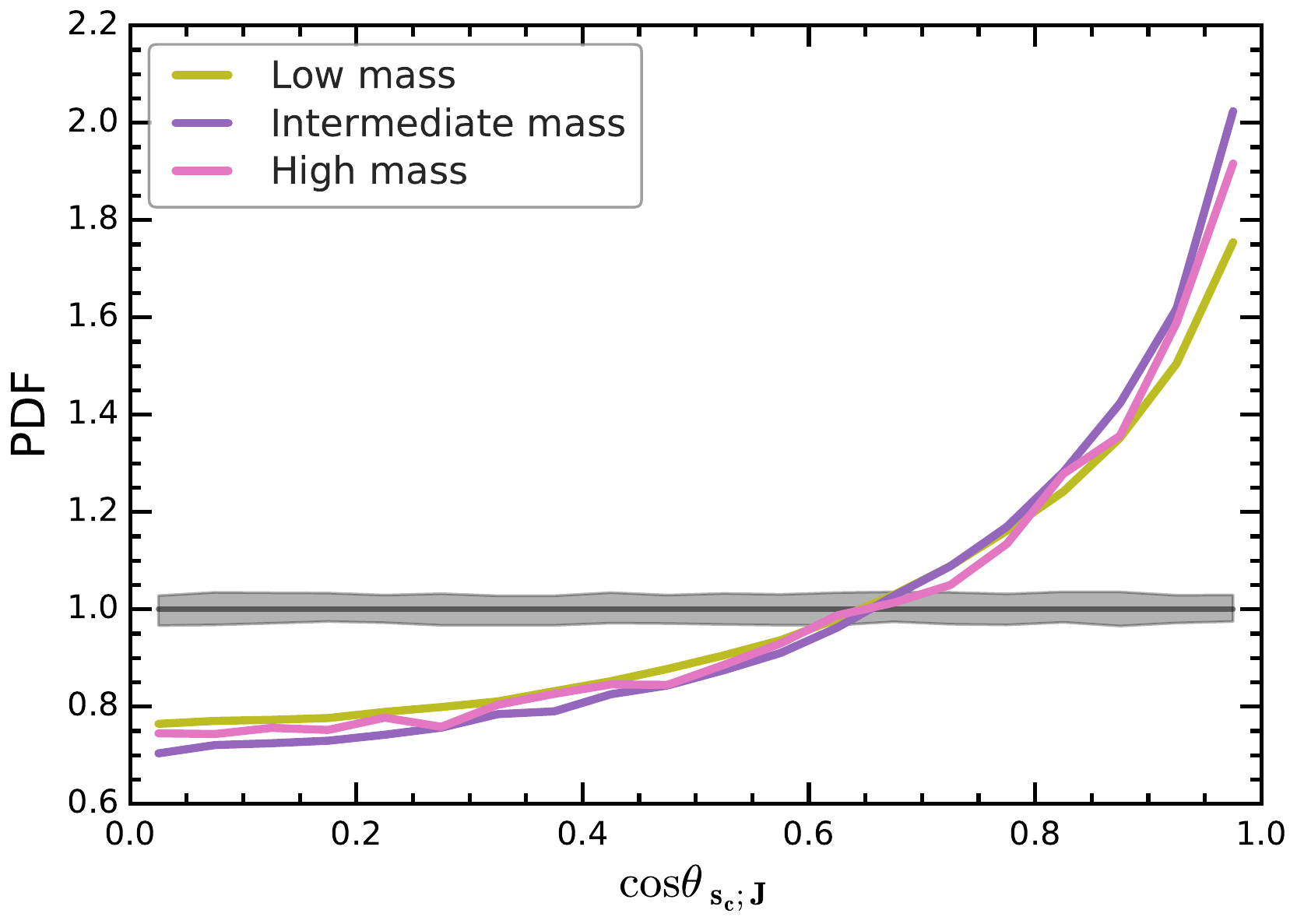}}\\
	\vskip -0.3cm
	\mbox{\hskip -0.4truecm\includegraphics[width = 1.05\columnwidth]{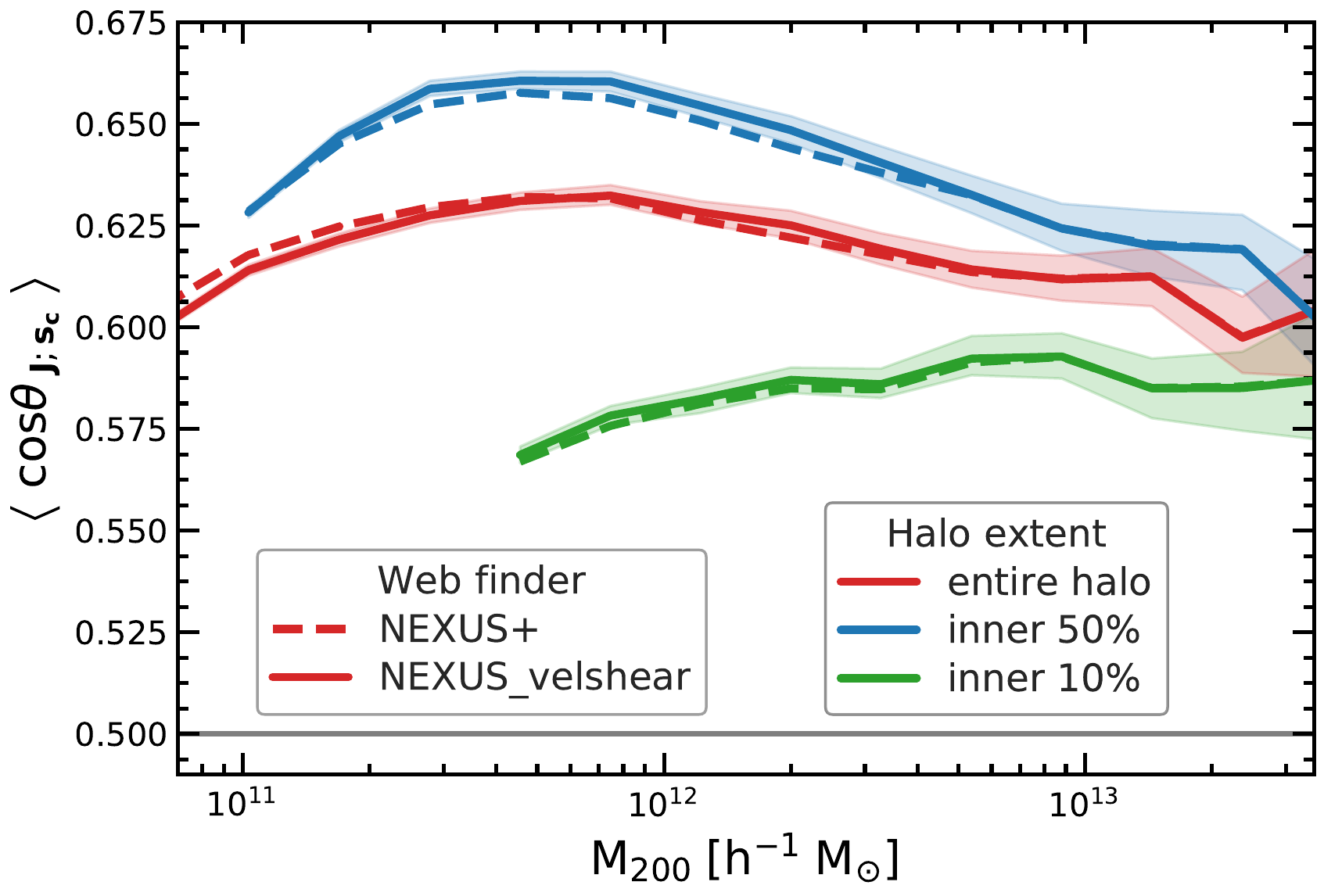}}
	\vskip -0.3cm
	\caption{ The alignment between the shape and spin of haloes. \textit{Top panel:} the distribution of the alignment angle, $\cos\theta_{\mathbf{s_c} ; \mathbf{J}}$, between halo short axis and halo spin for haloes of three different masses: low mass, $M_{200} = (5-9) \times 10^{10}$ \massUnit{}, intermediate mass, $(3-5) \times 10^{11}$ \massUnit{}, and high mass, $M_{200} = (3-5) \times 10^{12}$ \massUnit{}. 
 	\textit{Bottom panel:} the median alignment angle, $\left\langle \cos\theta_{\mathbf{s_c}; \mathbf{J}} \right\rangle$, between halo minor axis and halo spin as a function of halo mass. In both panels we show only filament haloes, which are the subject of this paper.
 	}
	\label{fig:haloshape-spin}
\end{figure} 

\subsection{Halo shape - halo spin alignment}
Several physical effects contribute to a preference of haloes to rotate along an axis that is close to their minor axis. First, the strong correlation between inertia tensor of a peak and the tidal field implies a spin direction 
that is closely aligned to the peak's minor axis \citep[see][]{lee2000}. Secondly, the peak collapses fastest along its shortest axis
\citep{icke1973}, and, moreover, a rotating self-gravitating isolated object is expected to contract to a larger extent along its rotation axis. 

\autoref{fig:haloshape-spin} shows that there is indeed a preference for the minor axis of a halo to be oriented along the spin axis.
Nonetheless, this tendency is rather weak \citep{Bailin2005,bett2007}. The distribution of alignment angles between rotation axis and the minor axis is very broad and, although it shows some dependence on halo mass, this variation is neither substantial nor systematic. 
Furthermore, the strength of the spin - minor axis alignment depends weakly on the radial extent of the halo: the inner 50\% of the halo is characterized by a stronger alignment than the inner 10\%, while, in the outer regions, the trend reverses, with the entire halo having a lower spin - minor axis alignment \citep[see][]{Bailin2005}. 

\begin{figure}
  	\mbox{\hskip -0.4truecm\includegraphics[width = 1.05\columnwidth]{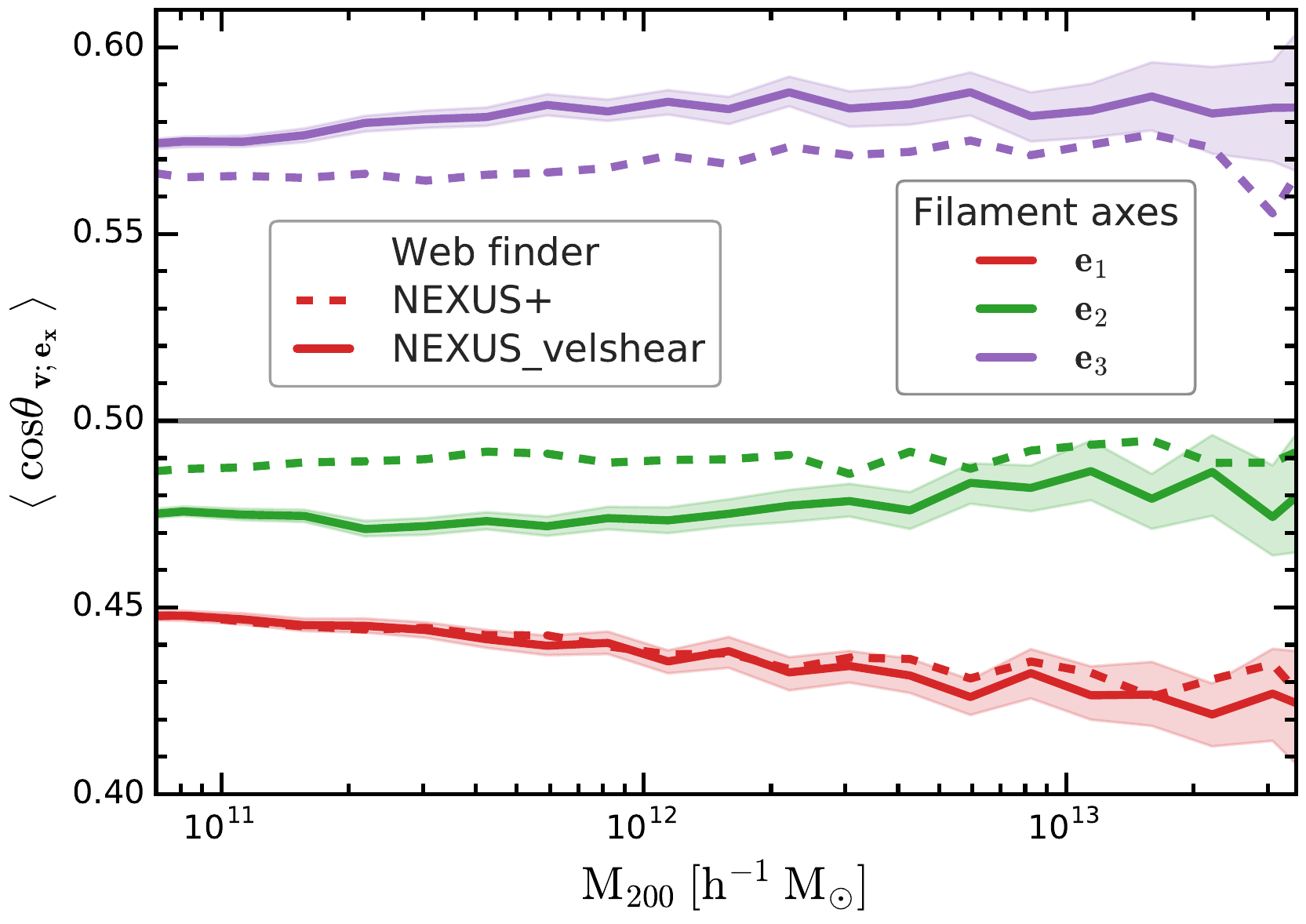}}
    \vskip -.3cm
	\caption{ The median alignment angle, $\left\langle \cos\theta_{v;e_x} \right\rangle$, between the halo bulk velocity and the preferential axes of filaments. It shows the alignment with the filament orientation, $\mathrm{e}_3$  (purple lines), and with the principal directions perpendicular to the filament, $\mathrm{e}_1$ and $\mathrm{e}_2$ (red and green lines, respectively). All haloes irrespective of their mass move preferentially along the direction of the filament and they show a coherent accretion inflow along the cross-sectional plane of filaments.
    }
	\label{fig:median_velocity}
\end{figure}

\section{Filamentary Accretion Flows \& Spin Flips}
\label{sec:accretion}
\revised{Numerical simulations reveal a complex mass dependence of the halo spin-filament alignment,} with the spin of high-mass haloes close to perpendicular to their host filament, while the spin of low-mass haloes showing the opposite result, being preferentially parallel to their host filament. The transition halo mass between the two configurations, i.e. preferentially perpendicular at high masses to preferentially parallel at low masses, is known as the spin flip mass. We found that the spin flip mass depends strongly on the nature of filaments, showing more than an order of magnitude variation between the thinnest and thickest filaments (see section ~\ref{sec:filament thickness}). In other words, same mass haloes are more likely to have perpendicular spin orientations with respect to their host filament if they are embedded in thinner filaments.  
The conventional TTT \revised{\citep[however see the latest predictions of][]{codis2015}} does not explain this trend, and \revised{previous works have} argued that the key element for understanding the spin flip phenomenon is the 
anisotropic accretion of mass and substructures along filaments \citep[see also][]{libeskind2013,Welker2014,Wang2017a,Wang2018a}. \revised{Our analysis agrees with this interpretation and, as we discuss shortly, provides additional evidence to support it.}

To obtain a detailed picture of the level of mass flow anisotropy in and around filaments, we use haloes as flow tracers and investigate the orientation of halo velocities with respect to the filaments in which they reside. To this end, we calculate the alignment angles between the halo bulk velocity and the three orthogonal directions that determine the principal axes of filaments: $\mathbf{e}_3$, which is the orientation of the filament ridge, and $\mathbf{e}_1$ and $\mathbf{e}_2$, which give the principal directions perpendicular to the filament.  

\autoref{fig:median_velocity} shows the median of the alignment angle between halo velocity and the three principal axes of filaments, as a function of halo mass. Overall, we find that the haloes flow preferentially parallel along the filament \citep{romero2014}.  While the velocity component along the filament represents the major share of the flow, the perpendicular components are a combination of the substantial level of mass accretion on to the filament and the velocity dispersion in the filament cross-sectional plane. Also, no bias is seen in flow properties between high-mass and low-mass haloes. The slight differences between \nexus{} and \Nexus{}$\_$velshear results may be ascribed to the fact that the \nexus{} filament population also includes dynamically weaker tendrils, with the haloes inside the tenuous filaments being slightly less likely to flow parallel to the filaments. 

\begin{figure}
	\includegraphics[width = .98\columnwidth]{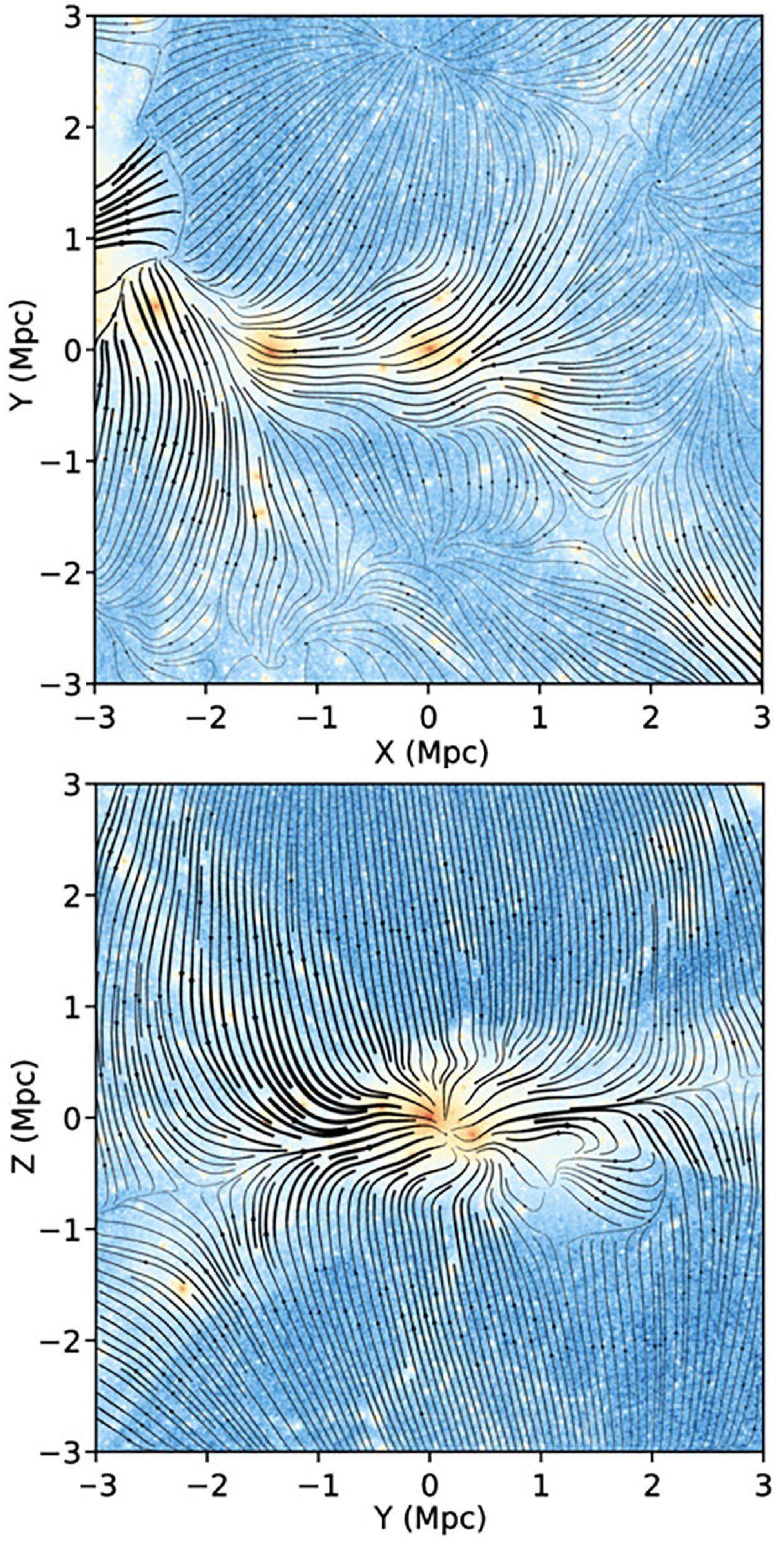}
    \vskip -.2cm
	\caption{Flow pattern along a filament in the cosmic web. The image shows the flow-lines in two mutually perpendicular planes centred on a galaxy sized halo in the \textsc{cosmogrid} simulation \citep[see e.g.][]{ishiyama2013}. The planes are defined by the eigenvectors of the inertia tensor of the mass distribution on a 2~Mpc scale. The first panel show the flow along the filament in which the halo is embedded, while the second panel offers a cross-section view, showing the accretion flow onto the filament.}
	\label{fig:cosmogrid}
\end{figure}
%%%%%%%%%%%%%%%%%%%%%%%%%%%%%%%%%%%%%%%%%%%%%%%%%%%%%%%%%
\begin{figure} % tikzpicture for halo accretion
\begin{center}
\begin{tikzpicture}[scale = 0.5]

\begin{scope}[yshift = 11cm]
\node at (-6,5) {{\LARGE Accreting Halo}};
\node at (0,0) {{\LARGE Halo}};
\draw [ultra thick] (0,0) circle [radius = 2];
\draw [line width=6, red, ->] (0,-4) -- (0,-2.5);
\draw [line width=6, red, ->] (0,4) -- (0,2.5);
\draw [line width=6, red, ->] (-4,0) -- (-2.5,0);
\draw [line width=6, red, ->] (4,0) -- (2.5,0);
\draw [line width=6, red, ->] (-3,3) -- (-2,2);
\draw [line width=6, red, ->] (3,3) -- (2,2);
\draw [line width=6, red, ->] (3,-3) -- (2,-2);
\draw [line width=6, red,->] (-3,-3) to (-2,-2);
\fill[pattern=north west lines,pattern color=red] (2,-1) rectangle (8,1);
\fill[pattern=north west lines,pattern color=red] (-2,-1) rectangle (-8,1);
 \fill[pattern=north west lines,pattern color=red] (1,2) to (2,1) to (5,4) to (4,5) to (1,2);
\draw (-8,-4.5) to (8,-4.5);
\end{scope}
%%%%%%%%%%%%%%%%%
\begin{scope}[yshift = -1cm]
\node at (-6,6.5) {{\LARGE Stalled Halo}};
\node at (0,5) {{\Large Absolute reference frame}};

\draw [ultra thick] (0,0) circle [radius = 2];

% the filament
\fill[pattern=north west lines,pattern color=red] (-7,4) to (0,4) to (0,2) arc (90:270:2) to (0,-4) to (-7,-4) to (-7,4);
\fill[pattern=north west lines,pattern color=red] (7,4) to (0,4) to (0,2) arc (90:-90:2) to (0,-4) to (7,-4) to (7,4);

\draw [line width=6,blue,->] (-3,0) -- (-6.5,0);
\draw [line width=6,blue,->] (1,0) -- (-1,0);
\draw [line width=6,blue,->] (5,0) -- (4,0);
 
\draw [line width=6,blue,->] (-3,2.75) -- (-6.5,2.75);
\draw [line width=6,blue,->] (1,2.75) -- (-1,2.75);
\draw [line width=6,blue,->] (5,2.75) -- (4,2.75);

\draw [line width=6,blue,->] (-3,-2.75) -- (-6.5,-2.75);
\draw [line width=6,blue,->] (1,-2.75) -- (-1,-2.75);
\draw [line width=6,blue,->] (5,-2.75) -- (4,-2.75);

\end{scope}
%%%%%%%%%%%%%%%%%
\begin{scope}[yshift = -11cm]
\node at (0,5) {{\Large Halo reference frame}};
\draw [ultra thick] (0,0) circle [radius = 2]; 
% the filament
\fill[pattern=north west lines,pattern color=red] (-7,4) to (0,4) to (0,2) arc (90:270:2) to (0,-4) to (-7,-4) to (-7,4);
\fill[pattern=north west lines,pattern color=red] (7,4) to (0,4) to (0,2) arc (90:-90:2) to (0,-4) to (7,-4) to (7,4);
%Arrows on top and bottom 
\draw [line width=6, red, ->] (0,-4) -- (0,-3);
\draw [line width=6, red, ->] (0,4) -- (0,3);
\draw [line width=6, red, ->] (2,-4) -- (1,-3);
\draw [line width=6, red, ->] (-2,4) -- (-1,3);
\draw [line width=6, red, ->] (-2,-4) -- (-1,-3);
\draw [line width=6, red, ->] (2,4) -- (1,3);

% Arrows left and right
\draw [line width=6,blue,->] (-3,0) -- (-5,0);
\draw [line width=6,blue,<-] (5,0) -- (3,0);
\draw [line width=6, blue,->] (-3,2.5) -- (-5,2.5);
\draw [line width=6,blue,<-] (5,2.5) -- (3,2.5);
\draw [line width=6,blue,->] (-3,-2.5) -- (-5,-2.5);
\draw [line width=6,blue,<-] (5,-2.5) -- (3,-2.5);
\end{scope}

\end{tikzpicture}
\end{center}

	\caption{ A schematic representation of the mass distribution around and the infall patterns of accreting and stalled haloes. In each panel, the circle represents the halo, the raster pattern indicates the position and extent of filaments, and the red and blue arrows show the direction and magnitude of the average velocity flow.
    Accreting haloes (top panel) are embedded in filaments that are thin compared to their radius and accrete matter from all directions. Due to the higher density of filaments, the majority of mass growth is due to infall along filaments and leads to a net increase in halo spin perpendicular to the filament.
    Stalled haloes typically reside in thick filaments with large velocity gradients (centre panel), which are indicated by longer arrows on the left-hand side of the panel than on the right-hand side. When viewed in the reference frame of the stalled halo (bottom panel), the surrounding matter flows away along the filament and infall can only take place from directions perpendicular to the filament. The inhomogeneities in the distribution of accreted mass impart a net spin that points preferentially along the filament.
	} 
	\label{fig:halo_accretion_schematic}
\end{figure}

Secondary accretion \citep{bertschinger1985} represents the key for understanding how the anisotropic filamentary shear inflow is
responsible for the observed spin flip of low mass galaxies \citep[see eg.][]{haarlem1993}. \autoref{fig:cosmogrid} provides
an impression of the typical flow patterns along and perpendicular to a filament. It shows the flow-lines in two perpendicular planes
centred on a galaxy-sized halo in the \textsc{cosmogrid} simulation \citep{ishiyama2013}.  
Compared to \autoref{fig:median_velocity}, which describe the flows of individual tracers, the flow lines characterize the mean flow at each point.
The flow in and around the filaments is a combination
of shear and divergent flow, which are themselves due to a combination of the outflow from neighbouring voids and the flow along the filament. In general, haloes accreted mass both along the filaments (e.g. see the top panel in \autoref{fig:cosmogrid}) and also perpendicular to their host filament. The former tends to preferentially increase the halo spin component that is perpendicular on the filament, while the latter increase the spin component parallel to the filament. Which of the two dominates depends on the balance between accretion along and perpendicular to the host filament. As we will discuss shortly, this balance depends on a combination of the mass and the local neighbourhood of a halo. 

The acquisition of halo angular momentum through secondary accretion results from the transfer of orbital angular momentum, which yields a non-zero residual spin for the halo. It is due to anisotropies in the distribution of accreted mass, such as spatial inhomogeneities (e.g. filamentary infall) as well as mergers with matter clumps. The majority of large and rapid changes in halo spin are caused by mass changes, minor mergers and flyby encounters, and not by major mergers \citep{bett2012,Bett2016,Contreras2017}.

\citet[][see also \citealt{Romano-Diaz2017,Garaldi2018}]{Borzyszkowski2017} describes how haloes can be divided in two groups: haloes that are still accenting and those that have stopped most of their mass accretion, so called stalled haloes. 
The large-scale mass distribution and velocity flow patterns around these two halo types are illustrated in \autoref{fig:halo_accretion_schematic}.
Accreting haloes typically consists of haloes that are the main perturber in their neighbourhood, they sit at the intersection of several filaments and accrete preferentially along these filaments. Thus, accreting haloes are expect to have their spin preferentially perpendicular on their host filament. The latter group of stalled haloes are found in regions of strong external tidal field, for example they are embedded in filaments much thicker than the halo size, and mostly accrete from directions perpendicular on their host filament orientation (see Figure 10 of \citeauthor{Borzyszkowski2017} for a visualization of the striking contrast between accreting and stalled haloes). Thus, the stalled haloes have spins mostly parallel to their host filament. The fraction of accreting versus stalled haloes is mass dependent, with the fraction of accreting haloes increasing rapidly with halo mass.

The dichotomy in terms of spin - filament alignment between accreting and stalled haloes provides a natural explanation for the trends we found in this work. While accreting haloes dominate the population of high-mass haloes, the converse is true for low-mass haloes. This suggests that the spin - filament alignment should vary smoothly from being preferentially perpendicular at high masses to preferentially parallel at low masses, which is exactly the trend we measure in \autoref{fig:median_ang_mom}. 

The fraction of accreting haloes varies with redshift and, at fixed halo mass, it was larger at higher redshift. It suggest that the spin flip mass should decrease with redshift, which is in very good agreement with previous studies \citep{codis2012,Wang2018a}. 
Furthermore, most of the recently accreted mass settles in the outer regions of the halo \citep{wang2011}, with the inner regions mostly maintaining the spin of the halo when they were assembled. Thus, the spin of the inner halo regions should be perpendicular to the host filament to a larger degree than the outer halo, which is what we observe in Figures~\ref{fig:multipanel_ang_mom_velshear} and \ref{fig:median_ang_mom}.

The fraction of accreting haloes depends on environment and at fixed halo mass is smaller in regions with strong external tidal fields, such as inside and around massive filaments (the tidal field is what leads to the formation of these filaments). Thus, same mass haloes should have a higher degree of parallel spin - filament configurations if the haloes are embedded in thicker filaments, which is what we find in \autoref{fig:filament thickness}. This trend also leads to the spin flip mass varying with filament thickness, with the transition mass being higher in thicker filaments.

%%%%%%%%%%%%%%%%%%%%%%%%%%%%%%%%%%%%%%%%%%%%%%%%%%%%%%%%%%%%%%

\section{Conclusions \& Discussion}
\label{sec:discussion}
In this study we have carried out a systematic investigation of the orientation of the spin, shape and peculiar velocity of haloes relative to the
filaments in which they are embedded. Our goal has been to elucidate one of the most outstanding manifestations of environmental influences on halo and galaxy formation, by specifically focussing on the connection between the generation of angular momentum on galactic scales \citep{lee2000,porciani2002,aragon2007, aragonPhd2007,jones2010, schaffer2009} and the dynamics of the
large-scale cosmic web \citep{bond1996,vdw2008,cautun2014}. 
\revised{Previous works, starting with \cite{aragon2007} and \cite{hahn2007}, have shown how cosmological simulations show a complex halo spin - filament alignment, with the mean orientation of halo spin changing from largely perpendicular for high-mass haloes to preferentially parallel for low-mass haloes, with the transition mass, typically ${\sim}10^{12}$\massUnit{}, known as the \emph{spin flip} mass.}

% According to TTT \citep{hoyle1949,peebles1969,white1984}, halo spins are orientated preferentially perpendicular to their host filaments \citep{lee2000,porciani2002}. The halo spin is largely imparted during the linear evolutionary phase of structure formation and thus TTT should provide a good description of how haloes acquire their spin. However, as pointed out by \cite{aragon2007} and almost simultaneously by \cite{hahn2007},
% cosmological simulations reveal a richer behaviour than that implied by the TTT prediction. The mean orientation of halo spin changes from largely perpendicular for high-mass haloes to preferentially parallel for low-mass haloes, with the transition mass being known as the \emph{spin flip} mass. \revised{\cite{codis2015} have shown how this dichotomy is expected in the TTT framework }

To study halo - filament alignments, we have used one of the largest cosmological N-body simulations available, P-Millennium. It has an impressive dynamic range, combining a large volume with a very high mass resolution, which makes it ideally suited for investigating the connection between halo formation and the large-scale structure. The halo - filament alignment can be a subtle and mass-dependent effect, even more so for the halo spin - filament alignment, and studying it needs a large number of haloes spanning a wide mass range. P-Millennium fulfils both requirements, having no less than 7.5 million well resolved haloes that span more than three orders of magnitude in halo mass. The large volume of P-Millennium is also critical, since it contains both the large-scale tidal forces responsible for the generation of halo spin and the diversity of environments in which haloes reside.

We have identified the filamentary network using the \Nexus{} multiscale
morphology filter \citep{aragon2007MMF,cautun2013,cautun2014}. To obtain further insight into the dynamical factors affecting the halo - filament alignments, we have studied the filament populations selected by two different versions of the \Nexus{} formalism.
The first, \nexus{}, extracts filaments on the basis of the density field and identifies a broad range of the filament spectrum,
from prominent arteries, which dominate the dynamics of the cosmic web, to tenuous tendrils, which branch off major arteries and reside in underdense regions. The second formalism, \Nexus{}$\_$velshear, is based on the velocity shear field; it mostly identifies the dynamically dominant filaments and typically assigns them a larger width than \nexus{}. As we discuss shortly, the contrast between \nexus{} and \Nexus{}$\_$velshear reveals key information about the processes behind the spin - filament alignment and its dependence on local environment.

In the current study we focus on the orientation of the spin, the shape and the peculiar velocities of the dark matter haloes relative to the filament in which they are embedded at the present epoch, $z=0$. The properties of the dark component have the advantage of being mostly determined purely by gravitational effects rather than the complex physical processes affecting the baryonic component. In subsequent studies, we will perform a detailed comparison of halo-by-halo evolution as a function of cosmic web environment, and we will investigate the alignments of the stellar and gas components of galaxies in the \textsc{eagle} project \citep{schaye2015}.

\bigskip
\noindent The following points summarize the main results of this paper concerning the alignments of halo spins and shape with their host large-scale filament:

\subsection*{1. Halo spin orientation}
In this study we have characterized how the spin of haloes is oriented with respect to their host filament to an unprecedented precision and over three orders of magnitude in halo mass. Overall, the orientation of the halo spin follows a wide distribution with a small, but statistically significant, preferential alignment with the direction of the filament (see Figures \ref{fig:multipanel_ang_mom_velshear} \& \ref{fig:median_ang_mom}). There is a clearly discernible systematic trend in the median of the spin orientation: high-mass haloes tend to have their spin perpendicular to their host filament, while low-mass haloes tend to have their spin parallel to their host filament \citep{ aragon2007,hahn2007,hahn2010,codis2012,trowland2013,romero2014}. We have found a transition mass of ${\sim}5 \times 10^{11} ~ h^{-1}\rm{{M_\odot}}$ between perpendicular and parallel alignments, which is in good agreement with the wide range of ``spin flip" masses, around $0.5$ to $5 \times 10^{12} ~{h^{-1}} M_{\odot}$, reported by previous studies.

\newcommand{\skipbefore}{\vskip .4cm}
\newcommand{\skipafter}{\vskip .2cm}

\skipbefore
\noindent\emph{1.1 Dependence on web finder}
\skipafter

\noindent Both the spin - filament alignment as well as the spin flip mass show a small, but systematic dependence on the method used to identify the cosmic web. For same mass haloes, the halo spin tends to be closer to perpendicular on \nexus{} filaments than on the \Nexus{}$\_$velshear ones. This is manifested as slightly different values for the spin flip mass, which we have found to be $4$ and $6\times10^{11} ~{h^{-1}} M_{\odot}$ for \nexus{} and \Nexus{}$\_$velshear filaments, respectively.
At high mass, the discrepancy is explained by the haloes themselves influencing the surrounding velocity shear field and thus limiting the extent to which \Nexus{}$\_$velshear can recover the direction of large-scale filaments. For masses lower than $10^{12}~{h^{-1}} M_{\odot}$, the discrepancy between the two web finders is mostly due to \nexus{} identifying a population of haloes associated to filamentary tendrils in low-density regions, which tend to have more perpendicular spin orientations.

Interestingly, for haloes with $M_{200}<10^{12}~{h^{-1}} M_{\odot}$, the differences in alignment between haloes in the \nexus{} and \Nexus{}$\_$velshear filament populations disappear when we study the common haloes identified by both web finders as residing in filaments (see \autoref{fig:common halos}). This implies that the discrepancy 
is due to differences in the halo population associated to filaments. Two outstanding differences are that the \nexus{} population contains a significant
fraction of thin filaments that are either branches of major filaments or tenuous tendrils inside underdense
regions. In contrast, the \Nexus{}$\_$velshear filaments consists of mostly the dynamically dominant arteries. As we discuss shortly, the variation between the two filament populations is mostly due to the dependence of the spin - filament alignment on filament properties. In short, the careful
comparison of halo alignments with both filament populations, in relation with the major visual differences
between the populations, casts a new light on the processes involved in the evolution of halo angular momentum and its environmental dependence.

\skipbefore
\noindent\emph{1.2 Dependence on filament properties}
\skipafter

\noindent We have also shown that the spin - filament alignment displays a strong systematic variation with the properties of filaments, in particular on the filament thickness. We have found that haloes of the same mass show a stronger trend to have their spin oriented perpendicular to their host filament if they are embedded in thinner filaments (see \autoref{fig:filament thickness} and \ref{fig:flip_mass_fila_thickness}, and \citealt{aragon2014}). The trend is strong enough to result in more than an order of magnitude variation in spin flip mass, from $0.1\times10^{12}~{h^{-1}} M_{\odot}$ for the thinnest filaments, with diameters below $1~h^{-1}\rm{Mpc}$, to $3.0\times10^{12}~{h^{-1}} M_{\odot}$ for the thickest filaments. The mass density and diameter of filaments shows a tight correlation \citep{cautun2014} and thus we expect that a similar trend would be visible as a function of filament mass density. We note that the multiscale character of \Nexus{} has been instrumental in identifying this trend, since the multiscale approach allows for the simultaneous identification of both thin and thick filaments.

The strong variation with filament properties explains many puzzling results of previous studies. For example, the discrepancy between alignment strengths and the spin flip values reported by previous studies is due to the variation in the characteristics of filaments identified by different web finders \citep[for a comparison of many web finders see][]{libeskind2018}. The same holds for the differences between the \nexus{} and \Nexus{}$\_$velshear methods which we have studied here. The dependence of spin orientation on filament thickness also explains the variation of spin - filament alignment on the smoothing scales used to identify filaments. For single scale web finders (which is not the case for the \Nexus{} formalism) increasing the smoothing scale  leads to identifying mostly thicker filaments \citep[see e.g.][]{cautun2013}, and thus results in halo spins that tend to be closer to perpendicular to their host filament, explaining the results of \citet{codis2012} and \citet{Wang2018a}.

\skipbefore
\noindent\emph{1.3 Dependence on halo radial extent}
\skipafter

\noindent We have studied for the first time how the spin - filament alignment depends on the radial position within the halo. For Milky Way mass haloes and below, the inner halo spin is more likely to be oriented perpendicular to filaments than the spin of the entire halo (see \autoref{fig:median_ang_mom}). The galaxies are more strongly aligned with the inner halo and thus, when compared to their entire host halo, we expect that galaxy spins are more likely to orient perpendicularly on their host filaments. This hypothesis is in good agreement with an upcoming analysis of galaxy spin - filament alignments in the \textsc{eagle} galaxy formation simulation (Ganeshaiah Veena et al., in prep.). For haloes more massive than ${\sim}5\times10^{12}~{h^{-1}} M_{\odot}$, the converse is true and the inner halo spin is less aligned with the host filament than the whole halo spin. 

Most of the recent mass accretion of a halo, especially if it is due to smooth accretion and minor mergers, is deposited in the outer regions and leaves the inner halo structure mainly intact \citep{wang2011}. Thus, by calculating the spin of different inner halo regions we have a window into the time evolution of halo spin. This suggests that the progenitors of haloes with present day mass, $M_{200}<2\times10^{12}~{h^{-1}} M_{\odot}$, had spins which were oriented perpendicular to filaments to a larger extent than their present day descendants. Thus, in low-mass haloes, recent accretion leads to a reorientation of halo spins to point preferentially along the filament. This trend is reversed for haloes more massive than ${\sim}5\times10^{12}~{h^{-1}} M_{\odot}$, whose progenitors spins were less likely to be oriented perpendicular to filaments than their present day descendants. Thus, in high-mass haloes, recent accretion leads to an increase in the halo spin tendency to be perpendicular on the host filament.

\subsection*{2. Halo shape orientation}
When considering the orientation of the halo's shape, i.e. of the inertia tensor, we find similar alignment results as for the halo spin. While the distribution of orientation angles is broad, we have found clear systematic alignment trends that are stronger than in the case of the halo spin - filament alignment (see \autoref{fig:median_shape_alignment}). For all mass ranges, the major axis of the halo points preferentially along its host filament ridge. On the other hand, the minor axis tends towards a perpendicular orientation with respect to the host filament. The alignment of both major and minor axes is larger for more massive haloes, which is most likely a manifestation of recent accretion processes that vary with halo mass. When analysing different halo radial ranges, we have found that the shape of the inner halo is less well aligned with the host filament than the shape of the full halo. The different behaviour of the spin - filament and shape - filament alignments is due to the weak alignment between halo spin and halo shape, with the spin showing a surprisingly wide range of orientations with respect to the shape minor axis (see \autoref{fig:haloshape-spin}).

\subsection*{3. Secondary accretion and filament flows}
The results we have presented here \revised{reinforce and provide additional evidence that secondary anisotropic accretion is a major driver} for the late time acquisition of halo spin and its orientation with respect to the large-scale filaments in which the haloes are embedded \citep{libeskind2013,Welker2014,codis2015,Laigle2015,Wang2018a}.  
%The late time acquisition of spin is associated to secondary accretion \citep[see e.g.][for a thorough discussion of the physics of secondary accretion]{bertschinger1985} whose dominant direction with respect to filaments depends on average on both halo mass and local environment. 
The change in halo spin is a residual effect due to the transfer of orbital angular momentum from accreted clumps and from anisotropies in the smoothly accreted component.
Low-mass haloes are more likely to accrete mass along directions perpendicular to their host filament, which results in their spins orienting preferentially along the filament spine. In contrast, high-mass haloes are more likely to accrete mass along their host filament, which ends up enhancing the tendency of their spin to be perpendicular to the host filament. Furthermore, haloes of the same mass are more likely to accrete mass along their host filament if they are embedded in thinner filaments.

This hypothesis is supported by the work of \citet{Borzyszkowski2017} which demonstrated a strong correlation between large-scale environment and the preferential directions of accretion. This is best understood in terms of halo types at opposite sides of the formation path spectrum: accreting versus stalled haloes. 
The typical mass distribution and velocity flow patterns around these two halo types are illustrated in \autoref{fig:halo_accretion_schematic}.
Accreting haloes represent the dominant mass concentration in their neighbourhood, are found at the intersection of several filaments, whose diameters are typically smaller than the halo size, and accrete most of their mass along these filaments. This represents the typical filamentary accretion picture, where filaments transport mass to the haloes at their endpoints. \citeauthor{Borzyszkowski2017} refer to these objects as accreting haloes since they have a large growth rate. 

Stalled haloes, on the other hand, are embedded in a strong external tidal field, such as inside a massive filament between two clusters, and, as their name suggests, have low growth rates. The growth of these haloes takes place through accretion mainly from directions perpendicular to their host filament, and thus their spin becomes more parallel to the filament as time goes by. To understand this, lets consider a low-mass halo embedded in a prominent filament between two massive clusters. Since the filament acts as a highway for channelling mass into the clusters at its endpoints, it is characterized by a large velocity gradient along its spine. This inhibits the growth of low-mass haloes embedded in the filament since, in the halo reference frame, the velocity gradient manifests itself as mass flowing away from the halo in both directions along the filament. If the halo has a low mass, it cannot overcome the velocity gradient and thus cannot accrete significantly along the filament direction, and can grow only by accreting mass from directions perpendicular to the filament. 

This hypothesis matches the results presented in this work as well as those of previous literature. 
The formation time of haloes depends on their mass, with massive haloes having formed only recently \citep[see e.g.][]{Davis1985,Hellwing2016}.
Thus, the fraction of accreting haloes increases with halo mass: from low-mass haloes that are mostly of the stalled type to high-mass haloes that are mostly of the accreting type \citep[e.g.][]{Ludlow2013}. 
This explains why the spin - filament alignment changes from preferentially parallel at low masses to preferentially perpendicular at high masses. 
The fraction of accreting haloes varies with redshift, with haloes of a given mass being more likely to be of the accreting type at high redshift. 
This describes why the spin - filament alignment changes with redshift, with the spin flip taking place at lower halo masses at high redshift. 
Furthermore, the fraction of accreting haloes is larger in thin filaments, like filamentary tendrils in underdense regions, since those filaments form in regions without massive haloes \citep{cautun2014}.
This observation reveals why haloes of the same mass are more likely to have their spins oriented perpendicularly when embedded in thinner filaments.

\bigskip

While the present study has concentrated on the present epoch, in an accompanying study we will investigate in detail the
build-up of halo angular momentum as haloes form and evolve during their complex hierarchical growth. We will investigate the 
processes that accompany the accretion on to and along filaments and in how far they augment the angular momentum imparted by tidal torquing during the early phases of structure formation. The redshift evolution will elucidate other aspects likely to affect the spin - filament alignment, such as the impact of the birth location of haloes (e.g. proto-haloes formed in voids versus filaments) and the role of their migration path. Tracing the detailed halo history will also reveal any differences in the evolution of haloes in various filament types, e.g. prominent versus minor filaments. 

For a full understanding of the impact of the cosmic web on the formation and evolution of galaxies, dark matter only
simulations as the one studied here are not sufficient. Gas, radiation and stellar (evolution) processes determine to a large extent the
outcome and morphology of the emerging galaxies, and the rotation properties of their gas and stellar content. \revised{For example, some models suggest that a significant fraction of the angular momentum of low mass galaxies is due to the accretion of cold gas streams, which can penetrate deeper in the halo than dark matter filaments \citep{Dekel2006, Pichon2011,Danovich2015,Stewart2017}. Before infall into the halo, gas and dark matter acquire angular momentum through the same processes, such as torquing due to the surrounding matter distribution. However, once the gas streams enters the inner fractions of the haloes, their angular momentum can change due to non-linear torques, dissipation, disc instabilities and feedback processes \citep[e.g. see][]{Danovich2015}.
Such processes might lead to a different galaxy spin - filament alignment than the one found for the inner region of haloes in dark matter only simulations.}  
%According to \citeauthor{Danovich2015}, the angular momentum acquisition in galaxies is a four phase process. Initially, during the linear phase of TTT, cold gas streams acquire angular momentum from torques due to the surrounding matter distribution. As these streams advance into the halo, they gain angular momentum from the outer regions of the halo. Finally as the streams enter the inner fractions of the halo, they gain most of their angular momentum through non-linear torques and dissipation. In the last phase, the galaxy spin is determined by disc instabilities and feedback. 
%Although we would expect the galaxies also to follow the same spin-filament alignment as the halos, there could be differences in their alignment pattern due to the last two phases of the spin acquisition processes mentioned above. }
In order to assess how far the
spin properties of the dark matter haloes are transferred to the gas and stars of the galaxy, we need to analyse galaxy formation simulations. 
In an accompanying paper, we will study spin - filament alignments in the \textsc{eagle} project \citep{schaye2015}. 
It will be a step towards understanding how the angular
momentum of gas and stars in galaxies is related to that of the parent dark halo, seeking to extend earlier studies along these
lines \revised{\citep[eg.][]{hahn2010,Dubois2014,Welker2014,Zavala2016}}.

\section*{Acknowledgements}
\revised{We are very grateful to the referee(s) for their constructive inputs, which helped us improve the paper substantially.}
PGV thanks the Institute for Computational Cosmology (ICC) in Durham for their support and hospitality during 2 long work visits during which
the major share of the work for this study was carried out. 
Also RvdW thanks the ICC for its hospitality and support during 2 short work visits. MC and CSF were supported by Science and Technology Facilities Council (STFC) [ST/P000541/1].
%; and by an ERC Advanced Investigator grant COSMIWAY [GA 267291].   
ET acknowledges the support by the ETAg grants IUT26-2, IUT40-2, and by the European Regional Development Fund (TK133, MOBTP86). 
This work used the DiRAC Data Centric system at Durham University, operated
by the Institute for Computational Cosmology on behalf of the STFC
DiRAC HPC Facility (\url{www.dirac.ac.uk}). This equipment was funded
by BIS National E-infrastructure capital grant ST/K00042X/1, STFC
capital grants ST/H008539/1 and ST/K00087X/1, STFC DiRAC Operations
grant ST/K003267/1 and Durham University. DiRAC is part of the
National E-Infrastructure. 

\bibliographystyle{mnras}
\bibliography{bibliography}

%%%%%%%%%%%%%%%%%%%%%%%%%%%%%%%%%%%%%%%%%%%%%%%%%%

\bsp	
\label{lastpage}

\end{document}